\documentclass{aa}  
\usepackage{graphicx}
\usepackage{subcaption}
\usepackage{multicol}
\usepackage{multirow}
\usepackage{natbib}

\usepackage{txfonts}

\usepackage[pdfencoding=auto,psdextra]{hyperref}
\hypersetup{
    colorlinks=true,
    linkcolor=red,
    anchorcolor=red,
    filecolor=magenta,      
    urlcolor=magenta,
    citecolor=blue
}
\usepackage[switch]{lineno}

\begin{document} 

   \title{A pilot sample of Planck-selected strongly lensed sub-mm galaxies: NOEMA observations and physical characterisation}

   \author{L. Trobbiani
          \inst{1,2,3}
          ,
          M. Bonato\inst{2,3}\fnmsep
          ,
          I. Baronchelli\inst{2,3}
          ,
          G. De Zotti\inst{4}
          ,
          M. Negrello\inst{5,6}
          ,
          M. Giulietti\inst{2}
          ,
          S. Berta\inst{7}
          ,
          \\C. Vignali\inst{1,8}
          \and
          M. Massardi\inst{2,3}
          }

    \institute{Dipartimento di Fisica e Astronomia "A. Righi", Alma Mater Studiorum Universit\`a di Bologna, Via Gobetti 93/2, I-40129 Bologna, Italy
        \and
             INAF, Istituto di Radioastronomia, Via Gobetti 101, Bologna I-40129, Italy;
        \and
             ALMA Regional Centre, Via Gobetti 101, Bologna I-40129, Italy;
        \and 
            INAF, Osservatorio Astronomico di Padova, Vicolo dell'Osservatorio 5, I-35122, Padova, Italy
        \and 
            Liceo Scientifico Paritario "Don Bosco", Via del Ghirlandaio 40, I-50121, Firenze, Italy
        \and 
            Liceo delle Scienze Umane Paritario "Giorgio La Pira", Via del Ghirlandaio 40, I-50121, Firenze, Italy
        \and
            Institut de Radioastronomie Millimétrique (IRAM), 300 Rue de la Piscine, 38400 Saint-Martin-d’Hères, France
        \and
            INAF, Osservatorio di Astrofisica e Scienza dello Spazio di Bologna, Via Gobetti 93/3, I-40129 Bologna, Italy
}

   \date{Received March 12, 2026; Accepted July 15, 2026}

  \abstract
  {}
   {The extreme brightness of high-$z$ strongly lensed galaxies detected by Planck surveys, thanks to their exceptionally large gravitational magnifications, offers a unique opportunity to investigate in extraordinary detail their structure and kinematics during their active star-formation phase. As a step in this direction, we present and analyse NOEMA imaging and spectroscopic data for four lensed high-z galaxy candidates. }
   {We performed continuum and line imaging of the sources at 2 and 3 mm bands down to $1''.2$ spatial resolution and 40 $\rm{km}\,{\rm s}^{-1}$ spectral resolution, reconstructed and fitted the line profiles, and produced moment maps of the line emission to investigate the spatial distribution and kinematics of the molecular gas. We also carried out the gravitational lens modelling for one of the sources.}
   {The continuum images showed multiple components for at least two of the sources, strongly supporting the strong lensing scenario. We detected with high S/N ratios two CO lines for all sources, at CO(3--2), CO(4--3) and CO(5--4) transitions; for one source, we also detected the [CI](1--0) line. We derived accurate spectroscopic redshifts $2.3 \lesssim z \lesssim 3.3$, with $1\sigma$ uncertainties $\delta z \approx 10^{-4}$ in redshift. All four sources lie well above the CO line luminosity--linewidth relation for unlensed galaxies, providing independent confirmation of their strongly lensed nature. Three sources exhibit broad (FWHM$\gtrsim 400\,{\rm km}\,{\rm s}^{-1}$), double-peaked line profiles and substantial velocity gradients, while the last one shows relatively narrow, single-peaked lines and no detectable velocity gradients, possibly indicating a nearly face-on geometry or intrinsically simple kinematics.}
   {}


    \authorrunning{Trobbiani et al.}
    \titlerunning{Strongly lensed SMGs with NOEMA}

   \maketitle

\section{Introduction}\label{Intro}

The Planck all-sky surveys at submillimetre (sub-mm) wavelengths have offered the unique opportunity to discover the brightest strongly lensed high-$z$ star-forming galaxies \citep{Negrello2007}. The combination of large gravitational magnification, by factors $\mu$ of 
up to several tens, with the associated stretching of images produces a boost in the apparent luminosity and area of the sources. This allows one 
to obtain exceptionally detailed information on the internal structure and kinematics of high-$z$ galaxies during their most active, dust-enshrouded 
star-formation phase \citep{rybak15, Massardi18, rybak20, rizzo21}.

The prediction was substantiated by the identification of tens of Planck-detected strongly-lensed galaxies using various approaches \citep{Fu2012, Herranz2013, Canameras2015,  Harrington16, Nesvadba2016, Negrello17, DiazSanchez2017, Harrington21, Trombetti21, Berman2022, Lammers22}. These galaxies were found to have $\mu$ in the range 10--50 and apparent far-infrared (FIR) luminosity up to $\mu L_{\rm FIR}\simeq 3\times 10^{14}\,L_\odot$. Most of these galaxies are found at redshifts in the range $2 \lesssim z \lesssim 4$. The potential of these extreme magnifications for the study of ultra-luminous FIR high-$z$ dusty galaxies is illustrated by ALMA observations of the strongly lensed galaxy PLCK\_$244.8+54.9$ at $z=3.055$ with $\mu=30$ \citep{Canameras2017ALMA}, selected in the Planck maps. With a $0\farcs1$ resolution, these observations reached spatial resolutions of about 60\,pc, similar to or smaller than the size of Galactic giant molecular clouds. Furthermore, CO spectroscopy enabled measurements of the molecular gas kinematics with uncertainties of 40--$50\,\hbox{km}\,\hbox{s}^{-1}$, making possible a direct investigation of massive outflows driven by AGN feedback at high $z$, with predicted 
velocities of $\sim 1000\,\hbox{km}\,\hbox{s}^{-1}$ 
\citep[e.g.][]{KingPounds2015}.

The complexity of the physical processes at work in high-$z$ dusty star-forming galaxies implies that a comprehensive view requires a large and statistically well-defined sample of strongly lensed systems. Motivated by this need, we have initiated a systematic search for strongly lensed sub-mm galaxies in the Planck maps. The strategy adopted for the identification of candidates, together with the selection procedure and the follow-up campaign aimed at securing accurate positions and photometry using ATCA, SCUBA-2, and NIKA2, is presented in a parent paper (Trobbiani et al., in prep.).

In this work, we focus on a pilot subsample of sources identified in that programme. A set of candidates detected with SCUBA-2 was further followed up with the Northern Extended Millimeter Array (NOEMA) in order to obtain high-resolution continuum imaging and spectroscopic observations of molecular gas tracers. These observations allow us to confirm the nature of the sources, derive accurate spectroscopic redshifts, and investigate the spatial distribution and kinematics of the molecular gas, providing a first assessment of the physical properties of the galaxies uncovered by our selection. 

The paper is organised as follows. In Sect.~\ref{sec:selection} we briefly summarise the selection procedure and the definition of the pilot sample. In Sect.~\ref{sec:data} we describe the NOEMA observations and data reduction. The continuum and spectroscopic properties of the four lens candidates are analysed in Sect.~\ref{sec:analysis}. Section~\ref{sec:lens_modelling} is dedicated to the lens modelling of the source for which sufficient data are available. The properties of the sources are discussed in Sect.~\ref{sec:stat_prop}. Finally, in Sect.~\ref{sec:discussion_conclusion} we summarise our results and conclusions.

\section{Sample selection}\label{sec:selection}

The overwhelming majority of extragalactic sources detected by Planck at sub-mm wavelengths are local dusty galaxies \citep[$z\lesssim 0.1$;][]{Negrello2013}. The remaining objects are a mixed bag of Galactic cold clumps \citep{PlanckCollaboration2016cold_cores}, proto-clusters of dusty galaxies \citep{PlanckCollaboration2016highz}, and blazars \citep{PlanckCollaboration2018PCNT}. 

Only a tiny fraction of them are high-$z$ strongly lensed galaxies; their search is further complicated by the fact that a substantial fraction of $\ge 5\,\sigma$ intensity peaks are cirrus or confusion fluctuations. The latter are dominated by high-$z$ galaxies due to the combination of strong cosmological evolution and negative K-correction, and have a highly super-Gaussian tail due to clustering, whose role is maximised by Planck's limited angular resolution (i.e. $\sim 5'.0$) \citep{DeZotti2015}. Thus, confusion peaks may have relatively high signal-to-noise ratios (SNRs) and sub-mm spectral energy distributions mimicking those of high-$z$ galaxies, making them difficult to distinguish from genuine strongly lensed sources. 

A great help to overcome the confusion problem is offered by cross-correlation of Planck catalogues with those from higher resolution surveys at mm/sub-mm wavelengths such as those carried out by \textit{Herschel} \citep{Eales2010, Oliver2012}, by the South Pole Telescope \citep[SPT;][]{Everett2020}, or by the Atacama Cosmology Telescope \citep[ACT;][]{Gralla2020}. Several Planck--detected strongly lensed galaxies have been identified in this way. However, searches are limited to the minor fraction of the sky covered by these surveys.  

The Wide-field Infrared Survey Explorer (WISE) all-sky survey \citep{Wright2010} at near/mid-infrared wavelengths has been successfully exploited to identify strongly lensed galaxies selected from the Planck Catalogue of Compact Sources, most notably by \citet{Berman2022} through a WISE colour--magnitude diagram that efficiently separates lensed DSFGs from other populations. However, even with such a colour selection, the completeness of the resulting sample remains limited. With $\sim 100$ WISE sources within the Planck beam, identifying the correct counterpart is non-trivial, and lensed galaxies whose WISE fluxes are below the detection threshold (in particular at W4), or whose photometry is contaminated by the foreground deflector, are inevitably missed. The loss in completeness is real but difficult to quantify.

The goal of our work was to build a sample as large as possible, and thus, given these limitations, we favoured completeness over selection efficiency. In this spirit, the adopted approach comprised the following steps:

\begin{enumerate}
    \item The sources were predominantly extracted from the second Planck Catalogue of Compact Sources \citep[PCCS2;][]{PCCS2}.
    \item To mitigate contamination from Galactic emission, only sources at high Galactic latitude ($|b| > 20^\circ$) were included. This Galactic latitude cut roughly aligns with the region outside the Planck G65 mask\footnote{Galactic foreground mask designed to retain 65\% of the sky \citep{planck13xv}}.
    \item A flux density threshold of $500$\,mJy at 545 GHz was applied. This threshold corresponds approximately to the 100\% completeness limit according to \citet{Maddox18} and \citet{PCCS2}. 
    \item Removal of contaminants (cross-matching performed within a $3'$ search radius):
    \begin{itemize}
        \item The selected sources were cross-matched with IRAS and radio catalogues to identify and exclude nearby star-forming galaxies and radio sources. Specifically, the IRAS PSC/FSC Combined Catalogue \citep{abrahamyan15} was used for local galaxies, and the Planck multi-frequency catalogue of non-thermal sources \citep[PCNT;][]{PlanckCollaboration2018PCNT} for radio sources.
        \item Additionally, the sample was cross-matched with the Planck Catalogue of Galactic cold clumps \citep{PlanckCollaboration2016cold_cores} to eliminate nearby objects whose sub-mm colours mimic those of high-redshift galaxies.
        \item Existing multi-wavelength data were inspected for all objects in the final sample to detect any Galactic sources, low-redshift objects, or radio counterparts that might have been missed by the automated catalogue cross-matching process.
    \end{itemize}
    \item The ``Matrix Filters'' \citep{herranz08} methodology was applied to the remaining sources to enhance their SNRs. This filtering process yielded a total of 228 candidate sources with $\hbox{SNR} \ge 3$ at 545 GHz and $\hbox{SNR} \ge 2$ in at least two other frequencies (857, 353, or 217\,GHz). 
\end{enumerate}

As a first step towards the identification of lensed galaxies, in 2019 a proposal to do snapshot observations at $850\,\mu$m with SCUBA-2 of 189 of our candidates was submitted (proposal M19BP010, P.I. M. Negrello). The main goals were:
\begin{itemize}
    \item to discriminate between point sources (likely lensed galaxies) and extended emission, like cirrus, cold clumps, confusion fluctuations, or proto-clusters of dusty galaxies; the core of some of the latter may also be strongly lensed \citep{Foo2025};
    \item to drastically improve the positional accuracy of our sources, to help the identification of counterparts in other surveys (e.g. WISE and optical/near-IR catalogues);
    \item to drive new follow-up observations, such as spectroscopy and high-resolution imaging. 
\end{itemize} 

Unfortunately, the observations had a noise level about a factor of 2 higher than expected. As a result, only 12 candidates were robustly detected with $\hbox{S/N} > 5$, all of which were found to be point-like. 
Among these, five have positive declination and are therefore accessible to NOEMA ($\delta \gtrsim -10^\circ$, Plateau de Bure, French Alps). One of them, Planck-3, corresponds to the well-known strongly lensed galaxy WISE\,J132934.18+224327.3 \citep[the ``Cosmic Eyebrow'';][]{DiazSanchez2017,iglesias_groth2017,Dannerbauer19, Harrington21}, which has already been extensively studied in the literature and was previously observed with NOEMA \citep{Dannerbauer2019}, and was therefore excluded from our NOEMA follow-up programme. The remaining four sources (Planck-41, Planck-68, Planck-89, and Planck-188), with estimated photometric redshifts $z\sim 2-3$, were selected for NOEMA follow-up observations, whose results are reported in this work. The remaining seven SCUBA-2 detections lie at negative declinations and are therefore inaccessible to NOEMA. An improved sample selection was later elaborated by \citet{Trombetti21}, and its efficiency is discussed in Trobbiani et al. (in prep.), in comparison to the one used for this work.

\section{Data}\label{sec:data}

\begin{table*}
\caption{Sources observed with NOEMA}
\label{table:pos}      
\centering          
\begin{tabular}{l c c c c c c c  }      
\hline\hline       
Source$^{c}$ & \multicolumn{2}{c}{Planck position}& \multicolumn{2}{c}{SCUBA-2 position} & \multicolumn{2}{c}{NOEMA position$^{a}$} & Redshift$^{b}$  \\
 & RA (deg) & DEC (deg) & RA (h:m:s) & DEC (d:m:s) &  RA (h:m:s) & DEC (d:m:s) & \\
\hline                    
Planck-41 & 266.306 & 40.5247 & 17:45:11.6 & +40:31:0.3 & 17:45:15.71 & +40:31:02.29 &  $ 2.3480\pm0.0007 $ \\
Planck-68 & 273.447 & 54.8952 & 18:13:59.4 & +54:53:27.0 & 18:13:58.78 & +54:53:26.95 &  $ 2.4363\pm0.0005 $   \\
Planck-89 & 204.1465 & 49.2204 & 13:36:34.9 & +49:14:13.0 & 13:36:34.91 & +49:13:13.88 & $ 3.2544\pm0.0007 $  \\
Planck-188 & 155.1342 & 42.5998 & 10:20:29.9 & +42:35:15.0 & 10:20:28.12 & +42:35:15.39 &$ 2.4895\pm0.00003 $  \\
\hline                  
\end{tabular}
\begin{flushleft}
\footnotesize
$^{a}$ The NOEMA positions are the average of the main peak positions in the observing bands, except for 
Planck-41 for which it is the expected position of the lens galaxy in the middle of the multiple images based on the lensing configuration.\\
$^{b}$ Spectroscopic redshift computed from the detected lines in this work.\\
$^c$ Source names follow the numbering convention of the parent catalogue of Planck-selected candidates (Trobbiani et al., in prep.), where sources are listed sequentially as Planck-$N$.
\end{flushleft}
\end{table*}

The SCUBA-2 observations are described in the parent paper (Trobbiani et al., in preparation). In Table~\ref{table:pos} we give the positions of the sources followed up with NOEMA. 
The sources were observed in bands~1 and~2 (3~mm and 2~mm, respectively), aiming at detecting $\rm ^{12}CO$ emission lines and securing their redshifts.

The program S20BQ (P.I. M.~Negrello) targeted each source between August and November 2020, using two or three spectral tunings. The NOEMA configuration D was employed, except for one setup observed in configuration C (Table\,\ref{tab:obsdetails}).

NOEMA is equipped with 2SB receivers that cover a spectral window of 7.744 GHz in each sideband and polarisation, with a gap of 7.744 GHz in between.

The initial setup was set at a local oscillator (LO) frequency of 96 GHz. When a line was detected, we tuned the instrument to a proper 2~mm setup aiming at a second possible $\rm ^{12}CO$ line. When not detected, a second 3~mm setup, tuned to fill the gaps between the lower and the upper sidebands, was used. An additional Director's Discretionary Time (DDT) project (E20AB, P.I. M.~Negrello) was carried out in May 2021 to measure an additional emission line for the target Planck-188 to complete the determination of its redshift.

The NOEMA data were reduced and calibrated in the standard way with the GILDAS\footnote{Grenoble Image and Line Data Analysis Software, \url{https://www.iram.fr/IRAMFR/GILDAS/}}, producing uv-tables of the lower and upper sidebands (LSB, USB), with the velocity scale centred on the detected spectral lines. The main calibrators adopted were MWC349 and LkH$\alpha$101. The absolute flux uncertainty is 10\% and the positional error is $\sim0.2$ arcsec.

Table \ref{tab:obsdetails} summarises the adopted NOEMA setups and the observation log.

\begin{table*}
\caption{\label{tab:obsdetails} Observational details of the NOEMA observations.}
\centering
\begin{tabular}{lcccccccc}
\hline\hline
Source & Setup & $\nu_{\rm LO}$ & $\sigma_{\rm cont}$ & $\sigma_{\rm line}$ & Date & Obs. Quality & \multicolumn{2}{c}{Antennas}\\
 & ID & [GHz] & [$\mu$Jy/beam] & [$\mu$Jy/beam] & [day/month/yr] & & Nr. & Config. \\
\hline
Planck-41  & 001 & 96.00ss & 34.0 & 425.4 & 10/08/2020 & excellent & 9  & D \\
                    &     &       &      &       & 12/08/2020 & mediocre  & 7  & D \\
                    & 006 & 146.50 & 66.5 & 666.1 & 11/09/2020 & ok        & 10 & D \\
                    &     &       &      &       & 08/10/2020 & ok        & 10 & D \\
\hline
Planck-68  & 002 & 96.00 & 34.0 & 425.4 & 15/08/2020 & excellent & 9  & D \\
                    & 005 & 139.01 & 66.5 & 666.1 & 11/09/2020 & ok        & 10 & D \\
\hline
Planck-89  & 003 & 96.00 & 64.6 & 672.6 & 31/08/2020 & poor      & 10 & D \\
                    &     &       &      &       & 05/09/2020 & ok        & 11 & D \\
                    & 007 & 103.50 & 34.0 & 425.4 & 13/09/2020 & excellent & 10 & D \\
                    & 009 & 140.01 &  66.5 & 666.1 & 31/10/2020 & ok & 9 & D \\
\hline
Planck-188 & 004 & 96.00 & 34.0 & 425.4 & 01/09/2020 & poor      & 10 & D \\
                    &     &       &      &       & 05/09/2020 & poor      & 11 & D \\
                    &     &       &      &       & 18/09/2020 & good      & 9  & D \\
                    & 008 & 103.50 & 64.6 & 672.6 & 18/11/2020 & ok        & 10 & C \\
                    &     &       &      &       & 19/11/2020 & poor      & 10 & C \\
                    &     &       &      &       & 24/11/2020 & excellent & 10 & C \\
                    &  E20AB & 138.6 & 29.6 & 243.1 & 27/05/2021 & excellent  & 9 & D\\
\hline
\end{tabular}
\tablefoot{Sensitivities $\sigma_{\rm cont}$ and $\sigma_{\rm line}$ are the values reached per setup. Only the days with satisfactory observation quality or which have been combined with others are shown.}
\end{table*}

\subsubsection*{Continuum imaging}

The NOEMA continuum imaging was performed using the Common Astronomy Software Applications (\texttt{CASA}) package \citep{mcmullin07}. A pixel size of 0.2 arcsec was adopted to finely sample the synthesised beam according to Nyquist's law ($\sim 5$ pixels per beam). A \texttt{uniform} weighting scheme was applied for the imaging of the sources Planck-68, 89 and 188 to obtain the highest resolution of the sources at the lowest scales that are not resolved. 

For the setup 009 of the observations of Planck-89, we decided to apply a \texttt{superuniform} weighting to further increase the resolution at the lowest scales already knowing the lensing nature of the source from \citet{Berman2022}. 

For Planck-41, a  \texttt{Briggs} \citep{briggs95} weighting scheme with a robust parameter of 0.5 was applied during the imaging process, as it showed multiple images already at the level of the dirty image. For all the setups of the two main bands, we merged the LSB and USB sidebands to retrieve the total continuum emission, and in the case of spectral windows with line emission, we identified the channels and removed them to measure only the continuum. 

To increase the number of visibilities we used the combination of all days of observation that reached a satisfying level of quality. Continuum images were generated using imaging tasks such as \texttt{tclean}, which handles the deconvolution of the observed visibilities. For the continuum imaging, a multi-frequency synthesis (MFS) method was employed. Furthermore, for all the images produced, we applied the primary beam correction to compute the flux density. 

The NOEMA observations for these sources typically yield angular resolutions ranging from $1''.2$ to $3''.5$ at 2\,mm, and $1''.7$ to $6''$ at 3\,mm. We show the positions of the sources detected in the continuum maps of NOEMA in Table~\ref{table:pos}, where we give the average position of the main peak in the observing bands, except for Planck-41, for which we used the expected position of the lens galaxy based on the estimated lens configuration. The analysis of the images is described in Sect.~\ref{sec:continuum}.

\subsubsection*{Lines detection and imaging}

The imaging of the line emission was performed by selecting the spectral setup sub-band in which a line was detected. First of all, we subtracted the continuum directly in the \textit{uv} plane using the \texttt{uvconstub} \texttt{CASA} task, selecting the line-free channels from the amplitude-channel distribution of the visibilities. 

The imaging was done in two phases using the \texttt{tclean} \texttt{CASA} task in the \texttt{cube} setting. A first imaging round was performed to estimate the noise level as the average rms level over all the channels in an emission-free region ($\sigma_{\rm noise}$). We then performed the second round of imaging employing $3\sigma_{\rm noise}$ as \texttt{clean} threshold for the line cleaning in each channel. 

\begin{figure}
\begin{tabular}{c}
\resizebox{\hsize}{!}{\includegraphics{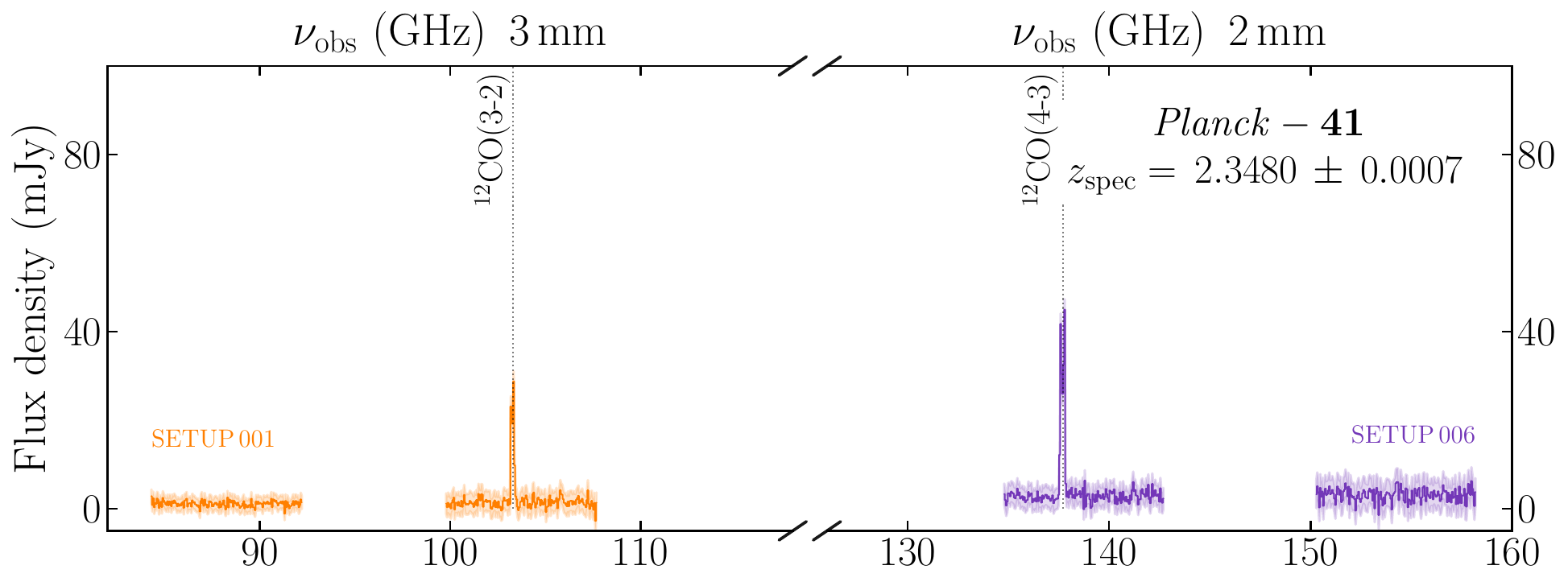}} \\
\resizebox{\hsize}{!}{\includegraphics{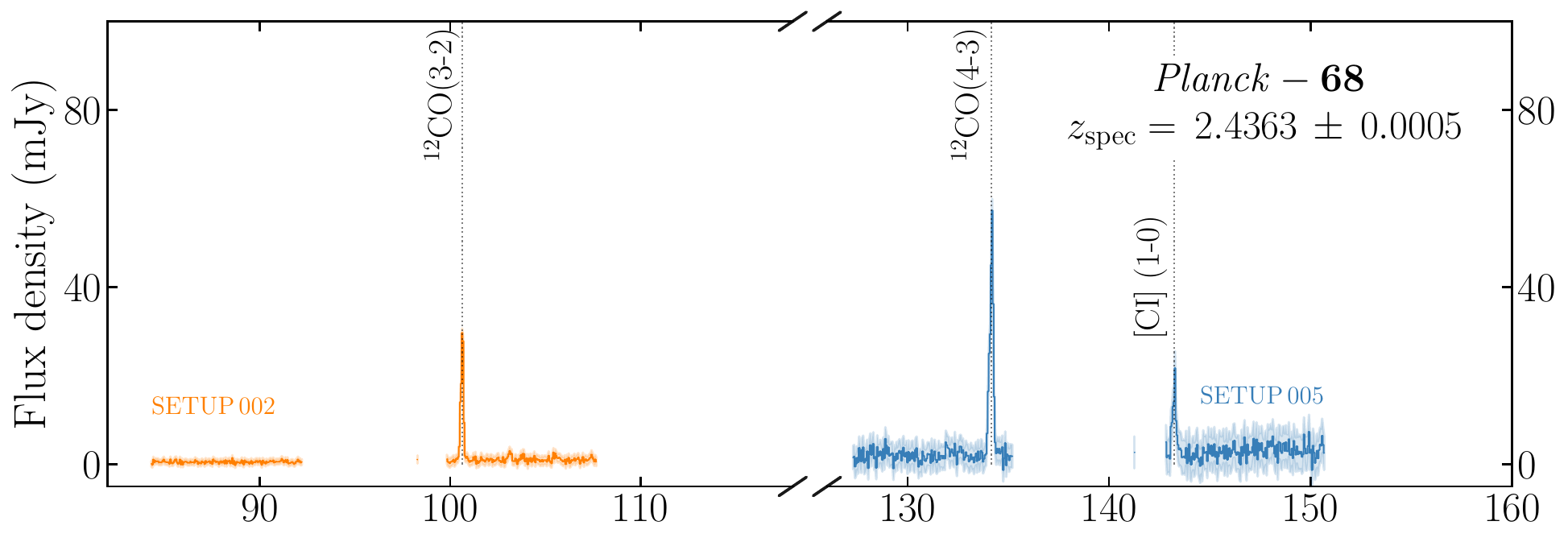}} \\
\resizebox{\hsize}{!}{\includegraphics{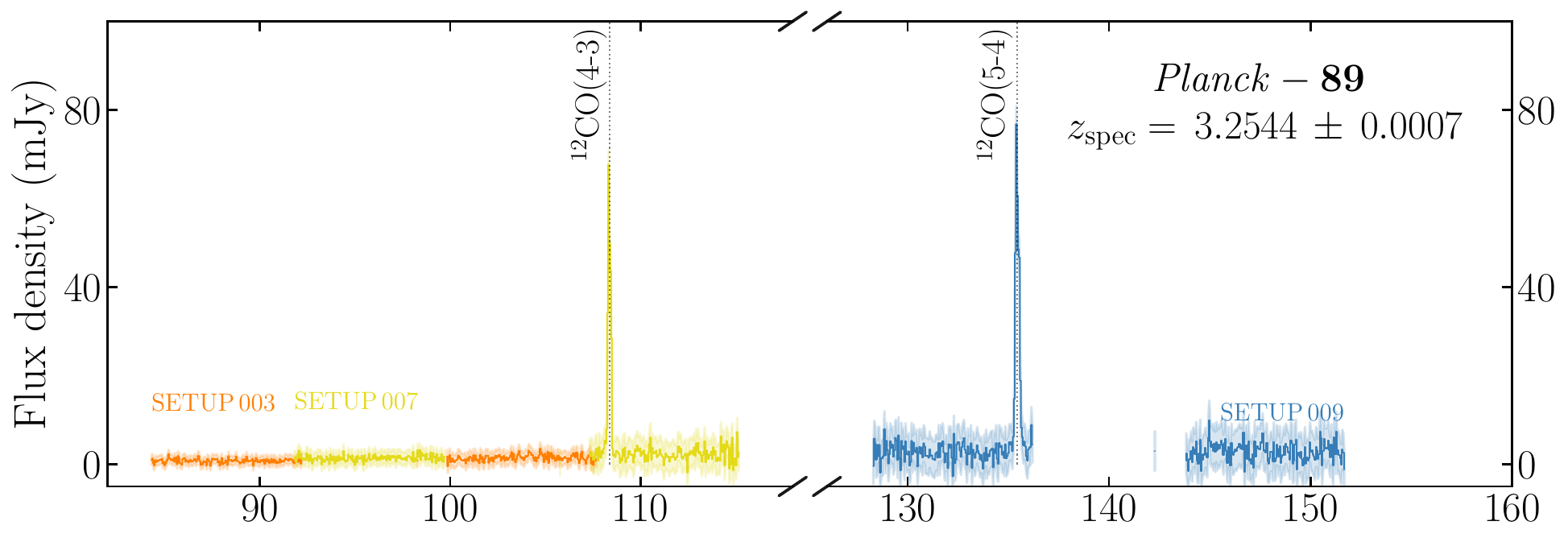}} \\
\resizebox{\hsize}{!}{\includegraphics{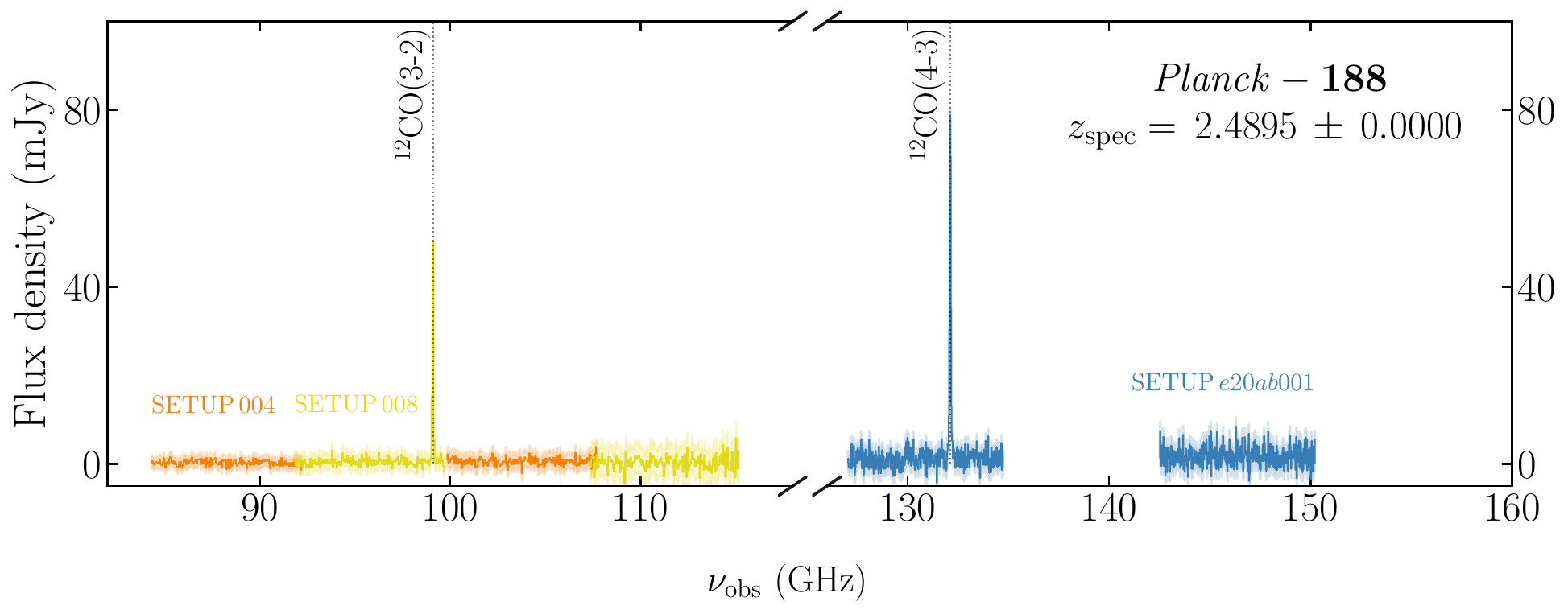}} \\
\end{tabular}
\caption{Observed 1-dimensional spectra of the NOEMA sources with the rebinning at 40 $\text{km}\,\text{s}^{-1}$. We highlight the observed $^{12}$CO and CI lines. }
\label{fig:spec_cov}
\end{figure}

We chose the same weighting schemes employed for the continuum imaging of the sources. Analogously, the imaging was done using a primary beam correction that allowed us to conservatively account for the slight separation from the phase centre of the FoV to obtain less noisy spectra. To robustly determine the line $\mathrm{S/N}$ and produce a uniform emission line analysis of the different sources, we applied a rebinning from $2\,\hbox{km}\,\hbox{s}^{-1}$ to $40\,\hbox{km}\,\hbox{s}^{-1}$. The detected lines are shown in Fig.~\ref{fig:spec_cov} and summarised below.

\paragraph*{Planck-41.} A clear emission line was detected in the USB of the first setup (track 001) at $103.28\,$GHz. The second setup (track 006) revealed a line in the LSB at 137.70\,GHz. Based on the observing strategy and the estimated photometric redshift, we identified the lines as CO(3--2) and CO(4--3).

\paragraph*{Planck-68.} The line detected in the first setup (track 002) in the USB at 100.65\,GHz was identified as CO(3--2). The second setup (track 005) revealed two spectral features: one in the LSB at 134.19\,GHz and another in the USB at 143.27\,GHz. They were identified as CO(4--3) and [CI](1--0), respectively.

\paragraph*{Planck-89.} The first setup (track 003) showed no significant lines. In the second (track 007), a line was found in the USB at 108.38~GHz. It was previously reported and identified as CO(4--3) by \citet{Berman2022}, who also detected the CO(3--2) line. The third setup (track 009) showed a line in the LSB at 135.47\,GHz; it was identified as CO(5--4). 

\paragraph*{Planck-188.} In the first setup (track 004), no line was clearly detected. The second setup (track 008) yielded a detection in the LSB at 99.10~GHz. The DDT program E20AB yielded the detection of a second emission line, at 127.06\,GHz. This allowed us to identify the two lines as CO(3--2) and CO(4--3). 

\section{Analysis of NOEMA data}\label{sec:analysis}

\subsection{Continuum observations}\label{sec:continuum}

\begin{figure*}[]
\centering

\begin{subfigure}{0.49\textwidth}
\centering
\includegraphics[width=\linewidth]{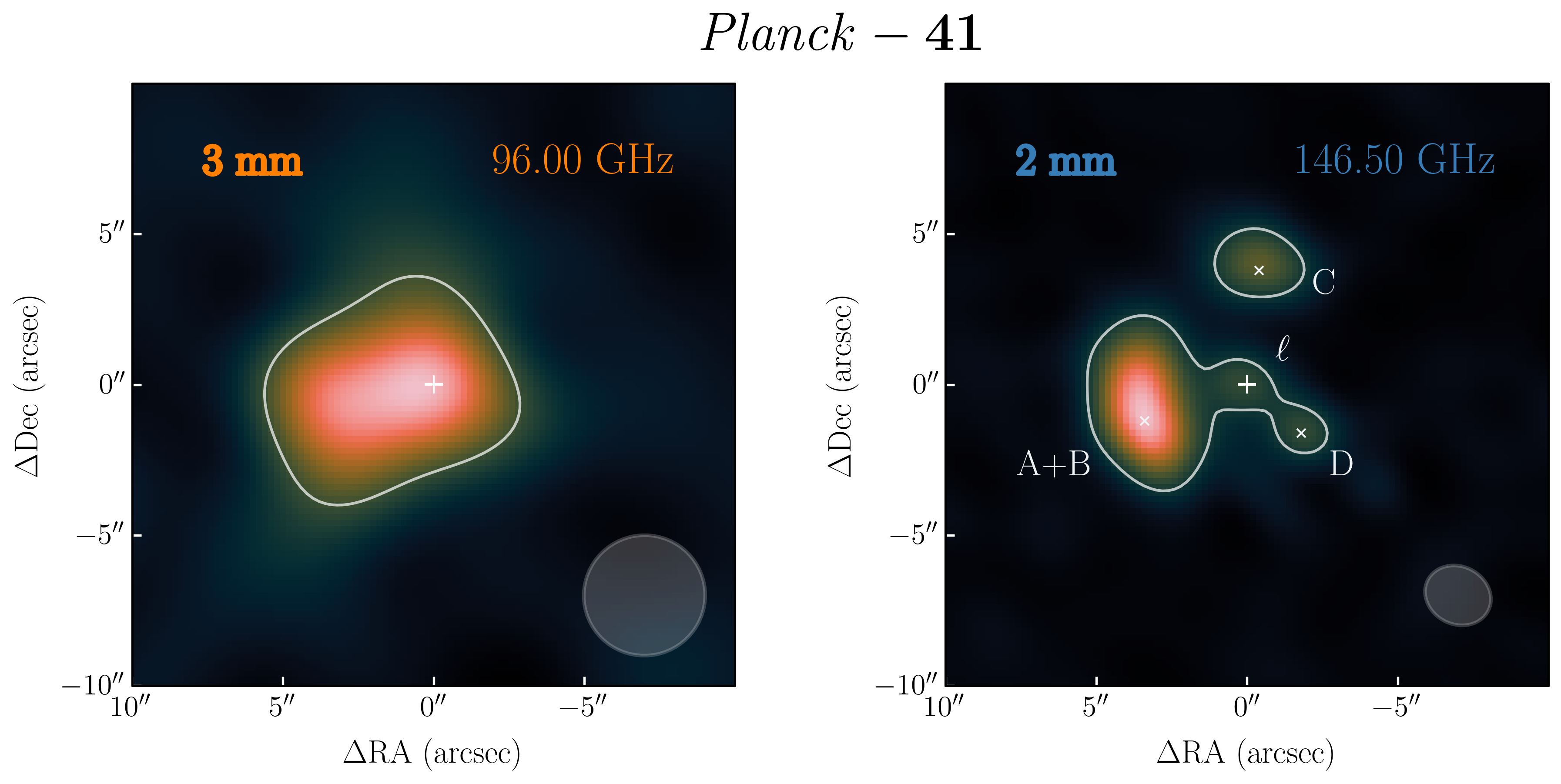}
\end{subfigure}
\hfill
\begin{subfigure}{0.49\textwidth}
\centering
\includegraphics[width=\linewidth]{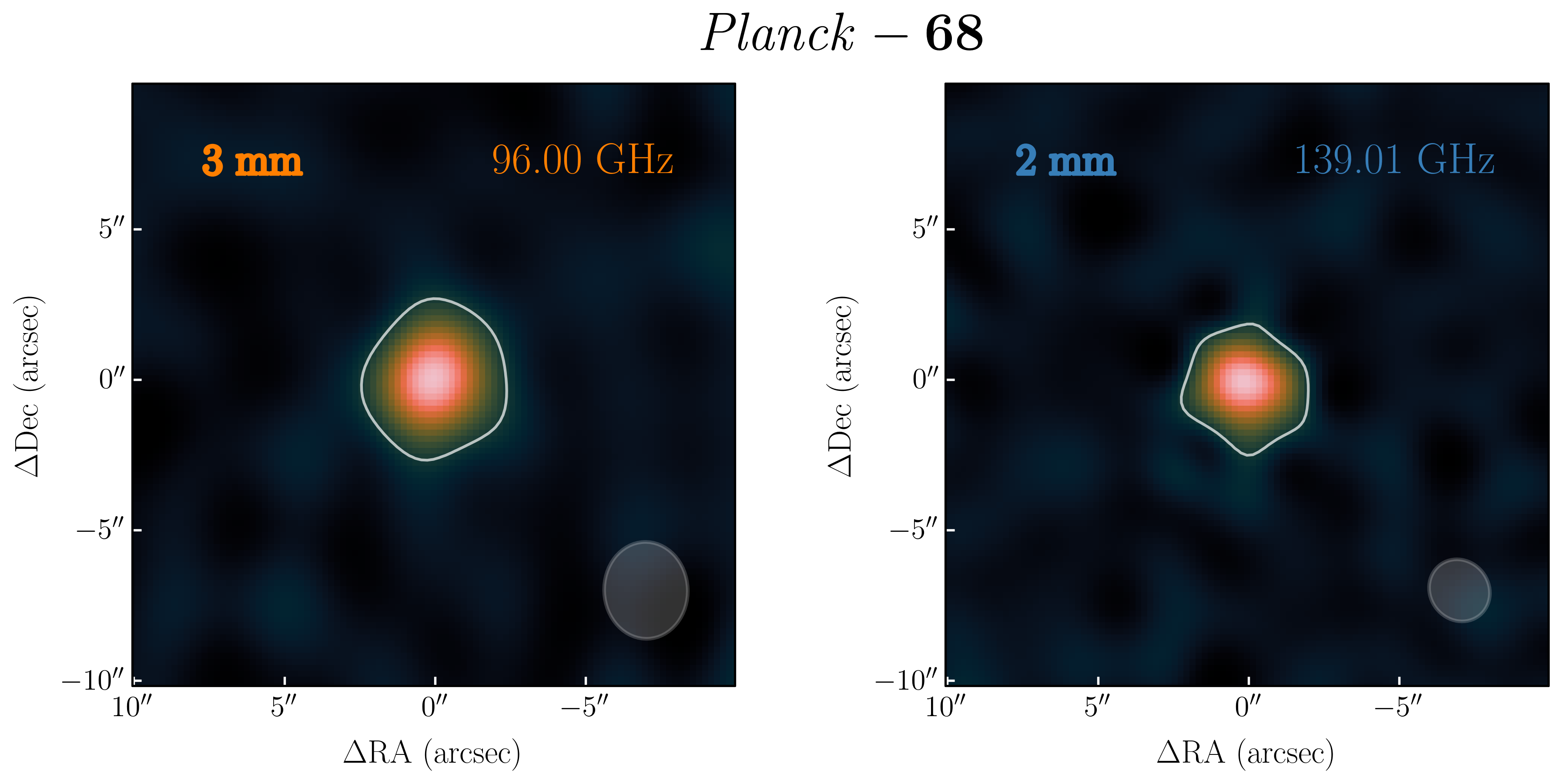}
\end{subfigure}

\vspace{0.25cm}
\hrule
\vspace{0.25cm}

\begin{subfigure}{0.49\textwidth}
\centering
\includegraphics[width=\linewidth]{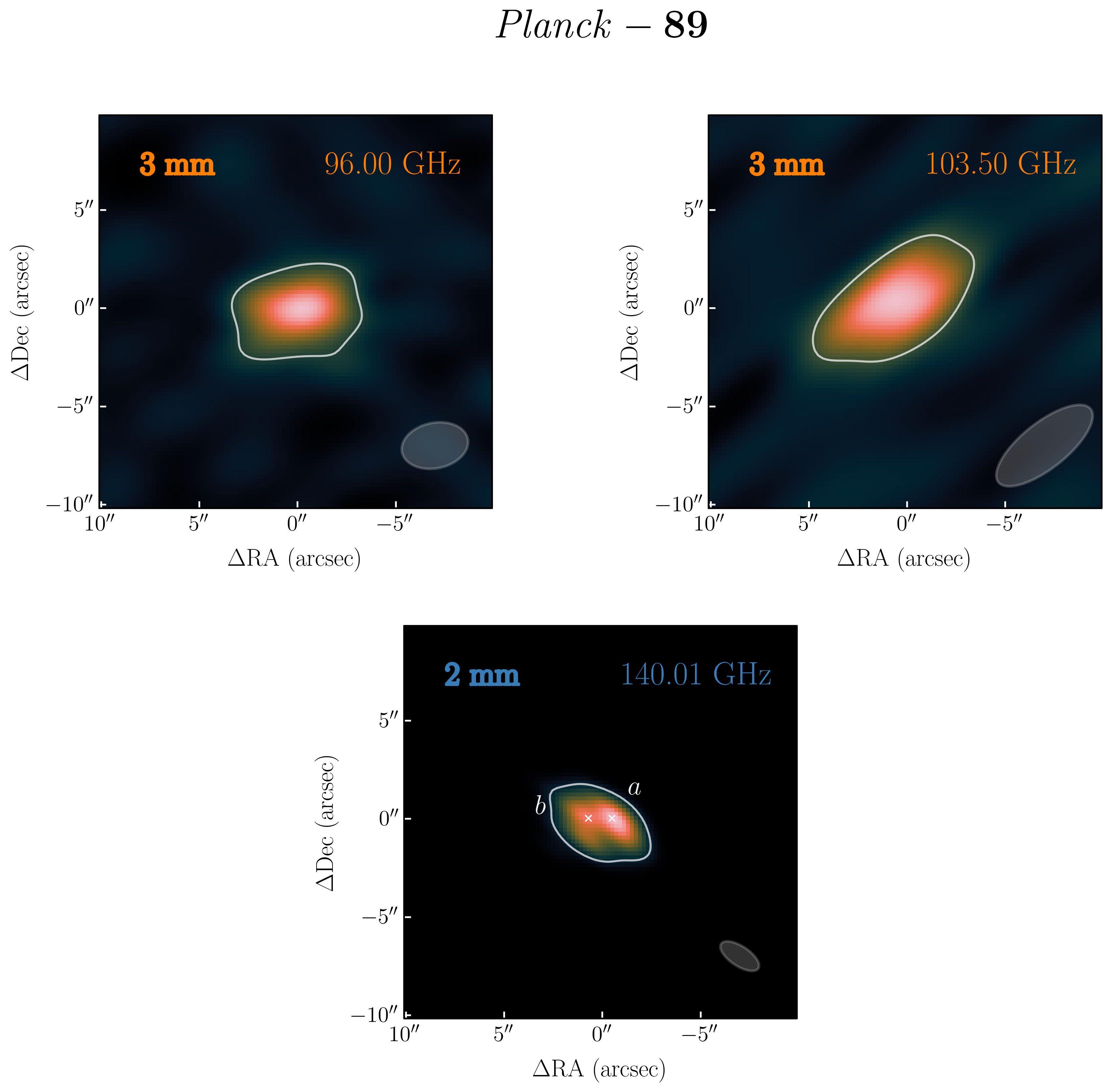}
\end{subfigure}
\hfill
\begin{subfigure}{0.49\textwidth}
\centering
\includegraphics[width=\linewidth]{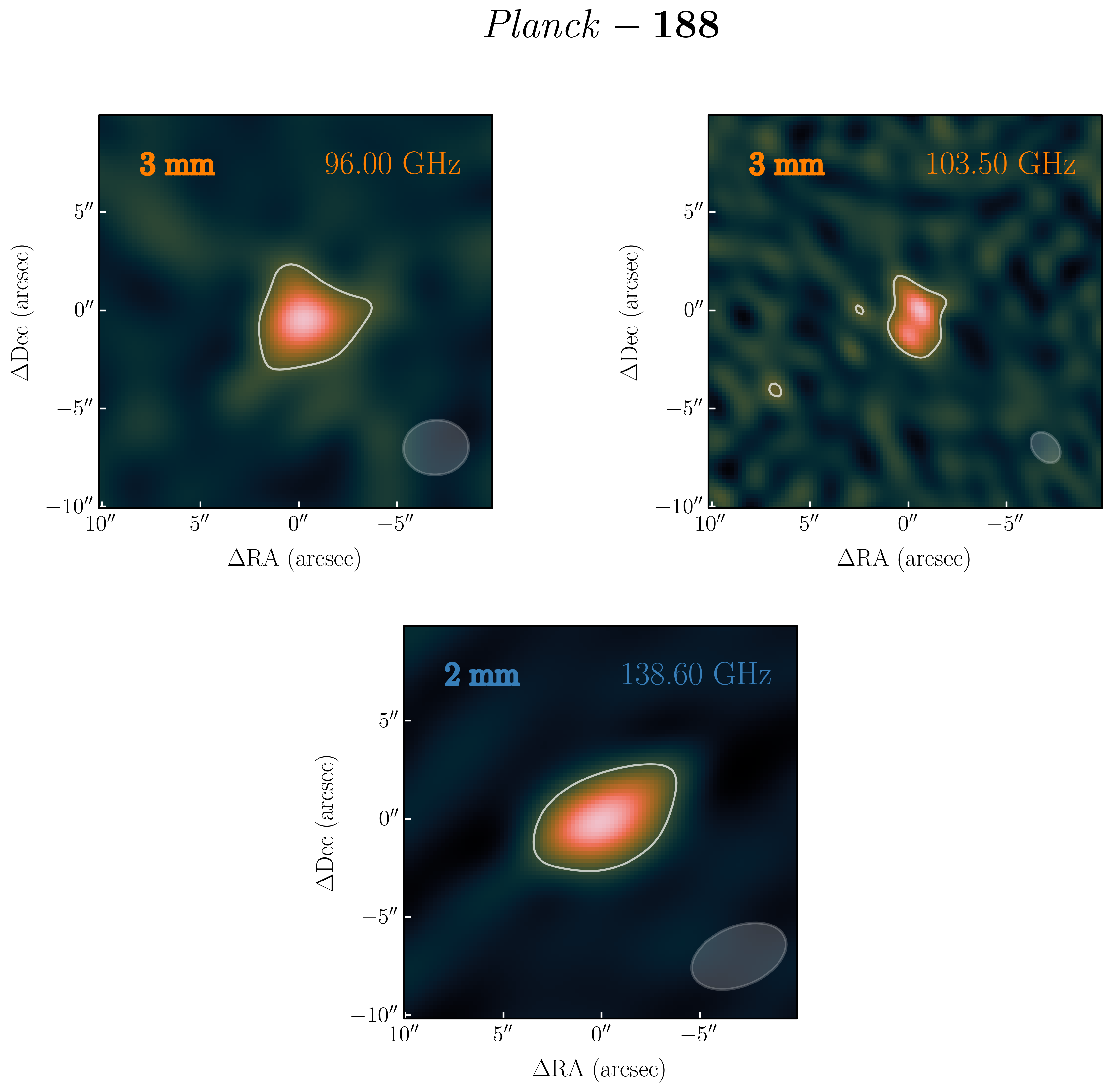}
\end{subfigure}

\caption{
NOEMA continuum maps of the four galaxies with 3$\sigma$ contours. 
The synthesised beam is displayed as a grey ellipse in the bottom-right corner of each map. 
The observing wavelength is reported in the top-left corner and the local oscillator frequency in the top-right. For 
Planck-41, the labels A, B (blended), C, and D indicate the 
multiple lensed images in a fold configuration, while $\ell$ marks the 
central component associated with the foreground lens galaxy. For 
Planck-89, labels a and b indicate the two main continuum components identified in the 2\,mm map.}
\label{fig:cont_maps}
\end{figure*}

Figure~\ref{fig:cont_maps} shows the NOEMA continuum maps of the four sources, produced with the imaging procedure detailed in Sect.~\ref{sec:data}. The continuum images provide crucial insights into the spatial extent and morphology of these lensed galaxies as all four sources were unresolved by the $13''$ beam of SCUBA-2. 

Planck-41 is resolved into multiple components at the NOEMA resolution in the 2\,mm band. In the figure, we label with A, B (blended), C and D the multiple images and with $\ell$ the central component, referring to a typical configuration in strong gravitational lensing morphology called `fold configuration' where two of the four images produced by the strong lensing are merged and extremely magnified compared to the other two. The angular distances of the three external components from the $\ell$ component are, respectively, $\sim 3''.0$ for A+B; $\sim 4''.0$ for C, and $\sim 2''.5$ for D. The central component can be associated either with the central image produced by the lensing and predicted by the lensing theory, or with the foreground object that acts as the lens. We refer to Sect.~\ref{sec:lens_modelling} for the detailed characterisation of the source in the lensing framework. On the other hand, the 3\,mm emission appears to be only partially resolved in an extended area where the dominant component is aligned with the expected position of the foreground galaxy.

Planck-68 and 188 show compact morphologies in the continuum and no direct hint of strong lensing. This is also the case for Planck-89, which instead showed an almost complete Einstein radius in  Karl G. Jansky Very Large Array (VLA) data at higher resolution, $\sim 0.3''$ \citep{Kamieneski24}. For this reason, as described in Sect.~\ref{sec:continuum}, we employed a \texttt{superuniform} weighting scheme ($2''.27\times1''.00$ beam) to extract all the information from the small scales in the 2\,mm band, the one with higher resolution among the Planck-89 setups. The image, in the left panel of the third row of Fig.~\ref{fig:cont_maps}, shows two closely blended clumps that we call $a$ and $b$, the first one dominating in flux.

The total continuum flux density is estimated from these maps by integrating the flux in the elliptical region encompassing the entire source defined by the Kron radius \citep{Kron80} for each map except for the 2\,mm emission of Planck-41, where the flux density was estimated by a circular region of $6''$ radius that encompasses all the images. The continuum fluxes are tabulated in Table~\ref{table:cont_flux_densities} along with Planck and SCUBA-2 flux densities. We estimated the uncertainties on these fluxes as:

\begin{equation}\label{eq:cont_flux_unc}
    \delta S = \sqrt{ N\sigma^2 + (0.10 \times S)^2}
\end{equation}

where $\sigma$ is the root mean square (rms) noise in empty regions with the same shape as that used for the estimation of the flux density. We then rescaled the noise by the number $N$ of synthesised beams in the region, and added the flux uncertainty from NOEMA calibration (10\%). The flux density values are shown alongside those of Planck and SCUBA-2 in Table~\ref{table:cont_flux_densities}.

\begin{table*}
\caption{Summary of continuum flux densities.}           
\label{table:cont_flux_densities}      
\centering          
\begin{tabular}{l c c c c c c c c c }     
\hline\hline       
 Source & \multicolumn{4}{c}{Planck} & \multicolumn{2}{c}{NOEMA (2 mm)} & \multicolumn{2}{c}{NOEMA (3 mm)} \\ 
 & $S_{857\,\mathrm{GHz}}$  & $S_{545\,\mathrm{GHz}}$ & $S_{353\,\mathrm{GHz}}$ & $S_{217\,\mathrm{GHz}}$ & $S_{139\,\mathrm{GHz}}$ & $S_{146\,\mathrm{GHz}}$ & $S_{96\,\mathrm{GHz}}$ & $S_{103\,\mathrm{GHz}}$  \\
  & (mJy)  & (mJy) & (mJy) & (mJy) & (mJy) & (mJy) & (mJy) & (mJy)  \\
\hline                    
Planck-41 & 1215$\pm$226 & 674$\pm$107 & 178$\pm$59 & 45$\pm$28 & -- & 7.7$\pm$0.80 & 2.3$\pm$0.25 & -- \\ 
Planck-68 & 983$\pm$179 & 553$\pm$100 & 252$\pm$63 & 65$\pm$25 & 4.0$\pm$0.89 & -- & 1.1$\pm$0.23 & -- \\
Planck-89 & 946$\pm$176 & 537$\pm$95 & 240$\pm$59 & 50$\pm$26 & 7.8$\pm$1.00$^a$ & -- & 2.0$\pm$0.45 & 2.7$\pm$0.66 \\
Planck-188 & 960$\pm$175 & 511$\pm$86 & 230$\pm$61 & 65$\pm$36 & 2.4$\pm$0.65 & -- & 0.7$\pm$0.31 & 1.2$\pm$0.42 \\
\hline                  
\end{tabular}
\begin{flushleft}
\footnotesize
NOEMA measurements include 10\% of calibration uncertainty. The flux densities are observed  (magnification-uncorrected) values.\\
$^a$ Flux density reported for the uniform-weighted image for consistency with the other measurements; the value for the superuniform-weighted image is consistent within $1\,\sigma$.
\end{flushleft}
\end{table*}

\begin{table*}
\caption{Properties of NOEMA detected lines.}
\label{tab:line_fluxes_Lprime}
\centering
\begin{tabular}{lccccccc}
\hline\hline
Source & Line & FWHM  & $\mu S_{\text{line}}\Delta v$ & $\mu L'_{\text{line}}$ & $\mu L_{\text{line}}$ & $\mu L'_{\text{CO(1-0)}}$ $^a$ & $\mu M_{\rm mol, line}$ $^b$ \\
 &  & (km\,s$^{-1}$) 
 & (Jy km\,s$^{-1}$) & ($10^{10}\,\mathrm{K\,km\,s^{-1}\,pc^{2}}$) & ($10^{8}\,L_\odot$) & ($10^{10}\,\mathrm{K\,km\,s^{-1}\,pc^{2}}$) & ($10^{11}\,M_\odot$) \\
\hline \\ [-8pt]
 Planck-41 & CO(3--2) & $454 \pm 22$ & $13.50 \pm 0.35$ & $39.1 \pm 1.0$ & $4.86 \pm 0.13$ & $56.7 \pm 10.0$ & $22.7 \pm 4.0$ \\ 
 & CO(4--3) & $427 \pm 20$ & $16.26 \pm 0.28$ & $26.5 \pm 0.5$ & $7.80 \pm 0.13$ & $51.0 \pm 13.8$ & $20.4 \pm 5.5$ \\
\hline \\ [-8pt]
Planck-68 & CO(3--2) & $409 \pm 25$ & $14.34 \pm 0.25$ & $44.3 \pm 0.8$ & $5.50 \pm 0.10$ & $64.2 \pm 11.2$ & $25.7 \pm 4.5$ \\
 & CO(4--3) & $416 \pm 25$ & $19.21 \pm 0.45$ & $33.4 \pm 0.8$ & $9.82 \pm 0.23$ & $64.2 \pm 17.4$ & $25.7 \pm 6.9$ \\
& $\rm{[CI]}$(1--0) & $202 \pm 56$ & $3.60 \pm 0.61$ & $5.5 \pm 0.9$ & $1.96 \pm 0.33$ & $--$ & $8.9 \pm 4.6$ \\
\hline\\ [-8pt]
Planck-89 & CO(4--3) & $410 \pm 25$ & $26.93 \pm 0.55$ $^c$ & $76.1 \pm 1.5$ & $22.36 \pm 0.46$ & $146.3 \pm 39.5$ & $58.5 \pm 15.8$ \\
 & CO(5--4) & $434 \pm 27$ & $35.86 \pm 0.52$ & $64.8 \pm 0.9$ & $37.22 \pm 0.54$ & $175.2 \pm 71.1$ & $70.1 \pm 28.4$ \\
\hline \\ [-8pt]
Planck-188 & CO(3--2) & $155 \pm 10$ & $7.55 \pm 0.11$ & $24.2 \pm 0.4$ & $3.01 \pm 0.04$ & $35.1 \pm 6.1$ & $14.0 \pm 2.5$ \\
 & CO(4--3) & $134 \pm 12$ & $11.69 \pm 0.29$ & $21.1 \pm 0.5$ & $6.20 \pm 0.15$ & $40.6 \pm 11.0$ & $16.2 \pm 4.4$ \\
\hline
\end{tabular}
\begin{flushleft}
\footnotesize
FWHM, fluxes, luminosities, and molecular gas masses uncorrected for the putative gravitational magnification.\\
$^a$ Converted using the \citet{Harrington21} PASSAGES excitation ratios $r_{J1}$ for DSFGs. \\
$^b$ Computed using $\alpha_{\rm CO}$ from \citet{Prajapati2026} for $^{12}$CO lines and $\alpha_{\rm [CI](1-0)}$ from \citet{Harrington21}. \\
$^c$ The CO(4--3) integrated line flux of Planck-89 is consistent within $2\sigma$ with the value of 
$21.9 \pm 1.1$\,Jy\,km\,s$^{-1}$ reported by \citet{Berman2022} for the same source (PJ133634.9).
\end{flushleft}
\end{table*}

\subsection{Spectroscopy}\label{sec:spectroscopy}

\subsubsection*{Integrated spectra}

The lines were extracted using the same regions as for the source flux (see Sect.~\ref{sec:continuum}). 
The line profiles were fitted with  Gaussian functions. When the line exhibited evidence for substructure or asymmetry, we adopted a double-Gaussian model and computed the line centroid as the intensity-weighted frequency centroid of the spectra in the line channels. This approach provided both the observed line frequency and its associated uncertainty. All lines from the first three sources were fitted with a double-Gaussian profile, those of the last one (Planck-188) with a single-Gaussian profile. From the fit, we computed the velocity line-widths of individual Gaussian components. In the case of double peaks, the total line width was computed, following \citet{Bothwell2013}, from the intensity-weighted second moment ($s_\nu$) of the spectra:
\begin{equation}
    s_v = \frac{\int (v - \bar v)^2I_v \text{d}v}{\int I_v\text{d}v},
\end{equation}
where $\bar v$ is the intensity-weighted velocity centroid of the line and $I_v$ is the flux as a function of frequency. The equivalent Gaussian FWHM of the line is then $\text{FWHM} = 2\sqrt{2\ln2}\,s_v$. To compute both the centroid and the second moment, we employed the spectroscopic Python package \texttt{specutils} with dedicated functions.  
A potential source of uncertainty for this approach is its sensitivity to noise spikes. To minimise this potential problem, we adopted a Monte Carlo approach to generate multiple copies of each spectrum with the inclusion of Gaussian noise. We adopted the median values of the distributions as our selected centroid and FWHM. In Table~\ref{tab:line_fluxes_Lprime} we show the FWHM estimates for the detected lines.

The line intensities were computed employing \texttt{specutils}. The associated uncertainties are given by $\sigma_{\rm flux} = \text{rms}\cdot \sqrt{N_{\rm chan}} \cdot\Delta v$, where the rms is computed on the line--free part of the spectrum, $N_{\rm chan}$ is the number of channels in the selected integration window, and $\Delta v$ is the velocity bin. In Table~\ref{tab:line_fluxes_Lprime} we show the flux estimates $S_{\text{line}}\Delta v$. 

\begin{figure*}
\centering

\begin{subfigure}{0.48\linewidth}
\centering
\includegraphics[width=\linewidth]{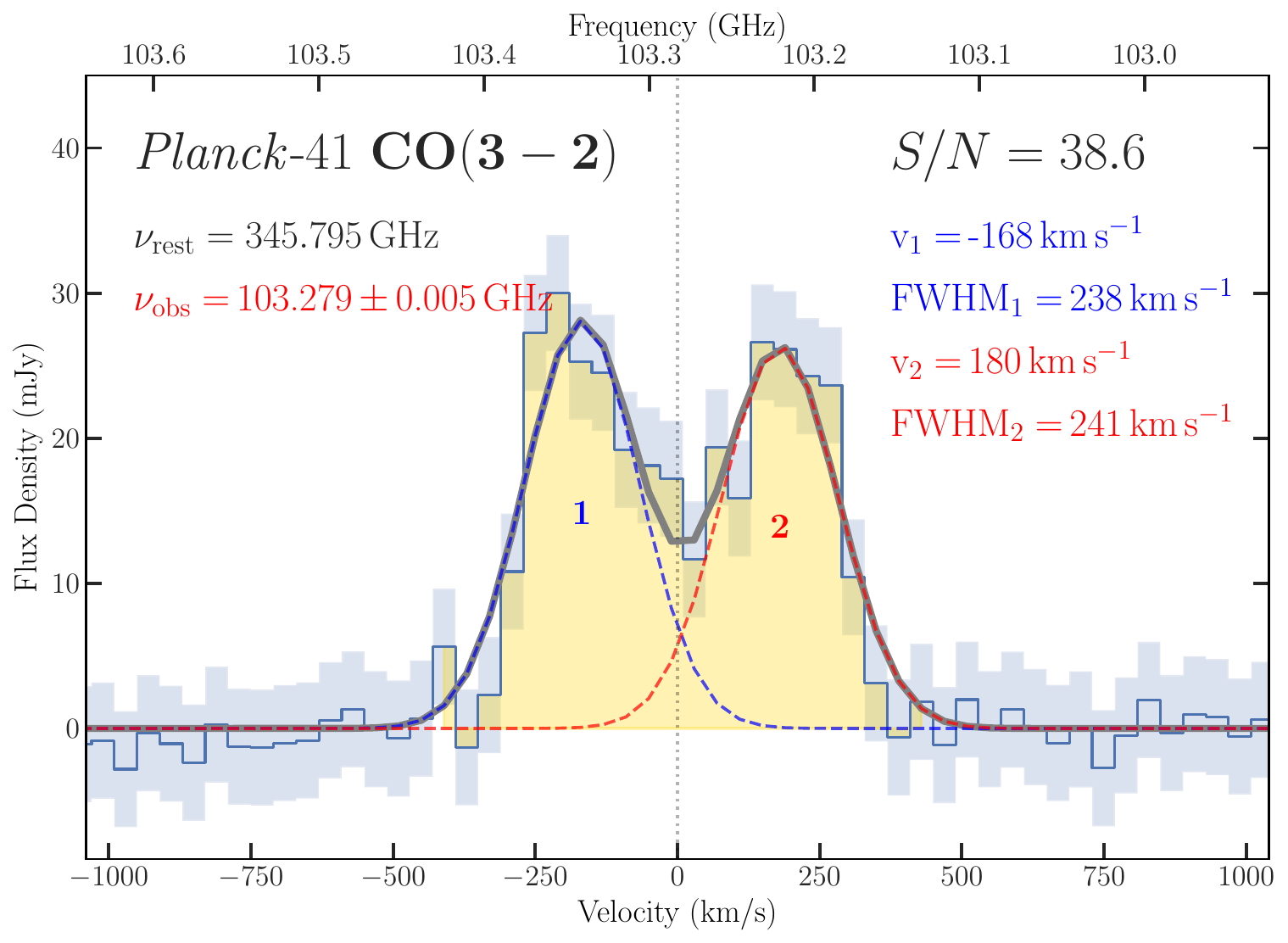}
\end{subfigure}
\hfill
\begin{subfigure}{0.48\linewidth}
\centering
\includegraphics[width=\linewidth]{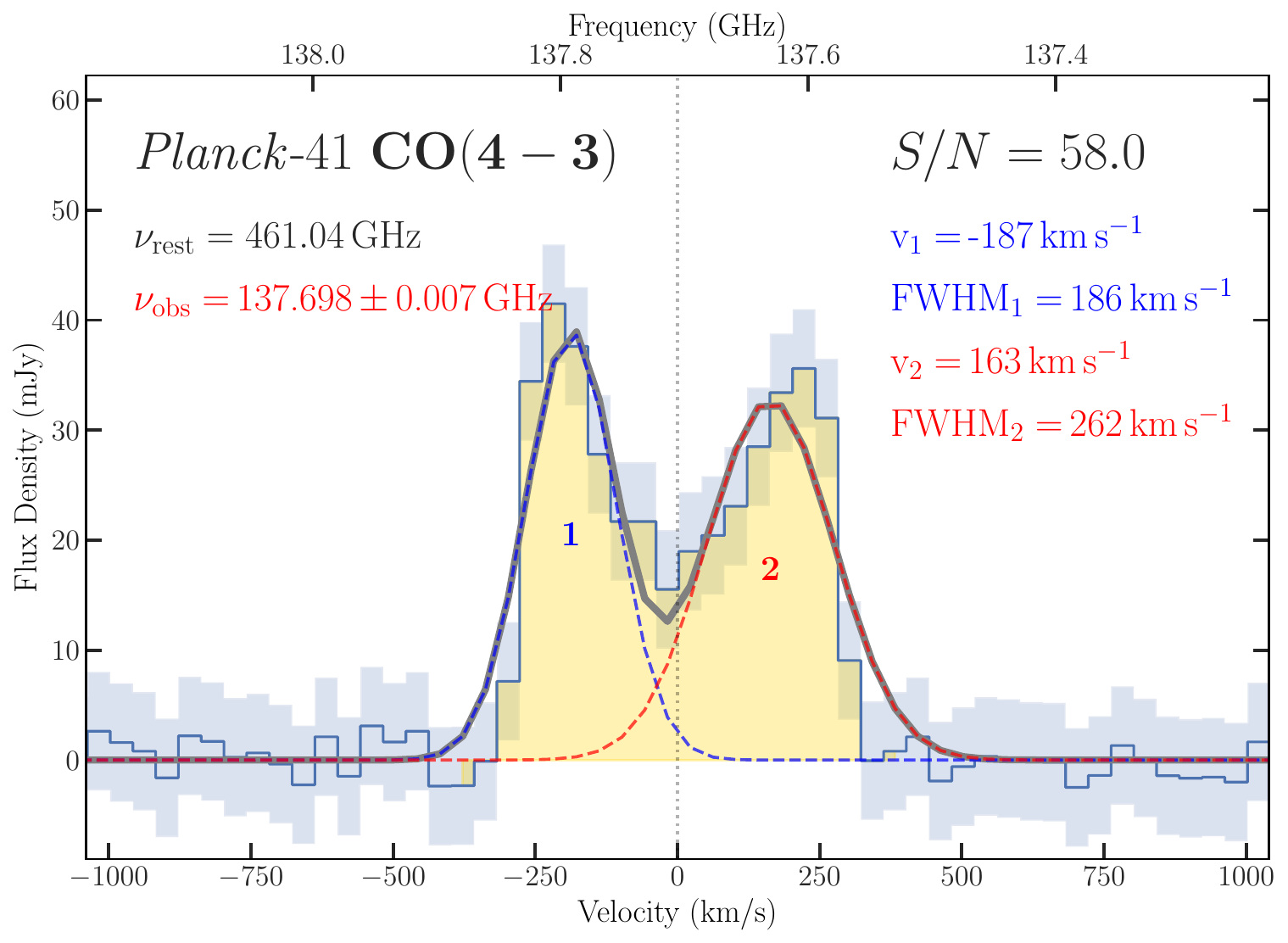}
\end{subfigure}

\vspace{0.3cm}

\begin{subfigure}{0.48\linewidth}
\centering
\includegraphics[width=\linewidth]{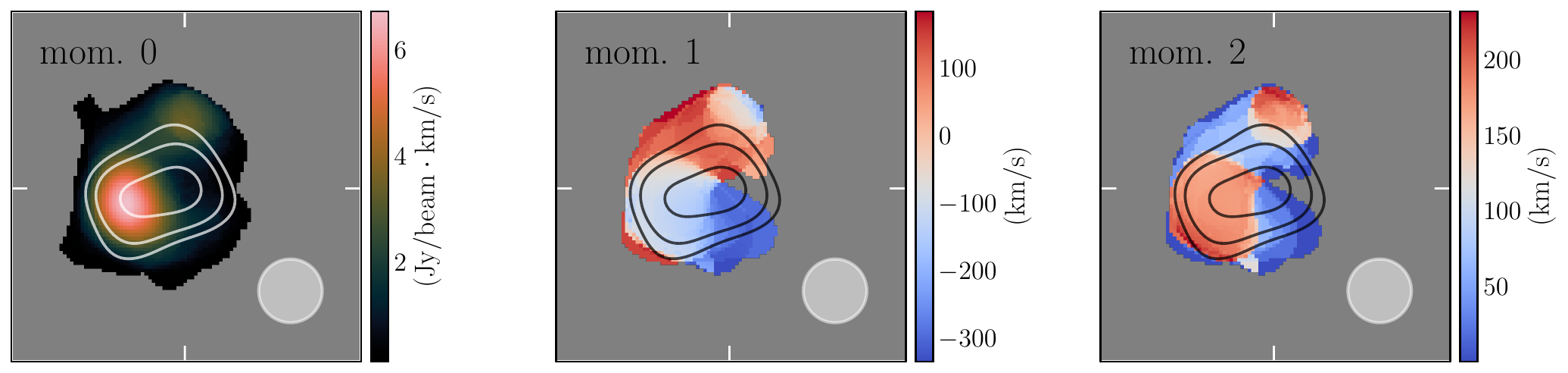}
\end{subfigure}
\hfill
\begin{subfigure}{0.48\linewidth}
\centering
\includegraphics[width=\linewidth]{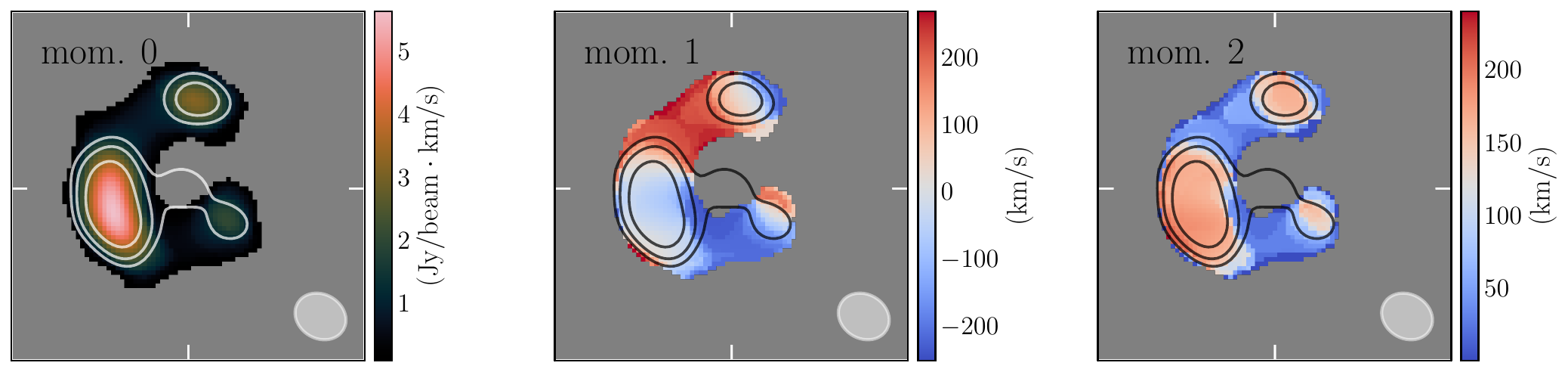}
\end{subfigure}

\caption{
Planck-41 line analysis: CO(3--2) on the left and CO(4--3) on the right.
Top panel: line profiles with derived spectroscopic quantities. Bottom panel: Moment 0th, 1st and 2nd maps. The white circles on the bottom right of the lower panels represent the synthesised beam, while the contours refer to the continuum emission in the same band at 3, 5 and 10$\sigma$.
}
\label{fig:Planck41_all}
\end{figure*}

All but one line have $S/N\gtrsim30$ (see below). Line fluxes and luminosities, uncorrected for gravitational magnification, are reported in Table~\ref{tab:line_fluxes_Lprime}.  Figure~\ref{fig:Planck41_all} shows the profiles of the detected CO lines of Planck-41, while we refer to Appendix~\ref{app:line_figs} for the profiles of the other sources; the shaded yellow areas show the integration windows for the derivation of the total flux.

The magnified line luminosities, $\mu L'_{\text{line}}$, in units of $\mathrm{K\,km\,s^{-1}\,pc^{2}}$, were calculated using the relation \citep{carilli&walter13, solomon92}:

\begin{equation}\label{eq:lum_prime}
\mu L'_{\text{line}} = 3.25 \times 10^{7} \times \mu S_{\text{line}} \Delta v \frac{D_{\text{L}}^{2}}{(1+z)^{3} \nu_{\text{obs}}^{2}}\,,
\end{equation}

where $\mu S_{\text{line}} \Delta v$ is the magnified flux of the line (in units of $\mathrm{Jy\,km\,s^{-1}}$), $D_{\text{L}}$ is the luminosity distance in $\mathrm{Mpc}$, $z$ is the redshift of the source, and $\nu_{\text{obs}}$ is the observed frequency of the line in $\mathrm{GHz}$. The luminosities expressed in solar luminosities ($\mu L_{\odot}$) were computed as $L_{\text{line}} = 3 \times 10^{-11} \nu_{\text{rest}}^{3} L'_{\text{line}}$, where $\nu_{\text{rest}}$ is the rest-frame frequency of the transition in $\mathrm{GHz}$. Below, we discuss the results for each source.

\paragraph*{Planck-41.} We obtained clear detections for both the CO(4--3) and the CO(3--2) line, with $\mathrm{S/N}$ ratios of 58.0 and 38.6, respectively. The CO(4--3) line was observed at $137.698\pm0.007~\text{GHz}$. The line shape suggests a complex and broad structure; it is fitted by two Gaussian components. The first one is blueshifted with a velocity offset of $-187~\text{km~s}^{-1}$ and an FWHM of $186~\text{km~s}^{-1}$. The second is redshifted by $163~\text{km~s}^{-1}$ with an FWHM of $262~\text{km~s}^{-1}$. The blueshifted peak is slightly higher (by $\sim5\,$mJy) than the redshifted one (see the top-right panel of Fig.~\ref{fig:Planck41_all}). Given the complex profile with broad, separated peaks, we used the intensity-weighted second moment method to compute the FWHM of the line, found to be $\sim430~\text{km~s}^{-1}$ (see Table~\ref{tab:line_fluxes_Lprime} for this and the other lines' FWHM with uncertainties).

The CO(3--2) line was detected at an observed frequency of $103.279\pm0.005~\text{GHz}$. Like the CO(4--3) line, its profile is also best described by a two-component Gaussian. The components are symmetrically distributed around zero velocity, with a blueshifted peak at $-168~\text{km~s}^{-1}$ (FWHM = $238~\text{km~s}^{-1}$) and a redshifted peak at $180~\text{km~s}^{-1}$ (FWHM = $241~\text{km~s}^{-1}$). Again, the blueshifted peak is more intense (see the top-left panel of Fig.~\ref{fig:Planck41_all}). The FWHM of both the CO(3--2) and the CO(4--3) line is of  $\sim460~\text{km~s}^{-1}$. 

\paragraph*{Planck-68.} We have detections of three lines, CO(3--2), CO(4--3), and [CI](1--0), with $\mathrm{S/N}$ ratios of $57.0$, 42.8, and 5.9, respectively. The observed frequency of CO(3--2) is $100.628\pm0.0043~\text{GHz}$. The spectrum shows two components: a redshifted peak at $243~\text{km~s}^{-1}$ (FWHM = $272\,\text{km~s}^{-1}$) and a blueshifted one of $-74\,\text{km~s}^{-1}$ (FWHM = $303\,\text{km~s}^{-1}$).

The CO(4--3) line was observed at $134.165\pm0.006\,\text{GHz}$ . The line profile is broad. Also this line shows a double peak, with the redshifted one at $263\,\text{km~s}^{-1}$ and an FWHM of $250\,\text{km~s}^{-1}$, and the blueshifted one at $-70\,\text{km~s}^{-1}$ with an FWHM of $289\,\text{km~s}^{-1}$. 

For both CO lines, differently from Planck-41, the redshifted peak is subdominant; its intensity is about half that of the blueshifted one. Like for Planck-41, we employed the non-parametric method for the estimation of the velocity width of the lines, allowing us to take into consideration the asymmetrical profile; the FWHMs of both lines are $\sim410\,\text{km~s}^{-1}$.

The [CI](1--0) line was observed at $143.25\pm0.03\,\text{GHz}$, at the border of the spectral window, where the noise is more significant. Although the $\mathrm{S/N}$ is relatively low, $5.9$, it was possible to find indications of two components: a blueshifted component at a velocity of $-91\,\text{km~s}^{-1}$ with a FWHM of $82\,\text{km~s}^{-1}$ and a redshifted component at $76\,\text{km~s}^{-1}$ with a FWHM of $264~\text{km~s}^{-1}$. 

The [CI](1--0) line profile is consistent, within the uncertainties, with those of the two measured CO lines. This correspondence of the line kinematics suggests that all these lines trace the same molecular gas. The [CI](1--0)/low-J CO ratio is a tracer of cold gas density \citep{Kaufman1999}. Our [CI](1--0)/CO(4--3) $=0.18\pm 0.03$ is in the bottom of the  range measured for high-$z$ submillimetre galaxies \citep{Bothwell2017, Birkin2021, FriasCastillo2025, Prajapati2026} and corresponds to a relatively high gas density \citep[$\log_{10}(n/\hbox{cm}^{-3})\simeq 5$, see Fig.\,6 of][]{Birkin2021}. 
On the other hand, the FWHM of [CI](1--0), $202 \pm 58\,\text{km~s}^{-1}$, is notably narrower than that of CO(4--3), $416 \pm 25\,\text{km~s}^{-1}$. This results in a FWHM ratio of $\text{FWHM}_{\text{[CI](1-0)}} / \text{FWHM}_{\text{CO(4-3)}} = 0.49 \pm 0.14$. This discrepancy can be partially attributed to the lower $\mathrm{S/N}$ ($\sim 5.9$) of the [CI](1--0) detection, as the faint, high-velocity wings of the line profile are likely lost within the noise, leading to an underestimation of the true velocity width. 

Alternatively, a narrower line width may reflect a more compact gas distribution. This kinematic indication is supported by the analysis of the \textit{moment 0} maps presented in Section~\ref{sec:moment_maps}, where the [CI](1--0) emission exhibits a slightly smaller spatial extension compared to the CO lines. Despite these differences in width, the shared velocity range suggests that both species trace the same global molecular reservoir, albeit with different spatial distributions or excitation conditions.

\paragraph*{Planck-89.} We detected the CO(4--3) and CO(5--4) lines, with $\mathrm{S/N}$ ratios of 49.1 and 69.0, respectively. The CO(4--3) line was previously detected by the Planck All-Sky Survey to analyse Gravitationally-lensed Extreme Starbursts (PASSAGES) observations \citep[][called PJ133634.9]{Berman2022} with the Large Millimeter Telescope (LMT). Their integrated line flux, $\mu S_{\text{CO(4--3)}}\Delta v = 21.9\pm 1.1\,\hbox{Jy}\,\hbox{km}\, \hbox{s}^{-1}$, is slightly lower than ours ($\mu S_{\text{CO(4--3)}}\Delta v = 26.99\pm 0.56\,\hbox{Jy}\,\hbox{km}\, \hbox{s}^{-1}$).

The profiles of both lines were fitted with double Gaussians. Unlike Planck-68, the blueshifted components are subdominant. The CO(4--3) line was observed at $108.368\pm0.005\,\text{GHz}$ and has a FWHM of $416~\text{km~s}^{-1}$. The redshifted component is at $96\,\text{km~s}^{-1}$ (FWHM = $278\,\text{km~s}^{-1}$). The blueshifted component is at $-208\,\text{km~s}^{-1}$ (FWHM = $214\,\text{km~s}^{-1}$).

The CO(5--4) line was detected at an observed frequency of $135.446\pm0.007\,\text{GHz}$. The overall FWHM of the line is $434\,\text{km~s}^{-1}$. The redshifted component has a velocity of $95\,\text{km~s}^{-1}$ and a FWHM = $280\,\text{km~s}^{-1}$. The blueshifted component is at $-214\,\text{km~s}^{-1}$ and has a FWHM = $258\,\text{km~s}^{-1}$. 

\subsubsection*{Spectroscopic redshift}\label{sec:spec_redshift}

To determine the spectroscopic redshift of the target sources, we exploited all line detections. 
To combine the information from multiple lines, we calculated a weighted average of redshift estimates, with weights given by the inverse square of the frequency uncertainties. We propagated uncertainties using standard error propagation and further refined the estimate via a Monte Carlo approach. For each line, we generated $10^5$ realisations of $\nu_{\rm obs}$ assuming Gaussian errors, computed the corresponding redshifts, and derived the final redshift distribution. The median of this distribution was adopted as the best estimate, while the $1\,\sigma$ uncertainty was derived from its standard deviation. The resulting distribution was fitted with a Gaussian model to verify consistency. 

For the three sources with double-peaked profiles, we estimated an additional systematic uncertainty due to differential magnification: in a fold-lens configuration, the two kinematic components of a rotating disk may be magnified by different factors, biasing the intensity-weighted centroid toward the more magnified component.
Adopting a conservative magnification gradient of $\mu_1/\mu_2 = 2$ between the two peaks \citep[as expected for fold-lens configurations;][]{Blain1999,Serjeant2012}, we computed the maximum resulting centroid shift and added it in quadrature to the formal Monte Carlo uncertainty.
All four sources have $z_{\rm spec} > 2.3$, confirming the photometric redshift estimates; the total $1\sigma$ absolute uncertainties in the dimensionless redshift is of $\delta z \approx 5 \times 10^{-4}$ for the double-peaks lines while $\delta z \approx 5 \times 10^{-5}$ for Planck-188 (see Table~\ref{table:pos}).
In the case of Planck-89, our redshift is in excellent agreement with that of \citet{Berman2022}.

\subsubsection*{Line morphology and image-plane kinematics}\label{sec:moment_maps}

To investigate the spatial distribution and the kinematics of the molecular gas, we produced moment maps from the calibrated, continuum-subtracted data cubes of each detected CO and [CI] transition. The zeroth-moment map ($\mathrm{mom.\,0}$) traces the integrated line intensity and provides the morphology and extent of the CO-emitting regions, which we compared with the continuum maps discussed in Sect.~\ref{sec:continuum}. This allowed us to assess whether dust and gas emission are co-spatial or show offsets that may indicate different distributions of star formation and molecular gas. 

The first- and second-moment maps ($\mathrm{mom.\,1}$ and $\mathrm{mom.\,2}$) were computed to characterize the line-of-sight velocity field and velocity dispersion, respectively. These maps offer a view of the image-plane kinematics of the lensed systems. While gravitational lensing prevents a direct interpretation in the source plane, the maps still reveal projected velocity gradients and kinematic structures that are essential for constraining lens models and will be used for source-plane reconstructions in future work. The moment maps were created using the \texttt{CASA} task \texttt{immoments}, integrating over the same velocity ranges used for the line flux extraction. For $\mathrm{mom.\,1}$ and $\mathrm{mom.\,2}$ we applied a $3\sigma$ mask, while just the $1\sigma$ mask for  $\mathrm{mom.\,0}$, using the rms estimated from line-free channels, to limit noise contamination.

The bottom panels of Figure~\ref{fig:Planck41_all} show the moment maps of the detected CO lines of Planck-41. The moment maps of the other sources are presented in  Appendix~\ref{app:line_figs}. In what follows, we describe the main features of each source.

\paragraph*{Planck-41.}
The integrated emission of CO(3--2) and CO(4--3) is as extended as the dust continuum and reveals more clearly the multiplicity of lensed images. In particular, the central continuum component ($\ell$) seen at 2\,mm disappears in the CO(4--3) $\mathrm{mom.\,0}$. The CO(3--2) $\mathrm{mom.\,0}$ map resolves the multiple images thanks to the absence of what we can now associate with the $\ell$ component, which dominates the continuum. 

The $\mathrm{mom.\,1}$ velocity maps display a sharp gradient between redshifted and blueshifted regions, in agreement with the double-peaked spectra. The CO(3--2) velocity map is nearly symmetric, while the higher resolution of the CO(4--3) map reveals multiple velocity components displaced across different lensed images. The $\mathrm{mom.\,2}$ maps show enhanced dispersion at the interface between the velocity components, consistent with strongly sheared kinematics. These features suggest that the double-peaked profiles are tracing spatially distinct kinematic components (e.g. a merger or a rotating disk) mapped into different image-plane positions by the lensing configuration.

\paragraph*{Planck-68 }
The $\mathrm{mom.\,0}$ maps of CO(3--2) and CO(4--3) lines largely coincide with the continuum emission, showing no clear substructure beyond what is visible in the dust emission. The [CI](1--0) line, detected at lower $\mathrm{S/N}$, appears compact and co-spatial with the CO, confirming that it traces the same molecular gas as the CO lines. The $\mathrm{mom.\,1}$ velocity maps show hints of the secondary, redshifted component seen in the spectra, but the resolution and the $\mathrm{S/N}$ limit further interpretation. The velocity dispersions in $\mathrm{mom.\,2}$ are modest and do not reveal strong gradients. Overall, the gas distribution appears to be less complex than in Planck-41, with asymmetry primarily evident in the line profiles. The [CI](1--0) line was the only one for which we applied a 2 $\sigma$ level mask for the computation of the moment maps, on account of its relatively weak detection.

\paragraph*{Planck-89.}
Given the extended Einstein-ring morphology previously identified in high-resolution VLA observations \citep{Berman2022}, our NOEMA data appear to indicate that this structure persists in the 2\,mm continuum and molecular gas. The 2\,mm continuum imaging shows a marginally resolved structure spanning $\sim1''$, which is broadly consistent with the presence of multiple components ($a \text{ and } b$) as seen in the radio. However, the current angular resolution ($\sim1.2''$) does not allow for a more detailed decomposition of these features. 

The $\mathrm{mom.\,0}$ maps of the CO lines reveal a dominant northern emission peak, spatially coincident with component $a$, and a fainter southern extension encompassing the region of $b$. The $\mathrm{mom.\,1}$ velocity maps show a smooth gradient from the redshifted northern peak toward the blueshifted southern structure, mirroring the morphology seen in the continuum. Both CO(4--3) and CO(5--4) exhibit similar spatial and kinematic distributions.

\paragraph*{Planck-188.}
The integrated emission of CO(3--2) and CO(4--3) lines is compact and has a morphology nearly identical to that of the dust continuum. The $\mathrm{mom.\,0}$ maps show no evidence for multiplicity or extended features. In agreement with the single-Gaussian line profiles, the $\mathrm{mom.\,1}$ maps reveal no velocity gradients, while the $\mathrm{mom.\,2}$ maps show uniformly low dispersions ($\sim 50\,{\rm km}\,{\rm s}^{-1}$). Planck-188 is therefore the only source consistent with a quiescent, dynamically simple system, lacking the complex kinematics observed in the other three sources, or with a face-on rotating disk, in which the projected rotation velocity is suppressed; this possibility is discussed further in Sect.~\ref{sec:double_peak}.

\subsubsection*{Double-peak profiles}\label{sec:double_peak}

The CO line profiles of three out of the four sources in our sample (Planck-41, Planck-68, and Planck-89) exhibit a clear double-peaked structure (see Fig.\,\ref{fig:Planck41_all}, \,\ref{fig:Planck68_all} and \,\ref{fig:Planck89_all} ). This prevalence ($\sim 75\%$) is notably high compared to results from larger surveys of unlensed sub-millimetre galaxies (SMGs); \citet{Bothwell2013} found that only 20--28\% of their 32 CO-detected SMGs exhibited double-peaked profiles.
With only four sources, however, this comparison carries limited statistical weight and no robust conclusion about the underlying prevalence can be drawn.

Double-peaked CO profiles are typically interpreted as signatures of ordered rotation in a massive disk or of an ongoing major merger, though disk inclination strongly modulates the observed profile: a rotating disk viewed close to face-on produces a single-peaked, narrow line because the projected line-of-sight velocity is small regardless of the true rotation speed.
Planck-188, the only single-peaked source, has the smallest FWHM in the sample and sits at the low-linewidth end of Fig.\,\ref{fig:tully_fischer}, consistent with a near face-on orientation.

Flux-limited lensed samples may also be biased toward double-peaked sources through a combination of effects: strongly lensed systems preferentially lie near fold or cusp caustics, where sources with elongated projected morphologies (high inclination of a disk or merger) subtend a larger solid angle in the high-magnification region and are thus more susceptible to strong lensing. On the other hand, an orientation bias in the opposite direction is expected based on the simulations by \citet{Lovell2022}, who predicted lowest sub-mm emission when the disc is viewed edge-on, and highest emission when it is face-on.
The finding of \citet{Bothwell2013} that double-peaked SMGs have far-infrared luminosities $\approx 20\%$ higher than single-peaked ones suggests that the first effect prevails.

\section{Lens modelling of Planck-41}
\label{sec:lens_modelling}

In this section, we present the gravitational lens modelling of the only NOEMA-detected system in our sample for which the lensing features are clearly resolved: Planck-41. High-resolution imaging of the lensed images is a fundamental requirement for reliable strong-lensing reconstruction, as insufficient angular resolution can lead to degenerate or poorly constrained mass and source models. For this reason, we restricted our analysis to Planck-41.

To perform the lens modelling, we employed the open-source Python package \texttt{PyAutoLens}\footnote{\url{https://github.com/Jammy2211/PyAutoLens}} \citep{pyautolens, Nightingale2015, Nightingale2018}. \texttt{PyAutoLens} is a flexible modelling framework designed to analyse both imaging and interferometric datasets through parametric and non-parametric approaches. It implements the regularised semi-linear inversion (SLI) technique \citep{SLI}, combined with an adaptive source-plane pixelization scheme \citep{Nightingale2015}, enabling robust reconstruction of the unlensed source while simultaneously constraining the lens mass distribution. For interferometric observations, such as those presented here, the modelling is performed directly in the visibility (\textit{uv}) plane, ensuring that the full information content of the data is preserved and avoiding systematic effects associated with image-plane deconvolution \citep[e.g.][]{powell21}. The extension of these techniques to interferometric datasets has been demonstrated in several previous works \citep{Massardi18, Dye18, Enia18, Dye22, Maresca22}.

In this work, we focus on modelling the total mass distribution of the lens galaxy and the light distribution of the lensed source.
We choose to model the 2\,mm continuum emission, which maximises the number of visibilities and provides the highest signal-to-noise ratio for robust parameter inference. A more comprehensive modelling, including the second continuum band and molecular line emission, will be presented in a dedicated future study.

\begin{figure}
    \includegraphics[width=\linewidth]{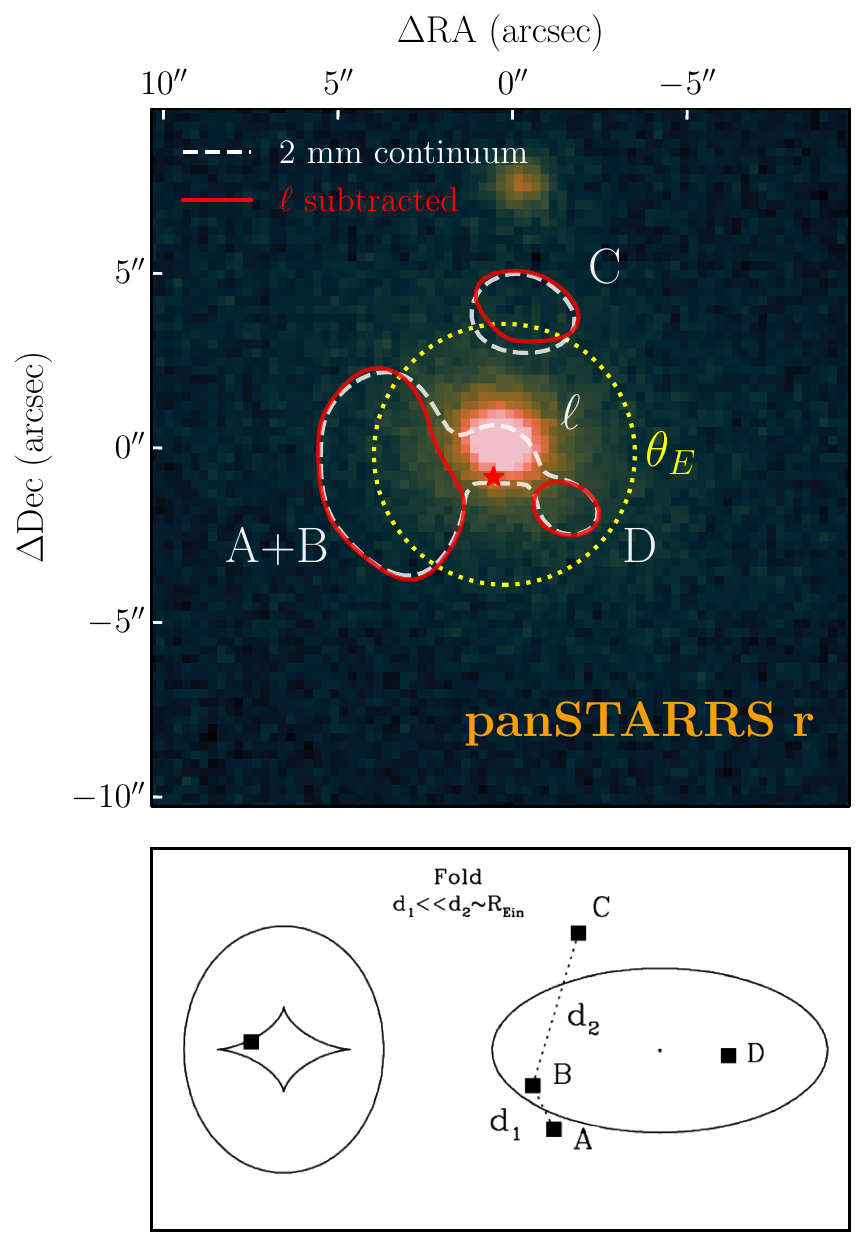}
    \caption{PanSTARRS $r$-band image of the field of Planck-41. The red curve is the $3\sigma$ contour of the 2\,mm continuum after subtraction of the central lens component ($\ell$) visibilities, while the dashed white curve is the total $3\sigma$ contour of the 2\,mm continuum. The dotted yellow circle shows the derived Einstein radius of the lens mass distribution. The red star marks the reconstructed position of Planck-41 in the source plane. Bottom panel: schematic of the fold caustic configuration \citep{keeton03} adapted to the observed image geometry of Planck-41.}
    \label{fig:2mass_lens_subtracted}
\end{figure}

As described in the previous sections, Planck-41 exhibits a complex and highly distorted lensed morphology, clearly visible in the NOEMA 2\,mm continuum map (Fig.~\ref{fig:cont_maps}) and in the moment maps of both detected CO transitions (bottom panels of Fig.~\ref{fig:Planck41_all}).
The observed configuration, consisting of the blended $A+B$ components, the isolated image $C$, and the counter-image $D$, is characteristic of a ``fold'' caustic geometry produced by an elliptical lens mass distribution, as shown schematically in the bottom panel of Fig.~\ref{fig:2mass_lens_subtracted}.
In such systems, the separation between images $A$ and $B$ is significantly smaller than the distance between $B$ and $C$, which is typically comparable to the Einstein radius of the lens.
Given this well-recognised morphology and the moderate angular resolution of the data, we adopted a parametric lens modelling approach to recover both the mass properties of the lens galaxy and the intrinsic structure of the background source.

The 2\,mm continuum emission also reveals a central component, labelled $\ell$, which is absent in the integrated CO line emission, particularly in the CO(3--2) moment-zero map where it would otherwise dominate (Fig.~\ref{fig:Planck41_all}), indicating that it is associated with a foreground source rather than with the lensed background galaxy.

We identified the central component $\ell$ as the foreground deflector galaxy 2MASX J17451577$+$4031028, detected in 2MASS and at higher resolution by PanSTARRS (Fig.~\ref{fig:2mass_lens_subtracted}, top panel).
Its position is consistent with the NOEMA continuum peak to within $0\farcs78$.
The photometric redshift $z_\ell = 0.216 \pm 0.015$ from the 2MPZ catalogue \citep{Bilicki2014} relies on near-infrared (NIR) photometry that may be partially contaminated by the lensed images. However, the NIR colours of a dusty galaxy at $z=2.348$  are very different from those of a galaxy at $z\simeq 0.2$, implying that the contamination by the background lensed galaxy must be small. A more refined photometric redshift $z_\ell = 0.291$, based on DESI Legacy Survey $g$, $r$, $z$ and WISE photometry using a Gaussian process machine-learning method \citep{Duncan2022}, is adopted for the lens modelling.
The galaxy is associated to a radio AGN in the LoTSS DR2 catalogue \citep[ILTJ174515.78$+$403102.7;][]{Hardcastle2025, Hardcastle2023}, with a 144\,MHz luminosity, $L_{144} \approx 10^{24}$\,W\,Hz$^{-1}$. 

The NOEMA continuum peak flux densities of the $\ell$ component, $S_{3\,\rm mm} = 1.02$\,mJy\,beam$^{-1}$ and $S_{2\,\rm mm} = 0.66$\,mJy\,beam$^{-1}$, yield a spectral index $\alpha = -1.0 \pm 0.5$ ($S_\nu \propto \nu^\alpha$), characteristic of optically thin synchrotron emission ($\alpha \sim -0.7$ to $-1.2$) from the radio AGN \citep{Hardcastle&Croston2020}, rather than free-free radiation ($\alpha \approx -0.1$).

Since our primary goal is to model the lensed emission of the background source without introducing an explicit model for the lens light distribution, we removed the foreground emission associated with the $\ell$ component directly in the visibility domain. This subtraction was performed by generating a clean model of the foreground galaxy using the \texttt{tclean} task and subtracting it from the data with the \texttt{uvsub} task in \texttt{CASA}. The resulting $3\sigma$ contours of the lens-subtracted 2\,mm continuum emission are shown in Fig.~\ref{fig:2mass_lens_subtracted} overlaid on the PanSTARRS $r$-band image of the foreground galaxy, confirming that the central component disappears after lens subtraction. 

The lens modelling was performed using the \texttt{PyAutoLens} package following the ``Source–Light–Mass'' (\texttt{SLaM}) pipeline strategy. In this framework, the foreground lens galaxy was modelled as a Singular Isothermal Ellipsoid (SIE; \citealt{kormann94}), implemented as an elliptical power-law mass profile characterised by the Einstein radius $\theta_{\rm E}$, the lens centre coordinates $(x^{\rm lens}, y^{\rm lens})$, the ellipticity components $(e_x, e_y)$, and the logarithmic slope $\gamma$. The lens redshift was fixed to the photometric value during the modelling.

The intrinsic light distribution of the background galaxy was described using a parametric Sérsic profile, characterised by the effective radius $R_{\rm eff}$, the Sérsic index $n$, the centroid position $(x^{\rm source}, y^{\rm source})$, and the ellipticity components. This parametric representation provides a physically motivated description of the source morphology that is adequate for the angular resolution and signal-to-noise ratio of the NOEMA observations. Further details of the modelling procedure and fitted parameters are presented in Appendix~\ref{app:lens_parameters}.

The magnification derived from the continuum model is $\mu_{\rm cont} = 10.8^{+1.7}_{-1.3}$, corresponding to a de-lensed effective radius of $R_{\rm eff} \simeq 0.8$\,arcsec, or $\simeq 6.4$\,kpc at the source redshift. The model yields an Einstein radius of $\theta_{\rm E} = 3.7 \pm 0.2\ \mathrm{arcsec}$. This reconstruction confirms the initial visual classification of Planck-41 as a fold-configuration lens system and demonstrates the ability of the \texttt{SLaM} parametric pipeline to robustly recover the intrinsic properties of strongly lensed dusty star-forming galaxies. 

The close spatial correspondence between the CO(4--3) emission, as seen in the moment-zero map (Fig.~\ref{fig:Planck41_all}), and the lens-subtracted 2\,mm continuum emission justifies the use of the continuum magnification factor to demagnify the molecular line emission. Applying this correction, we derive $L'_{\rm CO(4-3)} = 5.4^{+0.6}_{-0.8} \times 10^{10}\,\mathrm{K\,km\,s^{-1}\,pc^{2}}$ and $L_{\rm CO(4-3)} = 0.73^{+0.09}_{-0.11} \times 10^{8}\,L_\odot$.

The lens model presented here is limited by the angular resolution of the NOEMA observations: Planck-41 is marginally resolved at $1\farcs2$, with the full image configuration spanning roughly four synthesised beams.
While fitting directly in the visibility plane preserves the full interferometric information and avoids image-plane deconvolution artefacts, it does not compensate for this fundamental resolution limit.
With twelve free parameters across the mass and source models, residual degeneracies are unavoidable: most notably between $\theta_{\rm E}$ and $R_{\rm eff}$, since a larger Einstein radius paired with a larger source can produce image configurations similar to a more compact geometry at lower magnification, and between the power-law slope $\gamma$ and the assumed source brightness distribution, which cannot be independently constrained when the source is unresolved.
To minimise these effects, we deliberately kept both models as simple as possible, fixed the lens centre to the PanSTARRS $r$-band peak in the first \texttt{SLaM} stage, and progressively freed parameters across the two steps.
The physical consistency of the solution is supported by two independent checks: the fold morphology (images A+B, C, D) is correctly reproduced. The Einstein mass enclosed within $\theta_{\rm E}$, $M_{\rm E} \simeq 2.0 \times 10^{12}\,M_\odot$, is consistent with the stellar mass of the lens galaxy, $\log(M_\star/M_\odot) = 12.04^{+0.10}_{-0.10}$ \citep{Hardcastle2023}, the difference being expected from the dark matter contribution within the Einstein ring.

\section{Source properties}\label{sec:stat_prop}

This section compares the properties of our sources with those of other samples at similar redshifts, magnifications and luminosities. We consider primarily the PASSAGES (\citealp[]{Harrington21,Berman2022,Kamieneski24}) galaxies and the Planck Dusty GEMS (\citealp[]{Canameras2015}), also selected from Planck surveys. We also include in the comparison the \textit{Herschel}-selected high-$z$ dusty galaxies from the $z$-GAL survey (\citealp{cox23, ismail23, berta23}), including its VLA follow-up, V$z$-GAL (\citealp[]{Prajapati2026}).

The average redshift of our sources, $2.6 \pm 0.3$, aligns well with the median spectroscopic redshift of the $z$-GAL survey \citep[$z = 2.56 \pm 0.10$;][]{cox23}.
This value falls within the core distribution of the PASSAGES sample ($z=1.1-3.3$). It partially overlaps the redshift distribution of the slightly more redshifted Planck Dusty GEMS ($2.9 \pm 0.4$), which also utilised CO and [CI] detections for redshift estimates.
It is also consistent with the spectroscopic redshift distribution of the \textit{Herschel} Bright Sources sample \citep[HerBS; $z = 3.07 \pm 0.72$, 22 sources;][Table\,6]{Bakx2018}, despite the different selection function.

\subsection{The $L'_{CO(1-0)} - \mathrm{FWHM}$ relationship}\label{sec:tully_fischer}

The correlation between line luminosity ($\mu L'_{\rm{CO(1-0)}}$) and gas linewidth ($\rm{FWHM}$) serves as a CO-based proxy for the Tully-Fisher relation (\citealp[]{tully&fischer77, dick&kazes92,schoniger94}). While unlensed galaxies follow a roughly quadratic relationship between $L'_{\rm{CO(1-0)}}$ and FWHM, lensed sources appear as extreme outliers, typically 1--2 orders of magnitude above the baseline relation. This vertical offset occurs because gravitational lensing boosts the apparent luminosity without affecting the linewidth \citep{harris12}. This diagnostic effectively distinguishes lensed sources (high line luminosity with linewidths $200 - 700\,\rm{km}\,\rm{s}^{-1}$) from intrinsically hyper-luminous infrared galaxies (HyLIRGs), which exhibit exceptionally broad lines (${\rm FWHM} > 700\,\rm{km}\,\rm{s}^{-1}$) and follow the unlensed trend \citep{cox23}.

We computed the equivalent $\rm{FWHM}$ using the intensity-weighted second moment ($s_v^2$; see Sect.~\ref{sec:spectroscopy}). This profile-independent approach is more reliable than the single-Gaussian fits, used, e.g., in PASSAGES. The latter approach may oversimplify the kinematics and overestimate velocity dispersions. Nevertheless, our FWHMs lie within their range ($\sim 150-685\,\rm{km}\,\rm{s}^{-1}$). For the common source, Planck-89, our estimate ($416 \pm 29\,\rm{km}\,\rm{s}^{-1}$) is narrower than the PASSAGES value ($532 \pm 6\,\rm{km}\,\rm{s}^{-1}$). 
Conversely, the z-GAL sample exhibits a broader velocity range, much more extended towards large FWHM (150--$1750\,\rm{km}\,\rm{s}^{-1}$, with an average of $590 \pm 25\,\rm{km}\,\rm{s}^{-1}$). This is due to the fact that broad-line HyLIRGs and mergers were also targeted.

The ground-state luminosities were derived using the CO line ratios ($r_{J1}$) from PASSAGES \citep[][ $r_{31} = 0.69 \pm 0.12$, $r_{41}= 0.52\pm0.14$, and $r_{51}=0.37\pm0.15$]{Harrington21}, as our sources share the same selection for rare, hyper-luminous lensed galaxies. The luminosity was computed via:

\begin{equation}
   \mu L'_{\mathrm{CO(1-0)}} = \frac{\mu L'_{\mathrm{CO,J}}}{r_{J1}}.
\end{equation}\label{eq:lum_co10}

Our sources span the luminosity interval $\mu L'_{\mathrm{CO(1-0)}} \sim (37\text{--}166) \times 10^{10}~\mathrm{K~km~s^{-1}~pc^2}$ (Table~\ref{tab:line_fluxes_Lprime}). The luminosities of Planck-41, 68 and 188 are in the upper part of the z-GAL distribution, but still below the maximum luminosity, while Planck-89 is substantially above it. This reflects the fact that the Planck selection picks up extreme gravitational magnifications.

\begin{figure}
    \includegraphics[width=\linewidth]{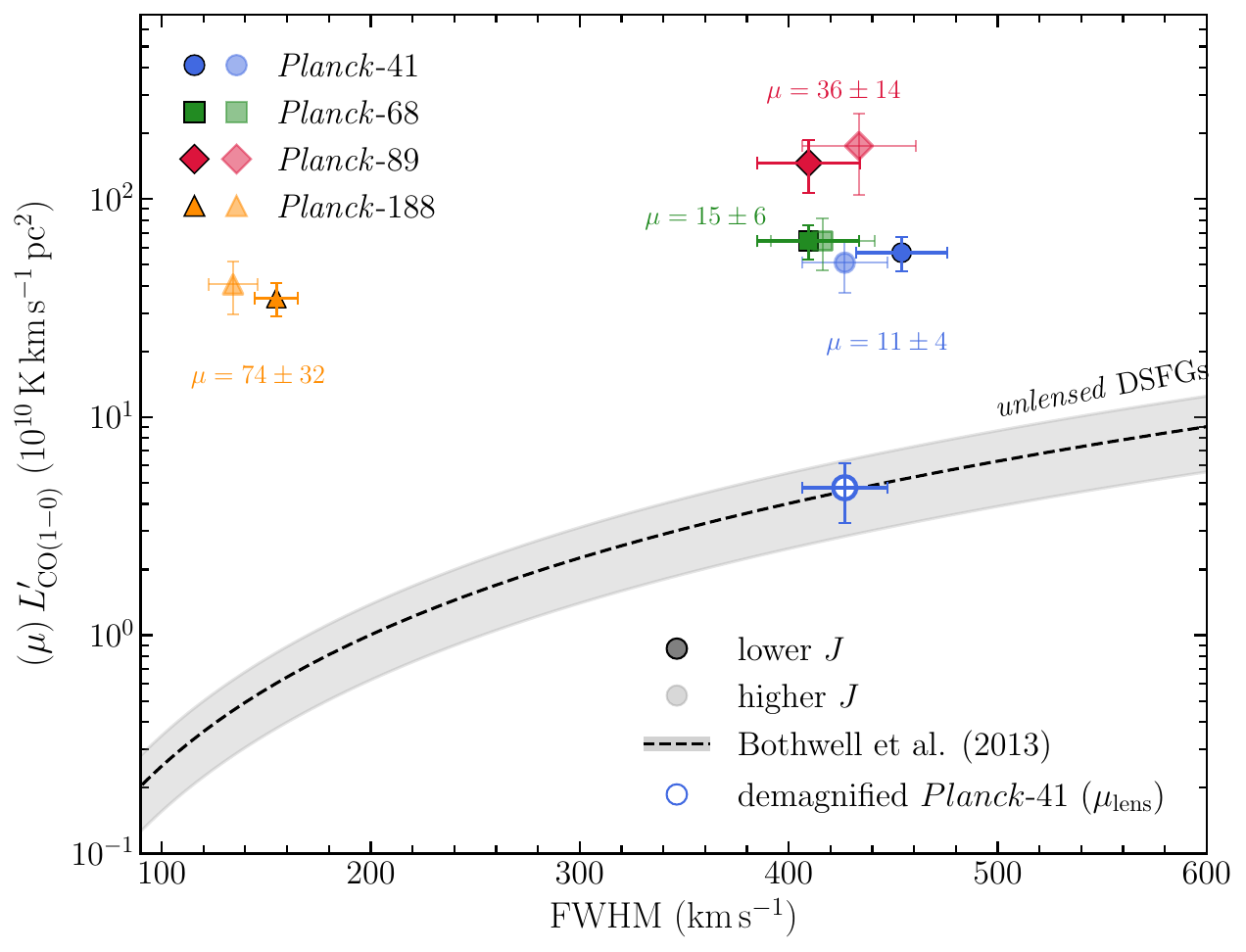}
    \caption{$\mu L'_{\rm CO(1-0)}$ versus line FWHM for our four sources, compared with the empirical relation of \citet{Bothwell2013} (dashed line) and its $\pm 38\%$ intrinsic scatter (grey band).
    For each source, filled and semi-transparent markers show the lower- and higher-$J$ transitions, respectively.
    The open circle shows the demagnified luminosity of Planck-41 using the lens-model magnification.
    }
    \label{fig:tully_fischer}
\end{figure}

Figure~\ref{fig:tully_fischer} shows $\mu L'_{\rm{CO(1-0)}}$ versus $\rm{FWHM}$ for our galaxies compared with the empirical relation from \cite{Bothwell2013}, $\mu L'_{\rm{CO(1-0)}}=10^{5.4}\times\Delta V^2$.
For each source, we computed $\mu L'_{\rm{CO(1-0)}}$ for each available CO transition, obtaining consistent estimates in all four cases.
All four sources sit well above the scatter band of the unlensed relation, confirming their strongly lensed nature.
As proposed by \citet{harris12}, the vertical offset yields an estimate of the magnification; combining the two available transitions per source, we obtain $\mu = 11 \pm 4$, $15 \pm 6$, $36 \pm 14$, and $74 \pm 32$ for Planck-41, -68, -89, and -188, respectively.
The uncertainties are dominated by the $38\%$ intrinsic scatter of the Bothwell relation, applied once as a correlated systematic on the combined estimate; inter-transition discrepancies contribute negligibly for Planck-41, -68, and -89, but are significant for Planck-188 (21\%).
For Planck-41 the CO-based estimate ($\mu = 11 \pm 4$) is in excellent agreement with the lens modelling result ($\mu_{\rm cont} = 10.8^{+1.7}_{-1.3}$, Sect.~\ref{sec:lens_modelling}), independently cross-validating both methods. Dividing the CO(4--3) luminosity by $\mu_{\rm cont}$ (the lens model was built on the continuum sub-band containing that line) places the demagnified point within 2\% of the Bothwell prediction at the same FWHM (Fig.~\ref{fig:tully_fischer}).
However, this concordance should not be over-emphasised: the large scatter of the relation means that CO-based $\mu$ estimates are unreliable in general, as illustrated by Planck-89, for which the relation gives $\mu = 36 \pm 14$, compared with $\mu_{\rm VLA} = 8.3$ from lens modelling \citep{Kamieneski24}.

\subsection{Molecular gas mass}\label{sec:gas_mass}

The molecular gas mass, $M_{\rm mol}$\footnote{We use the definition from \cite{berta23}: $M_{\rm mol} = M_{\rm H_2}+M_{\rm He} = 1.36 \times M_{\rm H_2}$ where $M_{\rm H_2}$ is the molecular hydrogen mass and $M_{\rm He}$ the helium mass.}, can be derived from the integrated CO line luminosity using the relation $M_{\rm mol} = \alpha_{\text{CO}} L'_{\text{CO}(1-0)}$ \citep{bolatto13}. Following \citet{berta23}, we adopt a conversion factor  $\alpha_{\text{CO}} = 4.0 \, M_{\odot} \, (\text{K km s}^{-1} \text{pc}^2)^{-1}$.

To mitigate conversion uncertainties, we use the value of $L'_{\text{CO}(1-0)}$ derived from data for the lowest-J transition for each source (see Sect.~\ref{sec:tully_fischer}). 

The resulting apparent (magnified) molecular gas masses for our sample span the interval $\mu M_{\rm mol} \sim (14 - 60) \times 10^{11} \, M_{\odot}$. This range is notably higher than that of the \textit{Herschel}--selected Vz-GAL sample ($2 - 20 \times 10^{11} \, M_{\odot}$; \citealp{Prajapati2026}), but is within that for the Planck-selected PASSAGES galaxies ($0.5 - 90 \times 10^{11} \, M_{\odot}$; \citealp{Berman2022}), reflecting the Planck selection of the most extreme lensed systems. Planck-89 stands out with a magnified mass of $\sim 58.5 \times 10^{11} \, M_{\odot}$, in excellent agreement with the value of $55.7 \pm 8.4 \times 10^{11} \, M_{\odot}$ reported by \cite{Berman2022}.

In the case of Planck-68, the [CI](1--0) transition provides an independent tracer of the total molecular gas mass. This line is typically optically thin, meaning that it is immune to saturation effects that can affect CO lines. We derive a magnified gas mass of $\mu M_{\text{mol}, [\text{CI}]} \simeq (8.9 \pm 4.6) \times 10^{11} \, M_{\odot}$, using the average conversion factor for Planck-selected sources, $\alpha_{[\rm{CI}]} = 16.2 \pm 7.9 \, M_{\odot} \, (\text{K km s}^{-1} \text{pc}^2)^{-1}$ \citep{Harrington21}. This [CI] estimate is a factor of $\sim 2.8$ lower than the CO-derived mass ($\mu M_{\rm mol} \simeq 25.7 \times 10^{11} \, M_{\odot}$). However, the discrepancy is not statistically significant, given the large uncertainties due to the relatively low $S/N$ ratio of the [CI] detection.

For Planck-41 we obtain $\mu M_{\rm mol} \simeq 22.7 \times 10^{11} \, M_{\odot}$. Applying the magnification factor $\mu_{\rm cont} \approx 10.8$ from our lens model, we derive an intrinsic gas mass of $M_{\rm mol} = 2.1^{+0.2}_{-0.3} \times 10^{11} \, M_{\odot}$, consistent with earlier results for massive, gas-rich starbursts at $z \sim 2-3$ \citep[e.g.,][]{harris12}. We also compare the molecular gas mass to the dynamical mass derived using the virial estimator \citep{Bothwell2013}: $M_{\rm dyn}/M_{\odot} = 1.56 \times 10^6 \sigma^2 R$. Here, $\sigma$ is the velocity dispersion ($s_v$) used to compute the FWHM (Sect.~\ref{sec:spectroscopy}), and $R$ is the source radius. Adopting the effective radius ($R_{\rm eff}$) of the S\'ersic profile derived from our lens modelling, we obtain $M_{\rm dyn}=3.7^{+1.2}_{-1.0} \times 10^{11} \, M_{\odot}$, so that the intrinsic molecular gas mass, $M_{\rm mol}$, hence the baryon fraction, represents a major fraction of the total dynamical mass within the CO emitting region. The interpretation of this result is complicated by the substantial uncertainty on both $M_{\rm mol}$ and $M_{\rm dyn}$. On one side, a commonly used value of 
$\alpha_{\text{CO}}$ is about a factor of 4 lower \citep[$\alpha_{\text{CO}}=0.8$;][]{DownesSolomon1998} than that used here. On the other side, the value of the virial coefficient used to compute $M_{\rm dyn}$ is also endowed with a substantial uncertainty. In the case of Planck-89, the gravitational lens modelling by \citet{Kamieneski24} yielded $\mu=8.3^{+4.2}_{-1.9}$ and $R_{\rm eff}=3.3\,$kpc. Combining with the measurement of the velocity dispersion, we get $M_{\rm dyn}\simeq 2 \times 10^{11} \, M_{\odot}$, lower than the de-lensed estimate of $M_{\rm mol}\simeq 7\times 10^{11} \, M_{\odot}$. 

\section{Results and conclusions}\label{sec:discussion_conclusion}

This work utilised NOEMA observations to robustly secure spectroscopic redshifts and to characterise morphology and kinematics of four ultra-bright dusty star-forming galaxies (DSFGs) initially selected from the Planck Catalogue of Compact Sources (PCCS2). All four targets were confirmed to be high-redshift systems, with spectroscopic $z_{\rm spec} > 2.3$ and with $1\sigma$ uncertainties $\delta z \approx 10^{-4}$. The mean redshift of the sample, $z = 2.6 \pm 0.3$, is consistent with the core of the PASSAGES redshift distribution \citep{Harrington21,Berman2022,Kamieneski24} as well as with the median redshift of the z-GAL sources \citep{cox23,berta23,Prajapati2026} and of the Planck Dusty GEMS sample \citep{Canameras2015}, indicating that our sources are representative of the broader population of strongly lensed DSFGs. It is also consistent with the spectroscopic redshift distribution of the \textit{Herschel} Bright Sources sample \citep[HerBS; $z = 3.07 \pm 0.72$, 22 sources;][Table\,6]{Bakx2018}, despite the different selection function.

Three out of the four systems (Planck-41, Planck-68, and Planck-89) exhibit complex kinematics characterised by broad, double-peaked CO line profiles, while Planck-188 shows comparatively narrow, single-peaked lines and no detectable velocity gradients. We note that our fraction of double-peaked CO lines is higher than that found by previous studies. \citet{Bothwell2013}  found that only seven, perhaps nine, of their 32 luminous sub-millimetre galaxies (SMGs) show double-peaked CO lines. A somewhat higher fraction (four to six out of 12) was reported by \citet{Greve2005}.   

The median linewidth of the sample ($\rm{FWHM} \simeq 410\,\rm{km\,s^{-1}}$) is consistent with values reported by other Planck-selected lensing surveys and significantly narrower than those typical of intrinsically hyper-luminous infrared galaxies, supporting the interpretation that these sources are strongly magnified rather than intrinsically extreme.

All four galaxies lie well above the unlensed $\mu L'_{\rm CO(1-0)}$--FWHM relation, providing an independent, population-level confirmation of their lensed nature. When multiple CO transitions were available, magnification estimates from individual transitions were combined to reduce the impact of excitation uncertainties. While magnification estimates derived from this relation are subject to systematic uncertainties, they are broadly consistent with detailed lens modelling where available: for Planck-41, the CO luminosity--linewidth estimate ($\mu = 11 \pm 4$) is in good agreement with the continuum lens model result ($\mu_{\rm cont} = 10.8^{+1.7}_{-1.3}$).

The apparent molecular gas masses inferred from CO and [CI] span $\mu M_{\rm mol} \sim (1.4$--$7.0)\times10^{12}\,M_\odot$, placing our targets among the most gas-rich systems found in current high-redshift surveys. 

A dedicated parametric lens modelling was also performed for Planck-41 with \texttt{PyAutoLens}, yielding a configuration consistent with strong gravitational magnification and reproducing the observed arrangement of the continuum components in a geometry compatible with a fold-type lens configuration. The central continuum component $\ell$ is identified as the foreground deflector galaxy, classified as a radio AGN in the LoTSS DR2 catalogue \citep{Hardcastle2023} with a photometric redshift $z_\ell = 0.291$ \citep{Duncan2022}; its mm spectral index $\alpha = -1.0 \pm 0.5$ is consistent with optically thin synchrotron emission, ruling out a free-free origin. The Einstein mass, $M_{\rm E} \sim 2.0 \times 10^{12} \,M_\odot$, is consistent with the stellar mass of the lens galaxy, $\log(M_\star/M_\odot) = 12.04\pm 0.1$ \citep{Hardcastle2023}, the difference being expected from the dark matter contribution within the Einstein ring.

While these results reinforce the lensing interpretation suggested by the CO linewidth-luminosity relation, the lens model is limited by the angular resolution of the NOEMA data and a definitive confirmation will require higher angular resolution observations.
In particular, high-resolution radio continuum imaging using Very Long Baseline Interferometry (VLBI) at 1.4 or 5 GHz (e.g., with the European VLBI Network, EVN) or sub-arcsecond interferometric imaging would allow the direct identification of multiple images or arcs. Complementary multi-wavelength observations, especially at optical and near-infrared wavelengths, would further enable the separation of the foreground lens galaxy from the background DSFG, providing a direct characterisation of the lens-source geometry.

\section*{Data availability}
 
Tables~\ref{table:pos}, \ref{table:cont_flux_densities}, and \ref{tab:line_fluxes_Lprime} are only available in electronic form at the CDS via anonymous ftp to \texttt{cdsarc.u-strasbg.fr} (130.79.128.5) or via \url{http://cdsweb.u-strasbg.fr/cgi-bin/qcat?J/A+A/}. The reduced NOEMA continuum images and CO/[CI] line cubes discussed in Sect.~\ref{sec:analysis} are also available at the CDS as associated FITS data.

\begin{acknowledgements}

M.B. acknowledges support from INAF under the mini-grant ``A systematic search for ultra-bright high-z strongly lensed galaxies in Planck catalogues''.
L.T. thanks James Nightingale, Ernest Rutherford Fellow at Newcastle University, developer of PyAutoLens, for the technical support with \texttt{PyAutoLens}. M. G. acknowledges support from INAF under the following funding schemes: Large Grant 2022 (project "MeerKAT and LOFAR Team up: a Unique Radio Window on Galaxy/AGN co-Evolution") and Large GO 2024 (project "MeerKAT and Euclid Team up: Exploring the galaxy-halo connection at cosmic noon").
This work uses the following software packages:
\href{https://github.com/astropy/astropy}{\texttt{Astropy}}
\citep{astropy1, astropy2}, 
\href{https://github.com/dfm/corner.py}{\texttt{corner.py}}
\citep{corner},
\href{https://github.com/joshspeagle/dynesty}{\texttt{dynesty}}
\citep{dynesty, 
dynesty4},
\href{https://github.com/dfm/emcee}{\texttt{emcee}}
\citep{emcee},
\href{https://github.com/matplotlib/matplotlib}{\texttt{matplotlib}}
\citep{matplotlib},
\href{numba` https://github.com/numba/numba}{\texttt{numba}}
\citep{numba},
\href{https://github.com/numpy/numpy}{\texttt{NumPy}}
\citep{numpy},
\href{https://github.com/rhayes777/PyAutoFit}{\texttt{PyAutoFit}}
\citep{pyautofit},
\href{https://github.com/Jammy2211/PyAutoGalaxy}{\texttt{PyAutoGalaxy}}
\citep{Nightingale2018, pyautogalaxy},
\href{https://github.com/Jammy2211/PyAutoLens}{\texttt{PyAutoLens}}
\citep{Nightingale2015, Nightingale2018, pyautolens},
\href{https://www.python.org/}{\texttt{Python}}
\citep{python},
\href{https://github.com/scipy/scipy}{\texttt{Scipy}}
\citep{scipy},
\href{https://github.com/JohannesBuchner/UltraNest}{\texttt{UltraNest}}
\citep{ultranest}

\end{acknowledgements}

\bibliography{globalbibs.bib}

@ARTICLE{Bothwell2013,
       author = {{Bothwell}, M.~S. and {Smail}, Ian and {Chapman}, S.~C. and {Genzel}, R. and {Ivison}, R.~J. and {Tacconi}, L.~J. and {Alaghband-Zadeh}, S. and {Bertoldi}, F. and {Blain}, A.~W. and {Casey}, C.~M. and {Cox}, P. and {Greve}, T.~R. and {Lutz}, D. and {Neri}, R. and {Omont}, A. and {Swinbank}, A.~M.},
        title = "{A survey of molecular gas in luminous sub-millimetre galaxies}",
      journal = {\mnras},
         year = 2013,
        month = mar,
       volume = {429},
       number = {4},
        pages = {3047-3067},
          doi = {10.1093/mnras/sts562},
archivePrefix = {arXiv},
       eprint = {1205.1511},
 primaryClass = {astro-ph.CO},
       adsurl = {https://ui.adsabs.harvard.edu/abs/2013MNRAS.429.3047B}
}

@ARTICLE{Berman2022,
       author = {{Berman}, Derek A. and {Yun}, Min S. and {Harrington}, K.~C. and {Kamieneski}, P. and {Lowenthal}, J. and {Frye}, B.~L. and {Wang}, Q.~D. and {Wilson}, G.~W. and {Aretxaga}, I. and {Chavez}, M. and {Cybulski}, R. and {De la Luz}, V. and {Erickson}, N. and {Ferrusca}, D. and {Hughes}, D.~H. and {Monta{\~n}a}, A. and {Narayanan}, G. and {S{\'a}nchez-Arg{\"u}elles}, D. and {Schloerb}, F.~P. and {Souccar}, K. and {Terlevich}, E. and {Terlevich}, R. and {Zavala}, J.~A.},
        title = "{PASSAGES: the Large Millimeter Telescope and ALMA observations of extremely luminous high-redshift galaxies identified by the Planck}",
      journal = {\mnras},
         year = 2022,
        month = sep,
       volume = {515},
       number = {3},
        pages = {3911-3937},
          doi = {10.1093/mnras/stac1494},
archivePrefix = {arXiv},
       eprint = {2206.00138},
 primaryClass = {astro-ph.GA},
       adsurl = {https://ui.adsabs.harvard.edu/abs/2022MNRAS.515.3911B}
}

@ARTICLE{Bilicki2014,
       author = {{Bilicki}, Maciej and {Jarrett}, Thomas H. and {Peacock}, John A. and {Cluver}, Michelle E. and {Steward}, Louise},
        title = "{Two Micron All Sky Survey Photometric Redshift Catalog: A Comprehensive Three-dimensional Census of the Whole Sky}",
      journal = {\apjs},
         year = 2014,
        month = jan,
       volume = {210},
       number = {1},
          eid = {9},
        pages = {9},
          doi = {10.1088/0067-0049/210/1/9},
archivePrefix = {arXiv},
       eprint = {1311.5246},
 primaryClass = {astro-ph.CO},
       adsurl = {https://ui.adsabs.harvard.edu/abs/2014ApJS..210....9B}
}

@ARTICLE{Birkin2021,
       author = {{Birkin}, Jack E. and {Weiss}, Axel and {Wardlow}, J.~L. and {Smail}, Ian and {Swinbank}, A.~M. and {Dudzevi{\v{c}}i{\={u}}t{\.{e}}}, U. and {An}, Fang Xia and {Ao}, Y. and {Chapman}, S.~C. and {Chen}, Chian-Chou and {da Cunha}, E. and {Dannerbauer}, H. and {Gullberg}, B. and {Hodge}, J.~A. and {Ikarashi}, S. and {Ivison}, R.~J. and {Matsuda}, Y. and {Stach}, S.~M. and {Walter}, F. and {Wang}, W.-H. and {van der Werf}, P.},
        title = "{An ALMA/NOEMA survey of the molecular gas properties of high-redshift star-forming galaxies}",
      journal = {\mnras},
         year = 2021,
        month = mar,
       volume = {501},
       number = {3},
        pages = {3926-3950},
          doi = {10.1093/mnras/staa3862},
archivePrefix = {arXiv},
       eprint = {2009.03341},
 primaryClass = {astro-ph.GA},
       adsurl = {https://ui.adsabs.harvard.edu/abs/2021MNRAS.501.3926B}
}

@ARTICLE{Bothwell2017,
       author = {{Bothwell}, M.~S. and {Aguirre}, J.~E. and {Aravena}, M. and {Bethermin}, M. and {Bisbas}, T.~G. and {Chapman}, S.~C. and {De Breuck}, C. and {Gonzalez}, A.~H. and {Greve}, T.~R. and {Hezaveh}, Y. and {Ma}, J. and {Malkan}, M. and {Marrone}, D.~P. and {Murphy}, E.~J. and {Spilker}, J.~S. and {Strandet}, M. and {Vieira}, J.~D. and {Wei{\ss}}, A.},
        title = "{ALMA observations of atomic carbon in z {\ensuremath{\sim}} 4 dusty star-forming galaxies}",
      journal = {\mnras},
         year = 2017,
        month = apr,
       volume = {466},
       number = {3},
        pages = {2825-2841},
          doi = {10.1093/mnras/stw3270},
archivePrefix = {arXiv},
       eprint = {1612.04380},
 primaryClass = {astro-ph.GA},
       adsurl = {https://ui.adsabs.harvard.edu/abs/2017MNRAS.466.2825B}
}

@ARTICLE{Canameras2017ALMA,
   author = {{Ca{\~n}ameras}, R. and {Nesvadba}, N. and {Kneissl}, R. and
	{Frye}, B. and {Gavazzi}, R. and {Koenig}, S. and {Le Floc'h}, E. and
	{Limousin}, M. and {Oteo}, I. and {Scott}, D.},
    title = "{Planck's dusty GEMS. IV. Star formation and feedback in a maximum starburst at z = 3 seen at 60-pc resolution}",
  journal = {\aap},
archivePrefix = "arXiv",
   eprint = {1704.05853},
     year = 2017,
    month = aug,
   volume = 604,
      eid = {A117},
    pages = {A117},
      doi = {10.1051/0004-6361/201630186},
   adsurl = {http://adsabs.harvard.edu/abs/2017A%26A...604A.117C}
}

@ARTICLE{Canameras2015,
   author = {{Ca{\~n}ameras}, R. and {Nesvadba}, N.~P.~H. and {Guery}, D. and
	{McKenzie}, T. and {K{\"o}nig}, S. and {Petitpas}, G. and {Dole}, H. and
	{Frye}, B. and {Flores-Cacho}, I. and {Montier}, L. and {Negrello}, M. and
	{Beelen}, A. and {Boone}, F. and {Dicken}, D. and {Lagache}, G. and
	{Le Floc'h}, E. and {Altieri}, B. and {B{\'e}thermin}, M. and
	{Chary}, R. and {de Zotti}, G. and {Giard}, M. and {Kneissl}, R. and
	{Krips}, M. and {Malhotra}, S. and {Martinache}, C. and {Omont}, A. and
	{Pointecouteau}, E. and {Puget}, J.-L. and {Scott}, D. and {Soucail}, G. and
	{Valtchanov}, I. and {Welikala}, N. and {Yan}, L.},
    title = "{Planck's dusty GEMS: The brightest gravitationally lensed galaxies discovered with the Planck all-sky survey}",
  journal = {\aap},
archivePrefix = "arXiv",
   eprint = {1506.01962},
     year = 2015,
    month = sep,
   volume = 581,
      eid = {A105},
    pages = {A105},
      doi = {10.1051/0004-6361/201425128},
   adsurl = {http://adsabs.harvard.edu/abs/2015A%26A...581A.105C}
}

@ARTICLE{DiazSanchez2017,
   author = {{D{\'{\i}}az-S{\'a}nchez}, A. and {Iglesias-Groth}, S. and {Rebolo}, R. and
	{Dannerbauer}, H.},
    title = "{Discovery of a Lensed Ultrabright Submillimeter Galaxy at z = 2.0439}",
  journal = {\apjl},
archivePrefix = "arXiv",
   eprint = {1707.02454},
     year = 2017,
    month = jul,
   volume = 843,
      eid = {L22},
    pages = {L22},
      doi = {10.3847/2041-8213/aa79ef},
   adsurl = {http://adsabs.harvard.edu/abs/2017ApJ...843L..22D}
}

@ARTICLE{DownesSolomon1998,
       author = {{Downes}, D. and {Solomon}, P.~M.},
        title = "{Rotating Nuclear Rings and Extreme Starbursts in Ultraluminous Galaxies}",
      journal = {\apj},
         year = 1998,
        month = nov,
       volume = {507},
       number = {2},
        pages = {615-654},
          doi = {10.1086/306339},
archivePrefix = {arXiv},
       eprint = {astro-ph/9806377},
 primaryClass = {astro-ph},
       adsurl = {https://ui.adsabs.harvard.edu/abs/1998ApJ...507..615D}
}

@ARTICLE{DeZotti2015,
       author = {{De Zotti}, G. and {Castex}, G. and {Gonz{\'a}lez-Nuevo}, J. and {Lopez-Caniego}, M. and {Negrello}, M. and {Cai}, Z.-Y. and {Clemens}, M. and {Delabrouille}, J. and {Herranz}, D. and {Bonavera}, L. and {Melin}, J.-B. and {Tucci}, M. and {Serjeant}, S. and {Bilicki}, M. and {Andreani}, P. and {Clements}, D.~L. and {Toffolatti}, L. and {Roukema}, B.~F.},
        title = "{Extragalactic sources in Cosmic Microwave Background maps}",
      journal = {\jcap},
         year = 2015,
        month = jun,
       volume = {2015},
       number = {6},
        pages = {018-018},
          doi = {10.1088/1475-7516/2015/06/018},
archivePrefix = {arXiv},
       eprint = {1501.02170},
 primaryClass = {astro-ph.CO},
       adsurl = {https://ui.adsabs.harvard.edu/abs/2015JCAP...06..018D}
}

@ARTICLE{Duncan2022,
       author = {{Duncan}, Kenneth J.},
        title = "{All-purpose, all-sky photometric redshifts for the Legacy Imaging Surveys Data Release 8}",
      journal = {\mnras},
         year = 2022,
        month = may,
       volume = {512},
       number = {3},
        pages = {3662-3683},
          doi = {10.1093/mnras/stac608},
archivePrefix = {arXiv},
       eprint = {2203.01949},
 primaryClass = {astro-ph.GA},
       adsurl = {https://ui.adsabs.harvard.edu/abs/2022MNRAS.512.3662D}
}

@ARTICLE{Eales2010,
       author = {{Eales}, S. and {Dunne}, L. and {Clements}, D. and {Cooray}, A. and {De Zotti}, G. and {Dye}, S. and {Ivison}, R. and {Jarvis}, M. and {Lagache}, G. and {Maddox}, S. and {Negrello}, M. and {Serjeant}, S. and {Thompson}, M.~A. and {Kampen}, E. Van and {Amblard}, A. and {Andreani}, P. and {Baes}, M. and {Beelen}, A. and {Bendo}, G.~J. and {Benford}, D. and {Bertoldi}, F. and {Bock}, J. and {Bonfield}, D. and {Boselli}, A. and {Bridge}, C. and {Buat}, V. and {Burgarella}, D. and {Carlberg}, R. and {Cava}, A. and {Chanial}, P. and {Charlot}, S. and {Christopher}, N. and {Coles}, P. and {Cortese}, L. and {Dariush}, A. and {da Cunha}, E. and {Dalton}, G. and {Danese}, L. and {Dannerbauer}, H. and {Driver}, S. and {Dunlop}, J. and {Fan}, L. and {Farrah}, D. and {Frayer}, D. and {Frenk}, C. and {Geach}, J. and {Gardner}, J. and {Gomez}, H. and {Gonz{\'a}lez-Nuevo}, J. and {Gonz{\'a}lez-Solares}, E. and {Griffin}, M. and {Hardcastle}, M. and {Hatziminaoglou}, E. and {Herranz}, D. and {Hughes}, D. and {Ibar}, E. and {Jeong}, Woong-Seob and {Lacey}, C. and {Lapi}, A. and {Lawrence}, A. and {Lee}, M. and {Leeuw}, L. and {Liske}, J. and {L{\'o}pez-Caniego}, M. and {M{\"u}ller}, T. and {Nandra}, K. and {Panuzzo}, P. and {Papageorgiou}, A. and {Patanchon}, G. and {Peacock}, J. and {Pearson}, C. and {Phillipps}, S. and {Pohlen}, M. and {Popescu}, C. and {Rawlings}, S. and {Rigby}, E. and {Rigopoulou}, M. and {Robotham}, A. and {Rodighiero}, G. and {Sansom}, A. and {Schulz}, B. and {Scott}, D. and {Smith}, D.~J.~B. and {Sibthorpe}, B. and {Smail}, I. and {Stevens}, J. and {Sutherland}, W. and {Takeuchi}, T. and {Tedds}, J. and {Temi}, P. and {Tuffs}, R. and {Trichas}, M. and {Vaccari}, M. and {Valtchanov}, I. and {van der Werf}, P. and {Verma}, A. and {Vieria}, J. and {Vlahakis}, C. and {White}, Glenn J.},
        title = "{The Herschel ATLAS}",
      journal = {\pasp},
         year = 2010,
        month = may,
       volume = {122},
       number = {891},
        pages = {499-515},
          doi = {10.1086/653086},
archivePrefix = {arXiv},
       eprint = {0910.4279},
 primaryClass = {astro-ph.CO},
       adsurl = {https://ui.adsabs.harvard.edu/abs/2010PASP..122..499E}
}

@ARTICLE{Everett2020,
       author = {{Everett}, W.~B. and {Zhang}, L. and {Crawford}, T.~M. and {Vieira}, J.~D. and {Aravena}, M. and {Archipley}, M.~A. and {Austermann}, J.~E. and {Benson}, B.~A. and {Bleem}, L.~E. and {Carlstrom}, J.~E. and {Chang}, C.~L. and {Chapman}, S. and {Crites}, A.~T. and {de Haan}, T. and {Dobbs}, M.~A. and {George}, E.~M. and {Halverson}, N.~W. and {Harrington}, N. and {Holder}, G.~P. and {Holzapfel}, W.~L. and {Hrubes}, J.~D. and {Knox}, L. and {Lee}, A.~T. and {Luong-Van}, D. and {Mangian}, A.~C. and {Marrone}, D.~P. and {McMahon}, J.~J. and {Meyer}, S.~S. and {Mocanu}, L.~M. and {Mohr}, J.~J. and {Natoli}, T. and {Padin}, S. and {Pryke}, C. and {Reichardt}, C.~L. and {Reuter}, C.~A. and {Ruhl}, J.~E. and {Sayre}, J.~T. and {Schaffer}, K.~K. and {Shirokoff}, E. and {Spilker}, J.~S. and {Stalder}, B. and {Staniszewski}, Z. and {Stark}, A.~A. and {Story}, K.~T. and {Switzer}, E.~R. and {Vanderlinde}, K. and {Wei{\ss}}, A. and {Williamson}, R.},
        title = "{Millimeter-wave Point Sources from the 2500 Square Degree SPT-SZ Survey: Catalog and Population Statistics}",
      journal = {\apj},
         year = 2020,
        month = sep,
       volume = {900},
       number = {1},
          eid = {55},
        pages = {55},
          doi = {10.3847/1538-4357/ab9df7},
archivePrefix = {arXiv},
       eprint = {2003.03431},
 primaryClass = {astro-ph.IM},
       adsurl = {https://ui.adsabs.harvard.edu/abs/2020ApJ...900...55E}
}

@ARTICLE{FriasCastillo2025,
       author = {{Frias Castillo}, Marta and {Rybak}, Matus and {Hodge}, Jacqueline A. and {van der Werf}, Paul and {Smail}, Ian and {Butterworth}, Joshua and {Jansen}, Jasper and {Topkaras}, Theodoros and {Chen}, Chian-Chou and {Chapman}, Scott C. and {Weiss}, Axel and {Algera}, Hiddo and {Birkin}, Jack E. and {da Cunha}, Elisabete and {Chen}, Jianhang and {Dannerbauer}, Helmut and {Ikarashi}, Soh and {Jim{\'e}nez-Andrade}, E.~F. and {Liao}, Cheng-Lin and {Murphy}, Eric J. and {Swinbank}, A.~M. and {Walter}, Fabian and {Calistro Rivera}, Gabriela and {Ivison}, R.~J. and {Lagos}, Claudia del P.},
        title = "{A Comparative Study of the Ground State Transitions of CO and C I as Molecular Gas Tracers at High Redshift}",
      journal = {\apj},
         year = 2025,
        month = jul,
       volume = {987},
       number = {2},
          eid = {158},
        pages = {158},
          doi = {10.3847/1538-4357/adc4e0},
archivePrefix = {arXiv},
       eprint = {2404.05596},
 primaryClass = {astro-ph.GA},
       adsurl = {https://ui.adsabs.harvard.edu/abs/2025ApJ...987..158F}
}

@ARTICLE{Fu2012,
   author = {{Fu}, H. and {Jullo}, E. and {Cooray}, A. and {Bussmann}, R.~S. and
	{Ivison}, R.~J. and {P{\'e}rez-Fournon}, I. and {Djorgovski}, S.~G. and
	{Scoville}, N. and {Yan}, L. and {Riechers}, D.~A. and {Aguirre}, J. and
	{Auld}, R. and {Baes}, M. and {Baker}, A.~J. and {Bradford}, M. and
	{Cava}, A. and {Clements}, D.~L. and {Dannerbauer}, H. and {Dariush}, A. and
	{De Zotti}, G. and {Dole}, H. and {Dunne}, L. and {Dye}, S. and
	{Eales}, S. and {Frayer}, D. and {Gavazzi}, R. and {Gurwell}, M. and
	{Harris}, A.~I. and {Herranz}, D. and {Hopwood}, R. and {Hoyos}, C. and
	{Ibar}, E. and {Jarvis}, M.~J. and {Kim}, S. and {Leeuw}, L. and
	{Lupu}, R. and {Maddox}, S. and {Mart{\'{\i}}nez-Navajas}, P. and
	{Micha{\l}owski}, M.~J. and {Negrello}, M. and {Omont}, A. and
	{Rosenman}, M. and {Scott}, D. and {Serjeant}, S. and {Smail}, I. and
	{Swinbank}, A.~M. and {Valiante}, E. and {Verma}, A. and {Vieira}, J. and
	{Wardlow}, J.~L. and {van der Werf}, P.},
    title = "{A Comprehensive View of a Strongly Lensed Planck-Associated Submillimeter Galaxy}",
  journal = {\apj},
archivePrefix = "arXiv",
   eprint = {1202.1829},
     year = 2012,
    month = jul,
   volume = 753,
      eid = {134},
    pages = {134},
      doi = {10.1088/0004-637X/753/2/134},
   adsurl = {http://adsabs.harvard.edu/abs/2012ApJ...753..134F}
}

@ARTICLE{Gralla2020,
       author = {{Gralla}, Megan B. and {Marriage}, Tobias A. and {Addison}, Graeme and {Baker}, Andrew J. and {Bond}, J. Richard and {Crichton}, Devin and {Datta}, Rahul and {Devlin}, Mark J. and {Dunkley}, Joanna and {D{\"u}nner}, Rolando and {Fowler}, Joseph and {Gallardo}, Patricio A. and {Hall}, Kirsten and {Halpern}, Mark and {Hasselfield}, Matthew and {Hilton}, Matt and {Hincks}, Adam D. and {Huffenberger}, Kevin M. and {Hughes}, John P. and {Kosowsky}, Arthur and {L{\'o}pez-Caraballo}, Carlos H. and {Louis}, Thibaut and {Marsden}, Danica and {Moodley}, Kavilan and {Niemack}, Michael D. and {Page}, Lyman A. and {Partridge}, Bruce and {Rivera}, Jesus and {Sievers}, Jonathan L. and {Staggs}, Suzanne and {Su}, Ting and {Swetz}, Daniel and {Wollack}, Edward J.},
        title = "{Atacama Cosmology Telescope: Dusty Star-forming Galaxies and Active Galactic Nuclei in the Equatorial Survey}",
      journal = {\apj},
         year = 2020,
        month = apr,
       volume = {893},
       number = {2},
          eid = {104},
        pages = {104},
          doi = {10.3847/1538-4357/ab7915},
archivePrefix = {arXiv},
       eprint = {1905.04592},
 primaryClass = {astro-ph.GA},
       adsurl = {https://ui.adsabs.harvard.edu/abs/2020ApJ...893..104G}
}

@ARTICLE{Greve2005,
       author = {{Greve}, T.~R. and {Bertoldi}, F. and {Smail}, Ian and {Neri}, R. and {Chapman}, S.~C. and {Blain}, A.~W. and {Ivison}, R.~J. and {Genzel}, R. and {Omont}, A. and {Cox}, P. and {Tacconi}, L. and {Kneib}, J.-P.},
        title = "{An interferometric CO survey of luminous submillimetre galaxies}",
      journal = {\mnras},
         year = 2005,
        month = may,
       volume = {359},
       number = {3},
        pages = {1165-1183},
          doi = {10.1111/j.1365-2966.2005.08979.x},
archivePrefix = {arXiv},
       eprint = {astro-ph/0503055},
 primaryClass = {astro-ph},
       adsurl = {https://ui.adsabs.harvard.edu/abs/2005MNRAS.359.1165G}
}

@ARTICLE{Herranz2013,
   author = {{Herranz}, D. and {Gonz{\'a}lez-Nuevo}, J. and {Clements}, D.~L. and
	{De Zotti}, G. and {Lopez-Caniego}, M. and {Lapi}, A. and {Rodighiero}, G. and
	{Danese}, L. and {Fu}, H. and {Cooray}, A. and {Baes}, M. and
	{Bendo}, G.~J. and {Bonavera}, L. and {Carrera}, F.~J. and {Dole}, H. and
	{Eales}, S. and {Ivison}, R.~J. and {Jarvis}, M. and {Lagache}, G. and
	{Massardi}, M. and {Micha{\l}owski}, M.~J. and {Negrello}, M. and
	{Rigby}, E. and {Scott}, D. and {Valiante}, E. and {Valtchanov}, I. and
	{Van der Werf}, P. and {Auld}, R. and {Buttiglione}, S. and
	{Dariush}, A. and {Dunne}, L. and {Hopwood}, R. and {Hoyos}, C. and
	{Ibar}, E. and {Maddox}, S.},
    title = "{Herschel-ATLAS: Planck sources in the phase 1 fields}",
  journal = {\aap},
archivePrefix = "arXiv",
   eprint = {1204.3917},
     year = 2013,
    month = jan,
   volume = 549,
      eid = {A31},
    pages = {A31},
      doi = {10.1051/0004-6361/201219435},
   adsurl = {http://adsabs.harvard.edu/abs/2013A%26A...549A..31H}
}

@ARTICLE{KingPounds2015,
       author = {{King}, Andrew and {Pounds}, Ken},
        title = "{Powerful Outflows and Feedback from Active Galactic Nuclei}",
      journal = {\araa},
         year = 2015,
        month = aug,
       volume = {53},
        pages = {115-154},
          doi = {10.1146/annurev-astro-082214-122316},
archivePrefix = {arXiv},
       eprint = {1503.05206},
 primaryClass = {astro-ph.GA},
       adsurl = {https://ui.adsabs.harvard.edu/abs/2015ARA&A..53..115K}
}

@ARTICLE{Kaufman1999,
       author = {{Kaufman}, Michael J. and {Wolfire}, Mark G. and {Hollenbach}, David J. and {Luhman}, Michael L.},
        title = "{Far-Infrared and Submillimeter Emission from Galactic and Extragalactic Photodissociation Regions}",
      journal = {\apj},
         year = 1999,
        month = dec,
       volume = {527},
       number = {2},
        pages = {795-813},
          doi = {10.1086/308102},
archivePrefix = {arXiv},
       eprint = {astro-ph/9907255},
 primaryClass = {astro-ph},
       adsurl = {https://ui.adsabs.harvard.edu/abs/1999ApJ...527..795K}
}

@article{keeton03,
	Adsurl = {http://adsabs.harvard.edu/abs/2003ApJ...584..664K},
	Author = {{Keeton}, C.~R.},
	Doi = {10.1086/345717},
	Eprint = {arXiv:astro-ph/0209040},
	Journal = {\apj},
	Month = feb,
	Pages = {664-674},
	Title = {{Analytic Cross Sections for Substructure Lensing}},
	Volume = 584,
	Year = 2003}

@ARTICLE{Negrello2007,
       author = {{Negrello}, M. and {Perrotta}, F. and {Gonz{\'a}lez-Nuevo}, J. and {Silva}, L. and {de Zotti}, G. and {Granato}, G.~L. and {Baccigalupi}, C. and {Danese}, L.},
        title = "{Astrophysical and cosmological information from large-scale submillimetre surveys of extragalactic sources}",
      journal = {\mnras},
         year = 2007,
        month = jun,
       volume = {377},
       number = {4},
        pages = {1557-1568},
          doi = {10.1111/j.1365-2966.2007.11708.x},
archivePrefix = {arXiv},
       eprint = {astro-ph/0703210},
 primaryClass = {astro-ph},
       adsurl = {https://ui.adsabs.harvard.edu/abs/2007MNRAS.377.1557N}
}

@ARTICLE{Negrello2013,
       author = {{Negrello}, M. and {Clemens}, M. and {Gonzalez-Nuevo}, J. and {De Zotti}, G. and {Bonavera}, L. and {Cosco}, G. and {Guarese}, G. and {Boaretto}, L. and {Serjeant}, S. and {Toffolatti}, L. and {Lapi}, A. and {Bethermin}, M. and {Castex}, G. and {Clements}, D.~L. and {Delabrouille}, J. and {Dole}, H. and {Franceschini}, A. and {Mandolesi}, N. and {Marchetti}, L. and {Partridge}, B. and {Sajina}, A.},
        title = "{The local luminosity function of star-forming galaxies derived from the Planck Early Release Compact Source Catalogue}",
      journal = {\mnras},
         year = 2013,
        month = feb,
       volume = {429},
       number = {2},
        pages = {1309-1323},
          doi = {10.1093/mnras/sts417},
archivePrefix = {arXiv},
       eprint = {1211.3832},
 primaryClass = {astro-ph.CO},
       adsurl = {https://ui.adsabs.harvard.edu/abs/2013MNRAS.429.1309N}
}

@ARTICLE{Nesvadba2016,
       author = {{Nesvadba}, N. and {Kneissl}, R. and {Ca{\~n}ameras}, R. and {Boone}, F. and {Falgarone}, E. and {Frye}, B. and {Gerin}, M. and {Koenig}, S. and {Lagache}, G. and {Le Floc'h}, E. and {Malhotra}, S. and {Scott}, D.},
        title = "{Planck's Dusty GEMS. II. Extended [CII] emission and absorption in the Garnet at z = 3.4 seen with ALMA}",
      journal = {\aap},
         year = 2016,
        month = aug,
       volume = {593},
          eid = {L2},
        pages = {L2},
          doi = {10.1051/0004-6361/201629037},
archivePrefix = {arXiv},
       eprint = {1610.01169},
 primaryClass = {astro-ph.GA},
       adsurl = {https://ui.adsabs.harvard.edu/abs/2016A&A...593L...2N}
}

@ARTICLE{Oliver2012,
       author = {{Oliver}, S.~J. and {Bock}, J. and {Altieri}, B. and {Amblard}, A. and {Arumugam}, V. and {Aussel}, H. and {Babbedge}, T. and {Beelen}, A. and {B{\'e}thermin}, M. and {Blain}, A. and {Boselli}, A. and {Bridge}, C. and {Brisbin}, D. and {Buat}, V. and {Burgarella}, D. and {Castro-Rodr{\'\i}guez}, N. and {Cava}, A. and {Chanial}, P. and {Cirasuolo}, M. and {Clements}, D.~L. and {Conley}, A. and {Conversi}, L. and {Cooray}, A. and {Dowell}, C.~D. and {Dubois}, E.~N. and {Dwek}, E. and {Dye}, S. and {Eales}, S. and {Elbaz}, D. and {Farrah}, D. and {Feltre}, A. and {Ferrero}, P. and {Fiolet}, N. and {Fox}, M. and {Franceschini}, A. and {Gear}, W. and {Giovannoli}, E. and {Glenn}, J. and {Gong}, Y. and {Gonz{\'a}lez Solares}, E.~A. and {Griffin}, M. and {Halpern}, M. and {Harwit}, M. and {Hatziminaoglou}, E. and {Heinis}, S. and {Hurley}, P. and {Hwang}, H.~S. and {Hyde}, A. and {Ibar}, E. and {Ilbert}, O. and {Isaak}, K. and {Ivison}, R.~J. and {Lagache}, G. and {Le Floc'h}, E. and {Levenson}, L. and {Faro}, B. Lo and {Lu}, N. and {Madden}, S. and {Maffei}, B. and {Magdis}, G. and {Mainetti}, G. and {Marchetti}, L. and {Marsden}, G. and {Marshall}, J. and {Mortier}, A.~M.~J. and {Nguyen}, H.~T. and {O'Halloran}, B. and {Omont}, A. and {Page}, M.~J. and {Panuzzo}, P. and {Papageorgiou}, A. and {Patel}, H. and {Pearson}, C.~P. and {P{\'e}rez-Fournon}, I. and {Pohlen}, M. and {Rawlings}, J.~I. and {Raymond}, G. and {Rigopoulou}, D. and {Riguccini}, L. and {Rizzo}, D. and {Rodighiero}, G. and {Roseboom}, I.~G. and {Rowan-Robinson}, M. and {S{\'a}nchez Portal}, M. and {Schulz}, B. and {Scott}, Douglas and {Seymour}, N. and {Shupe}, D.~L. and {Smith}, A.~J. and {Stevens}, J.~A. and {Symeonidis}, M. and {Trichas}, M. and {Tugwell}, K.~E. and {Vaccari}, M. and {Valtchanov}, I. and {Vieira}, J.~D. and {Viero}, M. and {Vigroux}, L. and {Wang}, L. and {Ward}, R. and {Wardlow}, J. and {Wright}, G. and {Xu}, C.~K. and {Zemcov}, M.},
        title = "{The Herschel Multi-tiered Extragalactic Survey: HerMES}",
      journal = {\mnras},
         year = 2012,
        month = aug,
       volume = {424},
       number = {3},
        pages = {1614-1635},
          doi = {10.1111/j.1365-2966.2012.20912.x},
archivePrefix = {arXiv},
       eprint = {1203.2562},
 primaryClass = {astro-ph.CO},
       adsurl = {https://ui.adsabs.harvard.edu/abs/2012MNRAS.424.1614O}
}

@ARTICLE{PlanckCollaboration2016cold_cores,
   author = {{Planck Collaboration} and {Ade}, P.~A.~R. and {Aghanim}, N. and
	{Arnaud}, M. and {Ashdown}, M. and {Aumont}, J. and {Baccigalupi}, C. and
	{Banday}, A.~J. and {Barreiro}, R.~B. and {Bartolo}, N. and et al.},
    title = "{Planck 2015 results. XXVIII. The Planck Catalogue of Galactic cold clumps}",
  journal = {\aap},
archivePrefix = "arXiv",
   eprint = {1502.01599},
     year = 2016,
    month = sep,
   volume = 594,
      eid = {A28},
    pages = {A28},
      doi = {10.1051/0004-6361/201525819},
   adsurl = {http://adsabs.harvard.edu/abs/2016A%26A...594A..28P}
}

@ARTICLE{PlanckCollaboration2016highz,
   author = {{Planck Collaboration Int. XXXIX} },
    title = "{Planck intermediate results. XXXIX. The Planck list of high-redshift source candidates}",
  journal = {\aap},
archivePrefix = "arXiv",
   eprint = {1508.04171},
     year = 2016,
    month = dec,
   volume = 596,
      eid = {A100},
    pages = {A100},
      doi = {10.1051/0004-6361/201527206},
   adsurl = {http://adsabs.harvard.edu/abs/2016A%26A...596A.100P}
}

@ARTICLE{PCCS2,
       author = {{Planck Collaboration XXVI} },
        title = "{Planck 2015 results. XXVI. The Second Planck Catalogue of Compact Sources}",
      journal = {\aap},
         year = 2016,
        month = sep,
       volume = {594},
          eid = {A26},
        pages = {A26},
          doi = {10.1051/0004-6361/201526914},
archivePrefix = {arXiv},
       eprint = {1507.02058},
 primaryClass = {astro-ph.CO},
       adsurl = {https://ui.adsabs.harvard.edu/abs/2016A&A...594A..26P}
}

@ARTICLE{PlanckCollaboration2018PCNT,
   author = {{Planck Collaboration Int. LIV}
	},
    title = "{Planck intermediate results. LIV. The Planck multi-frequency catalogue of non-thermal sources}",
  journal = {\aap},
archivePrefix = "arXiv",
   eprint = {1802.08649},
     year = 2018,
    month = nov,
   volume = 619,
      eid = {A94},
    pages = {A94},
      doi = {10.1051/0004-6361/201832888},
   adsurl = {http://adsabs.harvard.edu/abs/2018A%26A...619A..94P}
}

@ARTICLE{Sersic1963,
       author = {{S{\'e}rsic}, J.~L.},
        title = "{Influence of the atmospheric and instrumental dispersion on the brightness distribution in a galaxy}",
      journal = {Boletin de la Asociacion Argentina de Astronomia La Plata Argentina},
         year = 1963,
        month = feb,
       volume = {6},
        pages = {41-43},
       adsurl = {https://ui.adsabs.harvard.edu/abs/1963BAAA....6...41S}
}

@ARTICLE{Wright2010,
       author = {{Wright}, Edward L. and {Eisenhardt}, Peter R.~M. and {Mainzer}, Amy K. and {Ressler}, Michael E. and {Cutri}, Roc M. and {Jarrett}, Thomas and {Kirkpatrick}, J. Davy and {Padgett}, Deborah and {McMillan}, Robert S. and {Skrutskie}, Michael and {Stanford}, S.~A. and {Cohen}, Martin and {Walker}, Russell G. and {Mather}, John C. and {Leisawitz}, David and {Gautier}, III, Thomas N. and {McLean}, Ian and {Benford}, Dominic and {Lonsdale}, Carol J. and {Blain}, Andrew and {Mendez}, Bryan and {Irace}, William R. and {Duval}, Valerie and {Liu}, Fengchuan and {Royer}, Don and {Heinrichsen}, Ingolf and {Howard}, Joan and {Shannon}, Mark and {Kendall}, Martha and {Walsh}, Amy L. and {Larsen}, Mark and {Cardon}, Joel G. and {Schick}, Scott and {Schwalm}, Mark and {Abid}, Mohamed and {Fabinsky}, Beth and {Naes}, Larry and {Tsai}, Chao-Wei},
        title = "{The Wide-field Infrared Survey Explorer (WISE): Mission Description and Initial On-orbit Performance}",
      journal = {\aj},
         year = 2010,
        month = dec,
       volume = {140},
       number = {6},
        pages = {1868-1881},
          doi = {10.1088/0004-6256/140/6/1868},
archivePrefix = {arXiv},
       eprint = {1008.0031},
 primaryClass = {astro-ph.IM},
       adsurl = {https://ui.adsabs.harvard.edu/abs/2010AJ....140.1868W}
}

@ARTICLE{Kamieneski24,
       author = {{Kamieneski}, Patrick S. and {Yun}, Min S. and {Harrington}, Kevin C. and {Lowenthal}, James D. and {Wang}, Q. Daniel and {Frye}, Brenda L. and {Jim{\'e}nez-Andrade}, Eric F. and {Vishwas}, Amit and {Cooper}, Olivia and {Pascale}, Massimo and {Foo}, Nicholas and {Berman}, Derek and {Englert}, Anthony and {Garcia Diaz}, Carlos},
        title = "{PASSAGES: The Wide-ranging, Extreme Intrinsic Properties of Planck-selected, Lensed Dusty Star-forming Galaxies}",
      journal = {\apj},
         year = 2024,
        month = jan,
       volume = {961},
       number = {1},
          eid = {2},
        pages = {2},
          doi = {10.3847/1538-4357/acf930},
archivePrefix = {arXiv},
       eprint = {2301.09746},
 primaryClass = {astro-ph.GA},
       adsurl = {https://ui.adsabs.harvard.edu/abs/2024ApJ...961....2K}
}

@ARTICLE{Harrington21,
       author = {{Harrington}, Kevin C. and {Weiss}, Axel and {Yun}, Min S. and {Magnelli}, Benjamin and {Sharon}, C.~E. and {Leung}, T.~K.~D. and {Vishwas}, A. and {Wang}, Q.~D. and {Frayer}, D.~T. and {Jim{\'e}nez-Andrade}, E.~F. and {Liu}, D. and {Garc{\'\i}a}, P. and {Romano-D{\'\i}az}, E. and {Frye}, B.~L. and {Jarugula}, S. and {B{\u{a}}descu}, T. and {Berman}, D. and {Dannerbauer}, H. and {D{\'\i}az-S{\'a}nchez}, A. and {Grassitelli}, L. and {Kamieneski}, P. and {Kim}, W.~J. and {Kirkpatrick}, A. and {Lowenthal}, J.~D. and {Messias}, H. and {Puschnig}, J. and {Stacey}, G.~J. and {Torne}, P. and {Bertoldi}, F.},
        title = "{Turbulent Gas in Lensed Planck-selected Starbursts at z {\ensuremath{\sim}} 1-3.5}",
      journal = {\apj},
         year = 2021,
        month = feb,
       volume = {908},
       number = {1},
          eid = {95},
        pages = {95},
          doi = {10.3847/1538-4357/abcc01},
archivePrefix = {arXiv},
       eprint = {2010.16231},
 primaryClass = {astro-ph.GA},
       adsurl = {https://ui.adsabs.harvard.edu/abs/2021ApJ...908...95H}
}

@ARTICLE{Negrello17,
       author = {{Negrello}, M. and {Amber}, S. and {Amvrosiadis}, A. and {Cai}, Z. -Y. and {Lapi}, A. and {Gonzalez-Nuevo}, J. and {De Zotti}, G. and {Furlanetto}, C. and {Maddox}, S.~J. and {Allen}, M. and {Bakx}, T. and {Bussmann}, R.~S. and {Cooray}, A. and {Covone}, G. and {Danese}, L. and {Dannerbauer}, H. and {Fu}, H. and {Greenslade}, J. and {Gurwell}, M. and {Hopwood}, R. and {Koopmans}, L.~V.~E. and {Napolitano}, N. and {Nayyeri}, H. and {Omont}, A. and {Petrillo}, C.~E. and {Riechers}, D.~A. and {Serjeant}, S. and {Tortora}, C. and {Valiante}, E. and {Verdoes Kleijn}, G. and {Vernardos}, G. and {Wardlow}, J.~L. and {Baes}, M. and {Baker}, A.~J. and {Bourne}, N. and {Clements}, D. and {Crawford}, S.~M. and {Dye}, S. and {Dunne}, L. and {Eales}, S. and {Ivison}, R.~J. and {Marchetti}, L. and {Micha{\l}owski}, M.~J. and {Smith}, M.~W.~L. and {Vaccari}, M. and {van der Werf}, P.},
        title = "{The Herschel-ATLAS: a sample of 500 {\ensuremath{\mu}}m-selected lensed galaxies over 600 deg$^{2}$}",
      journal = {\mnras},
         year = 2017,
        month = mar,
       volume = {465},
       number = {3},
        pages = {3558-3580},
          doi = {10.1093/mnras/stw2911},
archivePrefix = {arXiv},
       eprint = {1611.03922},
 primaryClass = {astro-ph.GA},
       adsurl = {https://ui.adsabs.harvard.edu/abs/2017MNRAS.465.3558N}
}

@ARTICLE{Dannerbauer19,
       author = {{Dannerbauer}, H. and {Harrington}, K. and {D{\'\i}az-S{\'a}nchez}, A. and {Iglesias-Groth}, S. and {Rebolo}, R. and {Genova-Santos}, R.~T. and {Krips}, M.},
        title = "{Ultra-bright CO and [C I] Emission in a Lensed z = 2.04 Submillimeter Galaxy with Extreme Molecular Gas Properties}",
      journal = {\aj},
         year = 2019,
        month = jul,
       volume = {158},
       number = {1},
          eid = {34},
        pages = {34},
          doi = {10.3847/1538-3881/aaf50b},
archivePrefix = {arXiv},
       eprint = {1812.03845},
 primaryClass = {astro-ph.GA},
       adsurl = {https://ui.adsabs.harvard.edu/abs/2019AJ....158...34D}
}

@ARTICLE{Harrington16,
       author = {{Harrington}, K.~C. and {Yun}, Min S. and {Cybulski}, R. and {Wilson}, G.~W. and {Aretxaga}, I. and {Chavez}, M. and {De la Luz}, V. and {Erickson}, N. and {Ferrusca}, D. and {Gallup}, A.~D. and {Hughes}, D.~H. and {Monta{\~n}a}, A. and {Narayanan}, G. and {S{\'a}nchez-Arg{\"u}elles}, D. and {Schloerb}, F.~P. and {Souccar}, K. and {Terlevich}, E. and {Terlevich}, R. and {Zeballos}, M. and {Zavala}, J.~A.},
        title = "{Early science with the Large Millimeter Telescope: observations of extremely luminous high-z sources identified by Planck}",
      journal = {\mnras},
         year = 2016,
        month = jun,
       volume = {458},
       number = {4},
        pages = {4383-4399},
          doi = {10.1093/mnras/stw614},
archivePrefix = {arXiv},
       eprint = {1603.05622},
 primaryClass = {astro-ph.GA},
       adsurl = {https://ui.adsabs.harvard.edu/abs/2016MNRAS.458.4383H}
}

@ARTICLE{Trombetti21,
       author = {{Trombetti}, T. and {Burigana}, C. and {Bonato}, M. and {Herranz}, D. and {De Zotti}, G. and {Negrello}, M. and {Galluzzi}, V. and {Massardi}, M.},
        title = "{Search for candidate strongly lensed dusty galaxies in the Planck satellite catalogues}",
      journal = {\aap},
         year = 2021,
        month = sep,
       volume = {653},
          eid = {A151},
        pages = {A151},
          doi = {10.1051/0004-6361/202140830},
archivePrefix = {arXiv},
       eprint = {2108.01113},
 primaryClass = {astro-ph.CO},
       adsurl = {https://ui.adsabs.harvard.edu/abs/2021A&A...653A.151T}
}

@ARTICLE{Maddox18,
       author = {{Maddox}, S.~J. and {Valiante}, E. and {Cigan}, P. and {Dunne}, L. and {Eales}, S. and {Smith}, M.~W.~L. and {Dye}, S. and {Furlanetto}, C. and {Ibar}, E. and {de Zotti}, G. and {Millard}, J.~S. and {Bourne}, N. and {Gomez}, H.~L. and {Ivison}, R.~J. and {Scott}, D. and {Valtchanov}, I.},
        title = "{The Herschel-ATLAS Data Release 2. Paper II. Catalogs of Far-infrared and Submillimeter Sources in the Fields at the South and North Galactic Poles}",
      journal = {\apjs},
         year = 2018,
        month = jun,
       volume = {236},
       number = {2},
          eid = {30},
        pages = {30},
          doi = {10.3847/1538-4365/aab8fc},
archivePrefix = {arXiv},
       eprint = {1712.07241},
 primaryClass = {astro-ph.GA},
       adsurl = {https://ui.adsabs.harvard.edu/abs/2018ApJS..236...30M}
}

@ARTICLE{Kron80,
       author = {{Kron}, R.~G.},
        title = "{Photometry of a complete sample of faint galaxies.}",
      journal = {\apjs},
         year = 1980,
        month = jun,
       volume = {43},
        pages = {305-325},
          doi = {10.1086/190669},
       adsurl = {https://ui.adsabs.harvard.edu/abs/1980ApJS...43..305K}
}

@article{astropy1,
Adsurl = {http://adsabs.harvard.edu/abs/2013A%26A...558A..33A},
Archiveprefix = {arXiv},
Author = {{Astropy Collaboration} and {Robitaille}, T.~P. and {Tollerud}, E.~J. and {Greenfield}, P. and {Droettboom}, M. and {Bray}, E. and {Aldcroft}, T. and {Davis}, M. and {Ginsburg}, A. and {Price-Whelan}, A.~M. and {Kerzendorf}, W.~E. and {Conley}, A. and {Crighton}, N. and {Barbary}, K. and {Muna}, D. and {Ferguson}, H. and {Grollier}, F. and {Parikh}, M.~M. and {Nair}, P.~H. and {Unther}, H.~M. and {Deil}, C. and {Woillez}, J. and {Conseil}, S. and {Kramer}, R. and {Turner}, J.~E.~H. and {Singer}, L. and {Fox}, R. and {Weaver}, B.~A. and {Zabalza}, V. and {Edwards}, Z.~I. and {Azalee Bostroem}, K. and {Burke}, D.~J. and {Casey}, A.~R. and {Crawford}, S.~M. and {Dencheva}, N. and {Ely}, J. and {Jenness}, T. and {Labrie}, K. and {Lim}, P.~L. and {Pierfederici}, F. and {Pontzen}, A. and {Ptak}, A. and {Refsdal}, B. and {Servillat}, M. and {Streicher}, O.},
Doi = {10.1051/0004-6361/201322068},
Eid = {A33},
Eprint = {1307.6212},
Journal = {A\&A},
Month = oct,
Pages = {A33},
Primaryclass = {astro-ph.IM},
Title = {{Astropy: A community Python package for astronomy}},
Volume = 558,
Year = 2013}

@article{astropy2,
Adsurl = {https://ui.adsabs.harvard.edu/#abs/2018AJ....156..123T},
Author = {{Price-Whelan}, A.~M. and {Sip{\H{o}}cz}, B.~M. and {G{\"u}nther}, H.~M. and {Lim}, P.~L. and {Crawford}, S.~M. and {Conseil}, S. and {Shupe}, D.~L. and {Craig}, M.~W. and {Dencheva}, N. and {Ginsburg}, A. and {VanderPlas}, J.~T. and {Bradley}, L.~D. and {P{\'e}rez-Su{\'a}rez}, D. and {de Val-Borro}, M. and {Paper Contributors}, (Primary and {Aldcroft}, T.~L. and {Cruz}, K.~L. and {Robitaille}, T.~P. and {Tollerud}, E.~J. and {Coordination Committee}, (Astropy and {Ardelean}, C. and {Babej}, T. and {Bach}, Y.~P. and {Bachetti}, M. and {Bakanov}, A.~V. and {Bamford}, S.~P. and {Barentsen}, G. and {Barmby}, P. and {Baumbach}, A. and {Berry}, K.~L. and {Biscani}, F. and {Boquien}, M. and {Bostroem}, K.~A. and {Bouma}, L.~G. and {Brammer}, G.~B. and {Bray}, E.~M. and {Breytenbach}, H. and {Buddelmeijer}, H. and {Burke}, D.~J. and {Calderone}, G. and {Cano Rodr{\'\i}guez}, J.~L. and {Cara}, M. and {Cardoso}, J.~V.~M. and {Cheedella}, S. and {Copin}, Y. and {Corrales}, L. and {Crichton}, D. and {D{\textquoteright}Avella}, D. and {Deil}, C. and {Depagne}, {\'E}. and {Dietrich}, J.~P. and {Donath}, A. and {Droettboom}, M. and {Earl}, N. and {Erben}, T. and {Fabbro}, S. and {Ferreira}, L.~A. and {Finethy}, T. and {Fox}, R.~T. and {Garrison}, L.~H. and {Gibbons}, S.~L.~J. and {Goldstein}, D.~A. and {Gommers}, R. and {Greco}, J.~P. and {Greenfield}, P. and {Groener}, A.~M. and {Grollier}, F. and {Hagen}, A. and {Hirst}, P. and {Homeier}, D. and {Horton}, A.~J. and {Hosseinzadeh}, G. and {Hu}, L. and {Hunkeler}, J.~S. and {Ivezi{\'c}}, {\v{Z}}. and {Jain}, A. and {Jenness}, T. and {Kanarek}, G. and {Kendrew}, S. and {Kern}, N.~S. and {Kerzendorf}, W.~E. and {Khvalko}, A. and {King}, J. and {Kirkby}, D. and {Kulkarni}, A.~M. and {Kumar}, A. and {Lee}, A. and {Lenz}, D. and {Littlefair}, S.~P. and {Ma}, Z. and {Macleod}, D.~M. and {Mastropietro}, M. and {McCully}, C. and {Montagnac}, S. and {Morris}, B.~M. and {Mueller}, M. and {Mumford}, S.~J. and {Muna}, D. and {Murphy}, N.~A. and {Nelson}, S. and {Nguyen}, G.~H. and {Ninan}, J.~P. and {N{\"o}the}, M. and {Ogaz}, S. and {Oh}, S. and {Parejko}, J.~K. and {Parley}, N. and {Pascual}, S. and {Patil}, R. and {Patil}, A.~A. and {Plunkett}, A.~L. and {Prochaska}, J.~X. and {Rastogi}, T. and {Reddy Janga}, V. and {Sabater}, J. and {Sakurikar}, P. and {Seifert}, M. and {Sherbert}, L.~E. and {Sherwood-Taylor}, H. and {Shih}, A.~Y. and {Sick}, J. and {Silbiger}, M.~T. and {Singanamalla}, S. and {Singer}, L.~P. and {Sladen}, P.~H. and {Sooley}, K.~A. and {Sornarajah}, S. and {Streicher}, O. and {Teuben}, P. and {Thomas}, S.~W. and {Tremblay}, G.~R. and {Turner}, J.~E.~H. and {Terr{\'o}n}, V. and {van Kerkwijk}, M.~H. and {de la Vega}, A. and {Watkins}, L.~L. and {Weaver}, B.~A. and {Whitmore}, J.~B. and {Woillez}, J. and {Zabalza}, V. and {Contributors}, (Astropy},
Doi = {10.3847/1538-3881/aabc4f},
Eid = {123},
Journal = {AJ},
Month = Sep,
Pages = {123},
Primaryclass = {astro-ph.IM},
Title = {{The Astropy Project: Building an Open-science Project and Status of the v2.0 Core Package}},
Volume = {156},
Year = 2018
}

@article{corner,
  doi = {10.21105/joss.00024},
  url = {https://doi.org/10.21105/joss.00024},
  year  = {2016},
  month = {jun},
  publisher = {The Open Journal},
  volume = {1},
  number = {2},
  pages = {24},
  author = {Daniel Foreman-Mackey},
  title = {corner.py: Scatterplot matrices in Python},
  journal = {The J. Open Source Softw.}
}

@article{dynesty,
archivePrefix = {arXiv},
arxivId = {1904.02180},
author = {Speagle, Joshua S},
doi = {10.1093/mnras/staa278},
eprint = {1904.02180},
issn = {0035-8711},
journal = {MNRAS},
number = {3},
pages = {3132--3158},
title = {{dynesty: a dynamic nested sampling package for estimating Bayesian posteriors and evidences}},
volume = {493},
year = {2020}
}

@article{emcee,
archivePrefix = {arXiv},
arxivId = {1202.3665},
author = {Foreman-Mackey, Daniel and Hogg, David W. and Lang, Dustin and Goodman, Jonathan},
doi = {10.1086/670067},
eprint = {1202.3665},
issn = {00046280},
journal = {Publ. Astron. Soc. Pac.},
number = {925},
pages = {306--312},
title = {{emcee : The MCMC Hammer }},
volume = {125},
year = {2013}
}

@article{matplotlib,
  Author    = {Hunter, J. D.},
  Title     = {Matplotlib: A 2D graphics environment},
  Journal   = {Comput Sci Eng},
  Volume    = {9},
  Number    = {3},
  Pages     = {90--95},
  publisher = {IEEE COMPUTER SOC},
  doi       = {10.1109/MCSE.2007.55},
  year      = 2007
}

@ARTICLE{nautilus,
       author = {{Lange}, Johannes U.},
        title = "{NAUTILUS: boosting Bayesian importance nested sampling with deep learning}",
      journal = {\mnras},
         year = 2023,
        month = oct,
       volume = {525},
       number = {2},
        pages = {3181-3194},
          doi = {10.1093/mnras/stad2441},
archivePrefix = {arXiv},
       eprint = {2306.16923},
 primaryClass = {astro-ph.IM},
       adsurl = {https://ui.adsabs.harvard.edu/abs/2023MNRAS.525.3181L}
}

@article{numba,
author = {Lam, Siu Kwan and Pitrou, Antoine and Seibert, Stanley},
doi = {10.1145/2833157.2833162},
isbn = {9781450340052},
journal = {Proceedings of the Second Workshop on the LLVM Compiler Infrastructure in HPC - LLVM '15},
pages = {1--6},
title = {{Numba: a LLVM-based Python JIT compiler}},
url = {http://dl.acm.org/citation.cfm?doid=2833157.2833162},
year = {2015}
}

@article{numpy,
  author={S. {van der Walt} and S. C. {Colbert} and G. {Varoquaux}},
  doi={10.1109/MCSE.2011.37},
  journal={Comput Sci Eng},
  title={The NumPy Array2D: A Structure for Efficient Numerical Computation},
  year={2011},
  volume={13},
  number={2},
  pages={22-30},}

@article{pyautofit,
  doi = {10.21105/joss.02550},
  url = {https://doi.org/10.21105/joss.02550},
  year = {2021},
  publisher = {The Open Journal},
  volume = {6},
  number = {58},
  pages = {2550},
  author = {Nightingale, J. W. and Hayes, R. G. and Griffiths, M.},
  title = {`PyAutoFit`: A Classy Probabilistic Programming Language for Model Composition and Fitting},
  journal = {J. Open Source Softw.}
}

@article{pyautogalaxy,
  doi = {10.21105/joss.04475},
  url = {https://doi.org/10.21105/joss.04475},
  year = {2023},
  publisher = {The Open Journal},
  volume = {8},
  number = {81},
  pages = {4475},
  author = {James. W. Nightingale and Aristeidis Amvrosiadis and Richard G. Hayes and Qiuhan He and Amy Etherington and XiaoYue Cao and Shaun Cole and Jonathan Frawley and Carlos S. Frenk and Sam Lange and Ran Li and Richard J. Massey and Mattia Negrello and Andrew Robertson},
  title = {PyAutoGalaxy: Open-Source Multiwavelength Galaxy Structure & Morphology},
  journal = {J. Open Source Softw.}
 }

@article{pyautolens,
  doi = {10.21105/joss.02825},
  url = {https://doi.org/10.21105/joss.02825},
  year = {2021},
  publisher = {The Open Journal},
  volume = {6},
  number = {58},
  pages = {2825},
  author = {Nightingale, J. W. and Hayes, R. G. and Ashley Kelly and Aristeidis Amvrosiadis and Amy Etherington and Qiuhan He and Nan Li and XiaoYue Cao and Jonathan Frawley and Shaun Cole and Andrea Enia and Carlos S. Frenk and David R. Harvey and Ran Li and Richard J. Massey and Mattia Negrello and Andrew Robertson},
  title = {`PyAutoLens`: Open-Source Strong Gravitational Lensing},
  journal = {J. Open Source Softw.}
}

@book{python,
 author = {Van Rossum, Guido and Drake, Fred L.},
 title = {Python 3 Reference Manual},
 year = {2009},
 isbn = {1441412697},
 publisher = {CreateSpace},
 address = {Scotts Valley, CA}
}

@article{scipy,
       author = {{Virtanen}, Pauli and {Gommers}, Ralf and {Oliphant},
         Travis E. and {Haberland}, Matt and {Reddy}, Tyler and
         {Cournapeau}, David and {Burovski}, Evgeni and {Peterson}, Pearu
         and {Weckesser}, Warren and {Bright}, Jonathan and {van der Walt},
         St{\'e}fan J.  and {Brett}, Matthew and {Wilson}, Joshua and
         {Jarrod Millman}, K.  and {Mayorov}, Nikolay and {Nelson}, Andrew
         R.~J. and {Jones}, Eric and {Kern}, Robert and {Larson}, Eric and
         {Carey}, CJ and {Polat}, {\.I}lhan and {Feng}, Yu and {Moore},
         Eric W. and {Vand erPlas}, Jake and {Laxalde}, Denis and
         {Perktold}, Josef and {Cimrman}, Robert and {Henriksen}, Ian and
         {Quintero}, E.~A. and {Harris}, Charles R and {Archibald}, Anne M.
         and {Ribeiro}, Ant{\^o}nio H. and {Pedregosa}, Fabian and
         {van Mulbregt}, Paul and {Contributors}, SciPy 1. 0},
        title = "{SciPy 1.0: Fundamental Algorithms for Scientific
                  Computing in Python}",
      journal = {Nature Methods},
      year = "2020",
      volume={17},
      pages={261--272},
      adsurl = {https://rdcu.be/b08Wh},
      doi = {10.1038/s41592-019-0686-2},
}

@ARTICLE{ultranest,
       author = {{Buchner}, Johannes},
        title = "{UltraNest - a robust, general purpose Bayesian inference engine}",
      journal = {The J. Open Source Softw.},
         year = 2021,
        month = apr,
       volume = {6},
       number = {60},
          eid = {3001},
        pages = {3001},
          doi = {10.21105/joss.03001},
archivePrefix = {arXiv},
       eprint = {2101.09604},
 primaryClass = {stat.CO},
       adsurl = {https://ui.adsabs.harvard.edu/abs/2021JOSS....6.3001B}
}

@article{Nightingale2015,
archivePrefix = {arXiv},
arxivId = {1412.7436},
author = {Nightingale, J. W. and Dye, S.},
doi = {10.1093/mnras/stv1455},
eprint = {1412.7436},
issn = {13652966},
journal = {MNRAS},
month = {sep},
number = {3},
pages = {2940--2959},
title = {{Adaptive semi-linear inversion of strong gravitational lens imaging}},
volume = {452},
year = {2015}
}

@article{Nightingale2018,
archivePrefix = {arXiv},
arxivId = {1708.07377},
author = {Nightingale, J. W. and Dye, S. and Massey, Richard J.},
doi = {10.1093/mnras/sty1264},
eprint = {1708.07377},
issn = {13652966},
journal = {MNRAS},
number = {4},
pages = {4738--4784},
title = {{AutoLens: Automated modeling of a strong lens's light, mass, and source}},
url = {https://academic.oup.com/mnras/article/478/4/4738/5001434},
volume = {478},
year = {2018}
}

@article{dynesty4,
title={joshspeagle/dynesty: v2.1.2}, DOI={10.5281/zenodo.7995596},
abstractNote={<p>This is a bug release mostly concerning the checkpointing. Check the changelog for more details.</p>},
publisher={Zenodo},
author={Sergey Koposov and Josh Speagle and Kyle Barbary and Gregory Ashton and Ed Bennett and Johannes Buchner
and Carl Scheffler and Ben Cook and Colm Talbot and James Guillochon and et al.}, year={2023}, month={Jun} }

@ARTICLE{SLI,
       author = {{Warren}, S.~J. and {Dye}, S.},
        title = "{Semilinear Gravitational Lens Inversion}",
      journal = {\apj},
         year = 2003,
        month = jun,
       volume = {590},
       number = {2},
        pages = {673-682},
          doi = {10.1086/375132},
archivePrefix = {arXiv},
       eprint = {astro-ph/0302587},
 primaryClass = {astro-ph},
       adsurl = {https://ui.adsabs.harvard.edu/abs/2003ApJ...590..673W}
}

@ARTICLE{Massardi18,
       author = {{Massardi}, M. and {Enia}, A.~F.~M. and {Negrello}, M. and {Mancuso}, C. and {Lapi}, A. and {Vignali}, C. and {Gilli}, R. and {Burkutean}, S. and {Danese}, L. and {Zotti}, G. De},
        title = "{Chandra and ALMA observations of the nuclear activity in two strongly lensed star-forming galaxies}",
      journal = {\aap},
         year = 2018,
        month = feb,
       volume = {610},
          eid = {A53},
        pages = {A53},
          doi = {10.1051/0004-6361/201731751},
archivePrefix = {arXiv},
       eprint = {1709.10427},
 primaryClass = {astro-ph.GA},
       adsurl = {https://ui.adsabs.harvard.edu/abs/2018A&A...610A..53M}
}

@ARTICLE{Dye18,
       author = {{Dye}, S. and {Furlanetto}, C. and {Dunne}, L. and {Eales}, S.~A. and {Negrello}, M. and {Nayyeri}, H. and {van der Werf}, P.~P. and {Serjeant}, S. and {Farrah}, D. and {Micha{\l}owski}, M.~J. and {Baes}, M. and {Marchetti}, L. and {Cooray}, A. and {Riechers}, D.~A. and {Amvrosiadis}, A.},
        title = "{Modelling high-resolution ALMA observations of strongly lensed highly star-forming galaxies detected by Herschel}",
      journal = {\mnras},
         year = 2018,
        month = jun,
       volume = {476},
       number = {4},
        pages = {4383-4394},
          doi = {10.1093/mnras/sty513},
archivePrefix = {arXiv},
       eprint = {1705.05413},
 primaryClass = {astro-ph.GA},
       adsurl = {https://ui.adsabs.harvard.edu/abs/2018MNRAS.476.4383D}
}

@ARTICLE{Dye22,
       author = {{Dye}, S. and {Eales}, S.~A. and {Gomez}, H.~L. and {Jones}, G.~C. and {Smith}, M.~W.~L. and {Borsato}, E. and {Moss}, A. and {Dunne}, L. and {Maresca}, J. and {Amvrosiadis}, A. and {Negrello}, M. and {Marchetti}, L. and {Corsini}, E.~M. and {Ivison}, R.~J. and {Bendo}, G.~J. and {Bakx}, T. and {Cooray}, A. and {Cox}, P. and {Dannerbauer}, H. and {Serjeant}, S. and {Riechers}, D. and {Temi}, P. and {Vlahakis}, C.},
        title = "{A high-resolution investigation of the multiphase ISM in a galaxy during the first two billion years}",
      journal = {\mnras},
         year = 2022,
        month = mar,
       volume = {510},
       number = {3},
        pages = {3734-3757},
          doi = {10.1093/mnras/stab3569},
archivePrefix = {arXiv},
       eprint = {2112.03936},
 primaryClass = {astro-ph.GA},
       adsurl = {https://ui.adsabs.harvard.edu/abs/2022MNRAS.510.3734D}
}

@ARTICLE{Enia18,
       author = {{Enia}, A. and {Negrello}, M. and {Gurwell}, M. and {Dye}, S. and {Rodighiero}, G. and {Massardi}, M. and {De Zotti}, G. and {Franceschini}, A. and {Cooray}, A. and {van der Werf}, P. and {Birkinshaw}, M. and {Micha{\l}owski}, M.~J. and {Oteo}, I.},
        title = "{The Herschel-ATLAS: magnifications and physical sizes of 500-{\ensuremath{\mu}}m-selected strongly lensed galaxies}",
      journal = {\mnras},
         year = 2018,
        month = apr,
       volume = {475},
       number = {3},
        pages = {3467-3484},
          doi = {10.1093/mnras/sty021},
archivePrefix = {arXiv},
       eprint = {1801.01831},
 primaryClass = {astro-ph.GA},
       adsurl = {https://ui.adsabs.harvard.edu/abs/2018MNRAS.475.3467E}
}

@ARTICLE{Maresca22,
       author = {{Maresca}, Jacob and {Dye}, Simon and {Amvrosiadis}, Aristeidis and {Bendo}, George and {Cooray}, Asantha and {De Zotti}, Gianfranco and {Dunne}, Loretta and {Eales}, Stephen and {Furlanetto}, Cristina and {Gonz{\'a}lez-Nuevo}, Joaquin and {Greener}, Michael and {Ivison}, Robert and {Lapi}, Andrea and {Negrello}, Mattia and {Riechers}, Dominik and {Serjeant}, Stephen and {Tergolina}, M{\^o}nica and {Wardlow}, Julie},
        title = "{Modelling high-resolution ALMA observations of strongly lensed dusty star-forming galaxies detected by Herschel}",
      journal = {\mnras},
         year = 2022,
        month = may,
       volume = {512},
       number = {2},
        pages = {2426-2438},
          doi = {10.1093/mnras/stac585},
archivePrefix = {arXiv},
       eprint = {2111.09680},
 primaryClass = {astro-ph.GA},
       adsurl = {https://ui.adsabs.harvard.edu/abs/2022MNRAS.512.2426M}
}

@ARTICLE{kormann94,
       author = {{Kormann}, R. and {Schneider}, P. and {Bartelmann}, M.},
        title = "{Isothermal elliptical gravitational lens models.}",
      journal = {\aap},
         year = 1994,
        month = apr,
       volume = {284},
        pages = {285-299},
       adsurl = {https://ui.adsabs.harvard.edu/abs/1994A&A...284..285K}
}

@ARTICLE{carilli&walter13,
       author = {{Carilli}, C.~L. and {Walter}, F.},
        title = "{Cool Gas in High-Redshift Galaxies}",
      journal = {\araa},
         year = 2013,
        month = aug,
       volume = {51},
       number = {1},
        pages = {105-161},
          doi = {10.1146/annurev-astro-082812-140953},
archivePrefix = {arXiv},
       eprint = {1301.0371},
 primaryClass = {astro-ph.CO},
       adsurl = {https://ui.adsabs.harvard.edu/abs/2013ARA&A..51..105C}
}

@ARTICLE{abrahamyan15,
       author = {{Abrahamyan}, H.~V. and {Mickaelian}, A.~M. and {Knyazyan}, A.~V.},
        title = "{The IRAS PSC/FSC Combined Catalogue}",
      journal = {Astronomy and Computing},
         year = 2015,
        month = apr,
       volume = {10},
        pages = {99-106},
          doi = {10.1016/j.ascom.2014.12.002},
       adsurl = {https://ui.adsabs.harvard.edu/abs/2015A&C....10...99A}
}

@ARTICLE{herranz08,
       author = {{Herranz}, D. and {Sanz}, J.~L.},
        title = "{Matrix Filters for the Detection of Extragalactic Point Sources in Cosmic Microwave Background Images}",
      journal = {IEEE Journal of Selected Topics in Signal Processing},
         year = 2008,
        month = nov,
       volume = {2},
       number = {5},
        pages = {727-734},
          doi = {10.1109/JSTSP.2008.2005339},
archivePrefix = {arXiv},
       eprint = {0808.0300},
 primaryClass = {astro-ph},
       adsurl = {https://ui.adsabs.harvard.edu/abs/2008ISTSP...2..727H}
}

@INPROCEEDINGS{briggs95,
       author = {{Briggs}, D.~S.},
        title = "{High Fidelity Interferometric Imaging: Robust Weighting and NNLS Deconvolution}",
    booktitle = {American Astronomical Society Meeting Abstracts},
         year = 1995,
       series = {American Astronomical Society Meeting Abstracts},
       volume = {187},
        month = dec,
          eid = {112.02},
        pages = {112.02},
       adsurl = {https://ui.adsabs.harvard.edu/abs/1995AAS...18711202B}
}

@ARTICLE{Prajapati2026,
       author = {{Prajapati}, Prachi and {Riechers}, Dominik and {Cox}, Pierre and {Weiss}, Axel and {Saintonge}, Am{\'e}lie and {Jones}, Bethany and {Bakx}, Tom J.~L.~C. and {Berta}, Stefano and {van der Werf}, Paul and {Neri}, Roberto and {Butler}, Kirsty M. and {Cooray}, Asantha and {Ismail}, Diana and {Baker}, Andrew J. and {Borsato}, Edoardo and {Harris}, Andrew and {Ivison}, Rob and {Lehnert}, Matthew and {Marchetti}, Lucia and {Messias}, Hugo and {Omont}, Alain and {Vlahakis}, Catherine and {Yang}, Chentao},
        title = "{Vz-GAL: Probing Cold Molecular Gas in Dusty Star-forming Galaxies at z = 1─6}",
      journal = {\apjs},
         year = 2026,
        month = feb,
       volume = {282},
       number = {2},
          eid = {40},
        pages = {40},
          doi = {10.3847/1538-4365/ae27d4},
archivePrefix = {arXiv},
       eprint = {2509.25167},
 primaryClass = {astro-ph.GA},
       adsurl = {https://ui.adsabs.harvard.edu/abs/2026ApJS..282...40P}
}

@ARTICLE{cox23,
       author = {{Cox}, P. and {Neri}, R. and {Berta}, S. and {Ismail}, D. and {Stanley}, F. and {Young}, A. and {Jin}, S. and {Bakx}, T. and {Beelen}, A. and {Dannerbauer}, H. and {Krips}, M. and {Lehnert}, M. and {Omont}, A. and {Riechers}, D.~A. and {Baker}, A.~J. and {Bendo}, G. and {Borsato}, E. and {Buat}, V. and {Butler}, K. and {Chartab}, N. and {Cooray}, A. and {Dye}, S. and {Eales}, S. and {Gavazzi}, R. and {Hughes}, D. and {Ivison}, R. and {Jones}, B.~M. and {Marchetti}, L. and {Messias}, H. and {Nanni}, A. and {Negrello}, M. and {Perez-Fournon}, I. and {Serjeant}, S. and {Urquhart}, S. and {Vlahakis}, C. and {Wei{\ss}}, A. and {van der Werf}, P. and {Yang}, C.},
        title = "{z-GAL: A NOEMA spectroscopic redshift survey of bright Herschel galaxies. I. Overview}",
      journal = {\aap},
         year = 2023,
        month = oct,
       volume = {678},
          eid = {A26},
        pages = {A26},
          doi = {10.1051/0004-6361/202346801},
archivePrefix = {arXiv},
       eprint = {2307.15732},
 primaryClass = {astro-ph.GA},
       adsurl = {https://ui.adsabs.harvard.edu/abs/2023A&A...678A..26C}
}

@ARTICLE{ismail23,
       author = {{Ismail}, D. and {Beelen}, A. and {Buat}, V. and {Berta}, S. and {Cox}, P. and {Stanley}, F. and {Young}, A. and {Jin}, S. and {Neri}, R. and {Bakx}, T. and {Dannerbauer}, H. and {Butler}, K. and {Cooray}, A. and {Nanni}, A. and {Omont}, A. and {Serjeant}, S. and {van der Werf}, P. and {Vlahakis}, C. and {Wei{\ss}}, A. and {Yang}, C. and {Baker}, A.~J. and {Bendo}, G. and {Borsato}, E. and {Chartab}, N. and {Dye}, S. and {Eales}, S. and {Gavazzi}, R. and {Hughes}, D. and {Ivison}, R. and {Jones}, B.~M. and {Krips}, M. and {Lehnert}, M. and {Marchetti}, L. and {Messias}, H. and {Negrello}, M. and {Perez-Fournon}, I. and {Riechers}, D.~A. and {Urquhart}, S.},
        title = "{z-GAL: A NOEMA spectroscopic redshift survey of bright Herschel galaxies. II. Dust properties}",
      journal = {\aap},
         year = 2023,
        month = oct,
       volume = {678},
          eid = {A27},
        pages = {A27},
          doi = {10.1051/0004-6361/202346804},
archivePrefix = {arXiv},
       eprint = {2307.15747},
 primaryClass = {astro-ph.GA},
       adsurl = {https://ui.adsabs.harvard.edu/abs/2023A&A...678A..27I}
}

@ARTICLE{berta23,
       author = {{Berta}, S. and {Stanley}, F. and {Ismail}, D. and {Cox}, P. and {Neri}, R. and {Yang}, C. and {Young}, A.~J. and {Jin}, S. and {Dannerbauer}, H. and {Bakx}, T.~J.~L.~C. and {Beelen}, A. and {Wei{\ss}}, A. and {Nanni}, A. and {Omont}, A. and {van der Werf}, P. and {Krips}, M. and {Baker}, A.~J. and {Bendo}, G. and {Borsato}, E. and {Buat}, V. and {Butler}, K.~M. and {Chartab}, N. and {Cooray}, A. and {Dye}, S. and {Eales}, S. and {Gavazzi}, R. and {Hughes}, D. and {Ivison}, R.~J. and {Jones}, B.~M. and {Lehnert}, M. and {Marchetti}, L. and {Messias}, H. and {Negrello}, M. and {Perez-Fournon}, I. and {Riechers}, D.~A. and {Serjeant}, S. and {Urquhart}, S. and {Vlahakis}, C.},
        title = "{z-GAL: A NOEMA spectroscopic redshift survey of bright Herschel galaxies. III. Physical properties}",
      journal = {\aap},
         year = 2023,
        month = oct,
       volume = {678},
          eid = {A28},
        pages = {A28},
          doi = {10.1051/0004-6361/202346803},
archivePrefix = {arXiv},
       eprint = {2307.15748},
 primaryClass = {astro-ph.GA},
       adsurl = {https://ui.adsabs.harvard.edu/abs/2023A&A...678A..28B}
}

@ARTICLE{harris12,
       author = {{Harris}, A.~I. and {Baker}, A.~J. and {Frayer}, D.~T. and {Smail}, Ian and {Swinbank}, A.~M. and {Riechers}, D.~A. and {van der Werf}, P.~P. and {Auld}, R. and {Baes}, M. and {Bussmann}, R.~S. and {Buttiglione}, S. and {Cava}, A. and {Clements}, D.~L. and {Cooray}, A. and {Dannerbauer}, H. and {Dariush}, A. and {De Zotti}, G. and {Dunne}, L. and {Dye}, S. and {Eales}, S. and {Fritz}, J. and {Gonz{\'a}lez-Nuevo}, J. and {Hopwood}, R. and {Ibar}, E. and {Ivison}, R.~J. and {Jarvis}, M.~J. and {Maddox}, S. and {Negrello}, M. and {Rigby}, E. and {Smith}, D.~J.~B. and {Temi}, P. and {Wardlow}, J.},
        title = "{Blind Detections of CO J = 1-0 in 11 H-ATLAS Galaxies at z = 2.1-3.5 with the GBT/Zpectrometer}",
      journal = {\apj},
         year = 2012,
        month = jun,
       volume = {752},
       number = {2},
          eid = {152},
        pages = {152},
          doi = {10.1088/0004-637X/752/2/152},
archivePrefix = {arXiv},
       eprint = {1204.4706},
 primaryClass = {astro-ph.CO},
       adsurl = {https://ui.adsabs.harvard.edu/abs/2012ApJ...752..152H}
}

@ARTICLE{tully&fischer77,
       author = {{Tully}, R.~B. and {Fisher}, J.~R.},
        title = "{A new method of determining distances to galaxies.}",
      journal = {\aap},
         year = 1977,
        month = feb,
       volume = {54},
        pages = {661-673},
       adsurl = {https://ui.adsabs.harvard.edu/abs/1977A&A....54..661T}
}

@ARTICLE{dick&kazes92,
       author = {{Dickey}, John M. and {Kazes}, Ilya},
        title = "{The Tully-Fisher Relation for the CO Line}",
      journal = {\apj},
         year = 1992,
        month = jul,
       volume = {393},
        pages = {530},
          doi = {10.1086/171526},
       adsurl = {https://ui.adsabs.harvard.edu/abs/1992ApJ...393..530D}
}

@ARTICLE{schoniger94,
       author = {{Schoniger}, F. and {Sofue}, Y.},
        title = "{CO versus HI in the Tully-Fisher relation for a sample of 32 galaxies.}",
      journal = {\aap},
         year = 1994,
        month = mar,
       volume = {283},
        pages = {21-31},
       adsurl = {https://ui.adsabs.harvard.edu/abs/1994A&A...283...21S}
}

@ARTICLE{bolatto13,
       author = {{Bolatto}, Alberto D. and {Wolfire}, Mark and {Leroy}, Adam K.},
        title = "{The CO-to-H$_{2}$ Conversion Factor}",
      journal = {\araa},
         year = 2013,
        month = aug,
       volume = {51},
       number = {1},
        pages = {207-268},
          doi = {10.1146/annurev-astro-082812-140944},
archivePrefix = {arXiv},
       eprint = {1301.3498},
 primaryClass = {astro-ph.GA},
       adsurl = {https://ui.adsabs.harvard.edu/abs/2013ARA&A..51..207B}
}

@ARTICLE{powell21,
       author = {{Powell}, Devon and {Vegetti}, Simona and {McKean}, John P. and {Spingola}, Cristiana and {Rizzo}, Francesca and {Stacey}, Hannah R.},
        title = "{A novel approach to visibility-space modelling of interferometric gravitational lens observations at high angular resolution}",
      journal = {\mnras},
         year = 2021,
        month = feb,
       volume = {501},
       number = {1},
        pages = {515-530},
          doi = {10.1093/mnras/staa2740},
archivePrefix = {arXiv},
       eprint = {2005.03609},
 primaryClass = {astro-ph.IM},
       adsurl = {https://ui.adsabs.harvard.edu/abs/2021MNRAS.501..515P}
}

@ARTICLE{rybak15,
       author = {{Rybak}, M. and {Vegetti}, S. and {McKean}, J.~P. and {Andreani}, P. and {White}, S.~D.~M.},
        title = "{ALMA imaging of SDP.81 - II. A pixelated reconstruction of the CO emission lines}",
      journal = {\mnras},
         year = 2015,
        month = oct,
       volume = {453},
       number = {1},
        pages = {L26-L30},
          doi = {10.1093/mnrasl/slv092},
archivePrefix = {arXiv},
       eprint = {1506.01425},
 primaryClass = {astro-ph.GA},
       adsurl = {https://ui.adsabs.harvard.edu/abs/2015MNRAS.453L..26R}
}

@ARTICLE{rybak20,
       author = {{Rybak}, Matus and {Hodge}, J.~A. and {Vegetti}, S. and {van der Werf}, P. and {Andreani}, P. and {Graziani}, L. and {McKean}, J.~P.},
        title = "{Full of Orions: a 200-pc mapping of the interstellar medium in the redshift-3 lensed dusty star-forming galaxy SDP.81}",
      journal = {\mnras},
         year = 2020,
        month = jun,
       volume = {494},
       number = {4},
        pages = {5542-5567},
          doi = {10.1093/mnras/staa879},
archivePrefix = {arXiv},
       eprint = {1912.12538},
 primaryClass = {astro-ph.GA},
       adsurl = {https://ui.adsabs.harvard.edu/abs/2020MNRAS.494.5542R}
}

@ARTICLE{rizzo21,
       author = {{Rizzo}, Francesca and {Vegetti}, Simona and {Fraternali}, Filippo and {Stacey}, Hannah R. and {Powell}, Devon},
        title = "{Dynamical properties of z  4.5 dusty star-forming galaxies and their connection with local early-type galaxies}",
      journal = {\mnras},
         year = 2021,
        month = nov,
       volume = {507},
       number = {3},
        pages = {3952-3984},
          doi = {10.1093/mnras/stab2295},
archivePrefix = {arXiv},
       eprint = {2102.05671},
 primaryClass = {astro-ph.GA},
       adsurl = {https://ui.adsabs.harvard.edu/abs/2021MNRAS.507.3952R}
}

@ARTICLE{planck13xv,
       author = {{Planck Collaboration} and {Ade}, P.~A.~R. and {Aghanim}, N. and {Armitage-Caplan}, C. and {Arnaud}, M. and {Ashdown}, M. and {Atrio-Barandela}, F. and {Aumont}, J. and {Baccigalupi}, C. and {Banday}, A.~J. and {Barreiro}, R.~B. and {Bartlett}, J.~G. and {Battaner}, E. and {Benabed}, K. and {Beno{\^\i}t}, A. and {Benoit-L{\'e}vy}, A. and {Bernard}, J.-P. and {Bersanelli}, M. and {Bielewicz}, P. and {Bobin}, J. and {Bock}, J.~J. and {Bonaldi}, A. and {Bonavera}, L. and {Bond}, J.~R. and {Borrill}, J. and {Bouchet}, F.~R. and {Boulanger}, F. and {Bridges}, M. and {Bucher}, M. and {Burigana}, C. and {Butler}, R.~C. and {Calabrese}, E. and {Cardoso}, J.-F. and {Catalano}, A. and {Challinor}, A. and {Chamballu}, A. and {Chiang}, H.~C. and {Chiang}, L.-Y. and {Christensen}, P.~R. and {Church}, S. and {Clements}, D.~L. and {Colombi}, S. and {Colombo}, L.~P.~L. and {Combet}, C. and {Couchot}, F. and {Coulais}, A. and {Crill}, B.~P. and {Curto}, A. and {Cuttaia}, F. and {Danese}, L. and {Davies}, R.~D. and {Davis}, R.~J. and {de Bernardis}, P. and {de Rosa}, A. and {de Zotti}, G. and {Delabrouille}, J. and {Delouis}, J.-M. and {D{\'e}sert}, F.-X. and {Dickinson}, C. and {Diego}, J.~M. and {Dole}, H. and {Donzelli}, S. and {Dor{\'e}}, O. and {Douspis}, M. and {Dunkley}, J. and {Dupac}, X. and {Efstathiou}, G. and {Elsner}, F. and {En{\ss}lin}, T.~A. and {Eriksen}, H.~K. and {Finelli}, F. and {Forni}, O. and {Frailis}, M. and {Fraisse}, A.~A. and {Franceschi}, E. and {Gaier}, T.~C. and {Galeotta}, S. and {Galli}, S. and {Ganga}, K. and {Giard}, M. and {Giardino}, G. and {Giraud-H{\'e}raud}, Y. and {Gjerl{\o}w}, E. and {Gonz{\'a}lez-Nuevo}, J. and {G{\'o}rski}, K.~M. and {Gratton}, S. and {Gregorio}, A. and {Gruppuso}, A. and {Gudmundsson}, J.~E. and {Hansen}, F.~K. and {Hanson}, D. and {Harrison}, D. and {Helou}, G. and {Henrot-Versill{\'e}}, S. and {Hern{\'a}ndez-Monteagudo}, C. and {Herranz}, D. and {Hildebrandt}, S.~R. and {Hivon}, E. and {Hobson}, M. and {Holmes}, W.~A. and {Hornstrup}, A. and {Hovest}, W. and {Huffenberger}, K.~M. and {Hurier}, G. and {Jaffe}, A.~H. and {Jaffe}, T.~R. and {Jewell}, J. and {Jones}, W.~C. and {Juvela}, M. and {Keih{\"a}nen}, E. and {Keskitalo}, R. and {Kiiveri}, K. and {Kisner}, T.~S. and {Kneissl}, R. and {Knoche}, J. and {Knox}, L. and {Kunz}, M. and {Kurki-Suonio}, H. and {Lagache}, G. and {L{\"a}hteenm{\"a}ki}, A. and {Lamarre}, J.-M. and {Lasenby}, A. and {Lattanzi}, M. and {Laureijs}, R.~J. and {Lawrence}, C.~R. and {Le Jeune}, M. and {Leach}, S. and {Leahy}, J.~P. and {Leonardi}, R. and {Le{\'o}n-Tavares}, J. and {Lesgourgues}, J. and {Liguori}, M. and {Lilje}, P.~B. and {Linden-V{\o}rnle}, M. and {Lindholm}, V. and {L{\'o}pez-Caniego}, M. and {Lubin}, P.~M. and {Mac{\'\i}as-P{\'e}rez}, J.~F. and {Maffei}, B. and {Maino}, D. and {Mandolesi}, N. and {Marinucci}, D. and {Maris}, M. and {Marshall}, D.~J. and {Martin}, P.~G. and {Mart{\'\i}nez-Gonz{\'a}lez}, E. and {Masi}, S. and {Massardi}, M. and {Matarrese}, S. and {Matthai}, F. and {Mazzotta}, P. and {Meinhold}, P.~R. and {Melchiorri}, A. and {Mendes}, L. and {Menegoni}, E. and {Mennella}, A. and {Migliaccio}, M. and {Millea}, M. and {Mitra}, S. and {Miville-Desch{\^e}nes}, M.-A. and {Molinari}, D. and {Moneti}, A. and {Montier}, L. and {Morgante}, G. and {Mortlock}, D. and {Moss}, A. and {Munshi}, D. and {Murphy}, J.~A. and {Naselsky}, P. and {Nati}, F. and {Natoli}, P. and {Netterfield}, C.~B. and {N{\o}rgaard-Nielsen}, H.~U. and {Noviello}, F. and {Novikov}, D. and {Novikov}, I. and {O'Dwyer}, I.~J. and {Orieux}, F. and {Osborne}, S. and {Oxborrow}, C.~A. and {Paci}, F. and {Pagano}, L. and {Pajot}, F. and {Paladini}, R. and {Paoletti}, D. and {Partridge}, B. and {Pasian}, F. and {Patanchon}, G. and {Paykari}, P. and {Perdereau}, O. and {Perotto}, L. and {Perrotta}, F. and {Piacentini}, F. and {Piat}, M. and {Pierpaoli}, E. and {Pietrobon}, D. and {Plaszczynski}, S. and {Pointecouteau}, E. and {Polenta}, G. and {Ponthieu}, N.},
        title = "{Planck 2013 results. XV. CMB power spectra and likelihood}",
      journal = {\aap},
         year = 2014,
        month = nov,
       volume = {571},
          eid = {A15},
        pages = {A15},
          doi = {10.1051/0004-6361/201321573},
archivePrefix = {arXiv},
       eprint = {1303.5075},
 primaryClass = {astro-ph.CO},
       adsurl = {https://ui.adsabs.harvard.edu/abs/2014A&A...571A..15P}
}

@INPROCEEDINGS{mcmullin07,
       author = {{McMullin}, J.~P. and {Waters}, B. and {Schiebel}, D. and {Young}, W. and {Golap}, K.},
        title = "{CASA Architecture and Applications}",
    booktitle = {Astronomical Data Analysis Software and Systems XVI},
         year = 2007,
       editor = {{Shaw}, R.~A. and {Hill}, F. and {Bell}, D.~J.},
       series = {Astronomical Society of the Pacific Conference Series},
       volume = {376},
        month = oct,
        pages = {127},
       adsurl = {https://ui.adsabs.harvard.edu/abs/2007ASPC..376..127M}
}

@ARTICLE{Lammers22,
       author = {{Lammers}, Caleb and {Hill}, Ryley and {Lim}, Seunghwan and {Scott}, Douglas and {Ca{\~n}ameras}, Raoul and {Dole}, Herv{\'e}},
        title = "{Candidate high-redshift protoclusters and lensed galaxies in the Planck list of high-z sources overlapping with Herschel-SPIRE imaging}",
      journal = {\mnras},
         year = 2022,
        month = aug,
       volume = {514},
       number = {4},
        pages = {5004-5023},
          doi = {10.1093/mnras/stac1555},
archivePrefix = {arXiv},
       eprint = {2204.06752},
 primaryClass = {astro-ph.GA},
       adsurl = {https://ui.adsabs.harvard.edu/abs/2022MNRAS.514.5004L}
}

@ARTICLE{Foo2025,
       author = {{Foo}, Nicholas and {Harrington}, Kevin C. and {Frye}, Brenda L. and {Kamieneski}, Patrick S. and {Yun}, Min S. and {Pascale}, Massimo and {Yoon}, Ilsang and {Noble}, Allison and {Windhorst}, Rogier A. and {Cohen}, Seth H. and {Lowenthal}, James D. and {Kaasinen}, Melanie and {Alcalde Pampliega}, Bel{\'e}n and {Liu}, Daizhong and {Cooper}, Olivia and {Garcia Diaz}, Carlos and {D{\'\i}az-S{\'a}nchez}, Anastasio and {Diego}, Jose and {Garuda}, Nikhil and {Jim{\'e}nez-Andrade}, Eric F. and {Leimbach}, Reagen and {Vishwas}, Amit and {Wang}, Q. Daniel and {Zhou}, Dazhi and {Zitrin}, Adi},
        title = "{PASSAGES: The Discovery of a Strongly Lensed Protocluster Core Candidate at Cosmic Noon}",
      journal = {\apj},
         year = 2025,
        month = dec,
       volume = {995},
       number = {2},
          eid = {219},
        pages = {219},
          doi = {10.3847/1538-4357/adf4d5},
archivePrefix = {arXiv},
       eprint = {2504.05617},
 primaryClass = {astro-ph.GA},
       adsurl = {https://ui.adsabs.harvard.edu/abs/2025ApJ...995..219F}
}

@ARTICLE{iglesias_groth2017,
       author = {{Iglesias-Groth}, S. and {D{\'\i}az-S{\'a}nchez}, A. and {Rebolo}, R. and {Dannerbauer}, H.},
        title = "{A near/mid infrared search for ultra-bright submillimetre galaxies: Searching for Cosmic Eyelash Analogues}",
      journal = {\mnras},
         year = 2017,
        month = may,
       volume = {467},
       number = {1},
        pages = {330-339},
          doi = {10.1093/mnras/stx041},
archivePrefix = {arXiv},
       eprint = {1702.03206},
 primaryClass = {astro-ph.GA},
       adsurl = {https://ui.adsabs.harvard.edu/abs/2017MNRAS.467..330I}
}

@ARTICLE{solomon92,
       author = {{Solomon}, P.~M. and {Downes}, D. and {Radford}, S.~J.~E.},
        title = "{Warm Molecular Gas in the Primeval Galaxy IRAS 10214+4724}",
      journal = {\apjl},
         year = 1992,
        month = oct,
       volume = {398},
        pages = {L29},
          doi = {10.1086/186569},
       adsurl = {https://ui.adsabs.harvard.edu/abs/1992ApJ...398L..29S}
}

@ARTICLE{Serjeant2012,
       author = {{Serjeant}, Stephen},
        title = "{Strong biases in infrared-selected gravitational lenses}",
      journal = {\mnras},
         year = 2012,
        month = aug,
       volume = {424},
       number = {4},
        pages = {2429-2441},
          doi = {10.1111/j.1365-2966.2012.20761.x},
archivePrefix = {arXiv},
       eprint = {1203.2647},
 primaryClass = {astro-ph.CO},
       adsurl = {https://ui.adsabs.harvard.edu/abs/2012MNRAS.424.2429S}
}

@ARTICLE{Blain1999,
       author = {{Blain}, A.~W.},
        title = "{The differential magnification of high-redshift ultraluminous infrared galaxies}",
      journal = {\mnras},
         year = 1999,
        month = apr,
       volume = {304},
       number = {3},
        pages = {669-673},
          doi = {10.1046/j.1365-8711.1999.02426.x},
archivePrefix = {arXiv},
       eprint = {astro-ph/9903221},
 primaryClass = {astro-ph},
       adsurl = {https://ui.adsabs.harvard.edu/abs/1999MNRAS.304..669B}
}

@ARTICLE{Lovell2022,
       author = {{Lovell}, C.~C. and {Geach}, J.~E. and {Dav{\'e}}, R. and {Narayanan}, D. and {Coppin}, K.~E.~K. and {Li}, Q. and {Franco}, M. and {Privon}, G.~C.},
        title = "{An orientation bias in observations of submillimetre galaxies}",
      journal = {\mnras},
         year = 2022,
        month = sep,
       volume = {515},
       number = {3},
        pages = {3644-3655},
          doi = {10.1093/mnras/stac2008},
archivePrefix = {arXiv},
       eprint = {2106.11588},
 primaryClass = {astro-ph.GA},
       adsurl = {https://ui.adsabs.harvard.edu/abs/2022MNRAS.515.3644L}
}

@ARTICLE{Bakx2018,
       author = {{Bakx}, Tom J.~L.~C. and {Eales}, S.~A. and {Negrello}, M. and {Smith}, M.~W.~L. and {Valiante}, E. and {Holland}, W.~S. and {Baes}, M. and {Bourne}, N. and {Clements}, D.~L. and {Dannerbauer}, H. and {De Zotti}, G. and {Dunne}, L. and {Dye}, S. and {Furlanetto}, C. and {Ivison}, R.~J. and {Maddox}, S. and {Marchetti}, L. and {Micha{\l}owski}, M.~J. and {Omont}, A. and {Oteo}, I. and {Wardlow}, J.~L. and {van der Werf}, P. and {Yang}, C.},
        title = "{The Herschel Bright Sources (HerBS): sample definition and SCUBA-2 observations}",
      journal = {\mnras},
         year = 2018,
        month = jan,
       volume = {473},
       number = {2},
        pages = {1751-1773},
          doi = {10.1093/mnras/stx2267},
archivePrefix = {arXiv},
       eprint = {1709.01514},
 primaryClass = {astro-ph.GA},
       adsurl = {https://ui.adsabs.harvard.edu/abs/2018MNRAS.473.1751B}
}

@ARTICLE{Hardcastle2023,
       author = {{Hardcastle}, M.~J. and {Horton}, M.~A. and {Williams}, W.~L. and {Duncan}, K.~J. and {Alegre}, L. and {Barkus}, B. and {Croston}, J.~H. and {Dickinson}, H. and {Osinga}, E. and {R{\"o}ttgering}, H.~J.~A. and {Sabater}, J. and {Shimwell}, T.~W. and {Smith}, D.~J.~B. and {Best}, P.~N. and {Botteon}, A. and {Br{\"u}ggen}, M. and {Drabent}, A. and {de Gasperin}, F. and {G{\"u}rkan}, G. and {Hajduk}, M. and {Hale}, C.~L. and {Hoeft}, M. and {Jamrozy}, M. and {Kunert-Bajraszewska}, M. and {Kondapally}, R. and {Magliocchetti}, M. and {Mahatma}, V.~H. and {Mostert}, R.~I.~J. and {O'Sullivan}, S.~P. and {Pajdosz-{\'S}mierciak}, U. and {Petley}, J. and {Pierce}, J.~C.~S. and {Prandoni}, I. and {Schwarz}, D.~J. and {Shulewski}, A. and {Siewert}, T.~M. and {Stott}, J.~P. and {Tang}, H. and {Vaccari}, M. and {Zheng}, X. and {Bailey}, T. and {Desbled}, S. and {Goyal}, A. and {Gonano}, V. and {Hanset}, M. and {Kurtz}, W. and {Lim}, S.~M. and {Mielle}, L. and {Molloy}, C.~S. and {Roth}, R. and {Terentev}, I.~A. and {Torres}, M.},
        title = "{The LOFAR Two-Metre Sky Survey. VI. Optical identifications for the second data release}",
      journal = {\aap},
         year = 2023,
        month = oct,
       volume = {678},
          eid = {A151},
        pages = {A151},
          doi = {10.1051/0004-6361/202347333},
archivePrefix = {arXiv},
       eprint = {2309.00102},
 primaryClass = {astro-ph.GA},
       adsurl = {https://ui.adsabs.harvard.edu/abs/2023A&A...678A.151H}
}

@ARTICLE{Hardcastle2025,
       author = {{Hardcastle}, M.~J. and {Pierce}, J.~C.~S. and {Duncan}, K.~J. and {G{\"u}rkan}, G. and {Gong}, Y. and {Horton}, M.~A. and {Mingo}, B. and {R{\"o}ttgering}, H.~J.~A. and {Smith}, D.~J.~B.},
        title = "{Radio AGN selection in LoTSS DR2}",
      journal = {\mnras},
         year = 2025,
        month = may,
       volume = {539},
       number = {2},
        pages = {1856-1878},
          doi = {10.1093/mnras/staf622},
archivePrefix = {arXiv},
       eprint = {2504.09303},
 primaryClass = {astro-ph.GA},
       adsurl = {https://ui.adsabs.harvard.edu/abs/2025MNRAS.539.1856H}
}

@ARTICLE{Dannerbauer2019,
       author = {{Dannerbauer}, H. and {Harrington}, K. and {D{\'\i}az-S{\'a}nchez}, A. and {Iglesias-Groth}, S. and {Rebolo}, R. and {Genova-Santos}, R.~T. and {Krips}, M.},
        title = "{Ultra-bright CO and [C I] Emission in a Lensed z = 2.04 Submillimeter Galaxy with Extreme Molecular Gas Properties}",
      journal = {\aj},
         year = 2019,
        month = jul,
       volume = {158},
       number = {1},
          eid = {34},
        pages = {34},
          doi = {10.3847/1538-3881/aaf50b},
archivePrefix = {arXiv},
       eprint = {1812.03845},
 primaryClass = {astro-ph.GA},
       adsurl = {https://ui.adsabs.harvard.edu/abs/2019AJ....158...34D}
}

@ARTICLE{Hardcastle&Croston2020,
       author = {{Hardcastle}, M.~J. and {Croston}, J.~H.},
        title = "{Radio galaxies and feedback from AGN jets}",
      journal = {\nar},
         year = 2020,
        month = jun,
       volume = {88},
          eid = {101539},
        pages = {101539},
          doi = {10.1016/j.newar.2020.101539},
archivePrefix = {arXiv},
       eprint = {2003.06137},
 primaryClass = {astro-ph.HE},
       adsurl = {https://ui.adsabs.harvard.edu/abs/2020NewAR..8801539H}
}

\begin{appendix} 

\section{Line profiles and moment maps}\label{app:line_figs}

This appendix provides the full set of line profiles and moment maps for Planck-68 (Fig.~\ref{fig:Planck68_all}), Planck-89 (Fig.~\ref{fig:Planck89_all}), and Planck-188 (Fig.~\ref{fig:Planck188_all}), excluding Planck-41, which is presented in the main text as a representative case (Fig.~\ref{fig:Planck41_all}).

\begin{figure*}[p]
\centering

\begin{subfigure}{0.30\textwidth}\includegraphics[width=\linewidth]{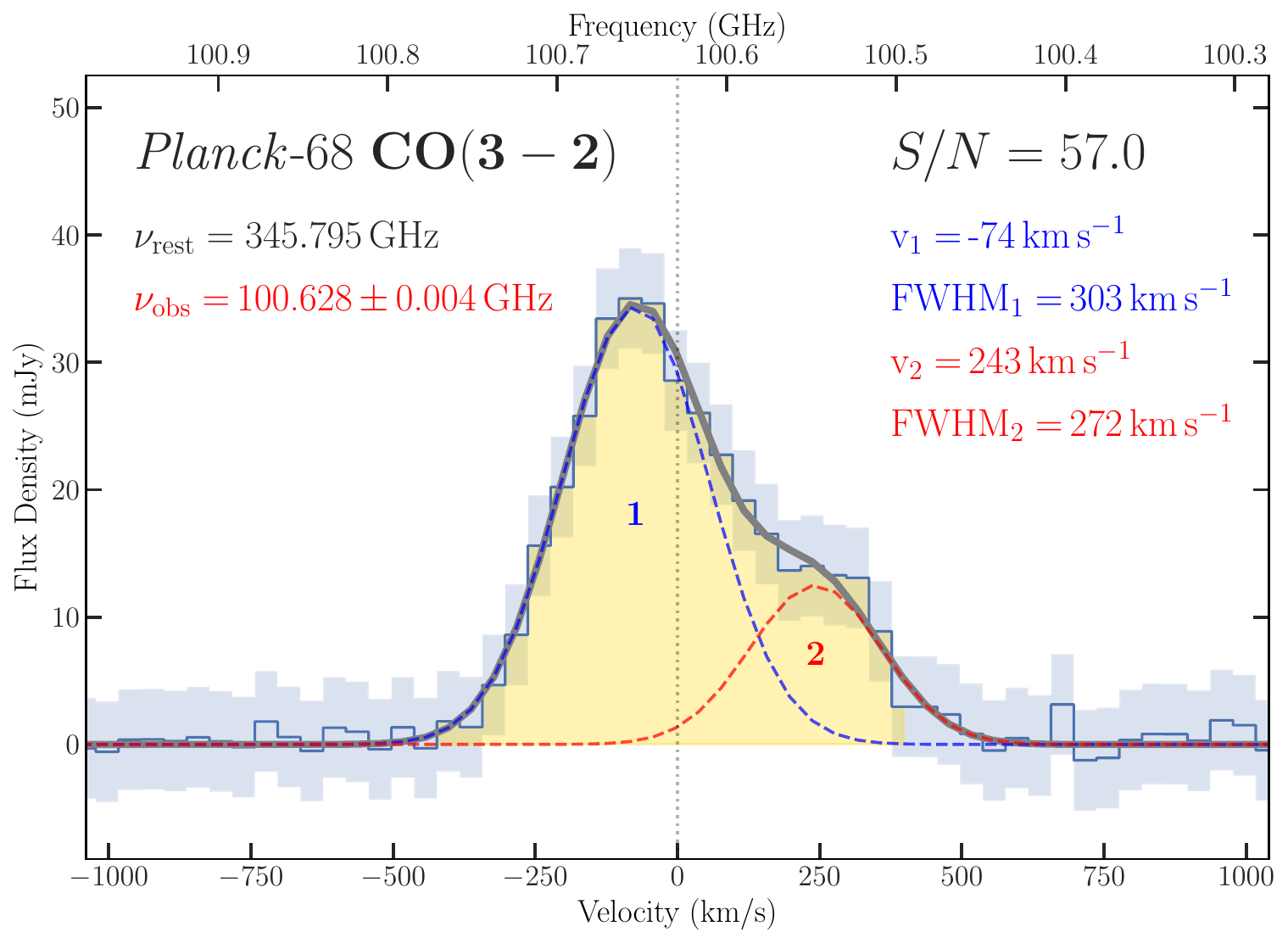}\end{subfigure}\hfill
\begin{subfigure}{0.30\textwidth}\includegraphics[width=\linewidth]{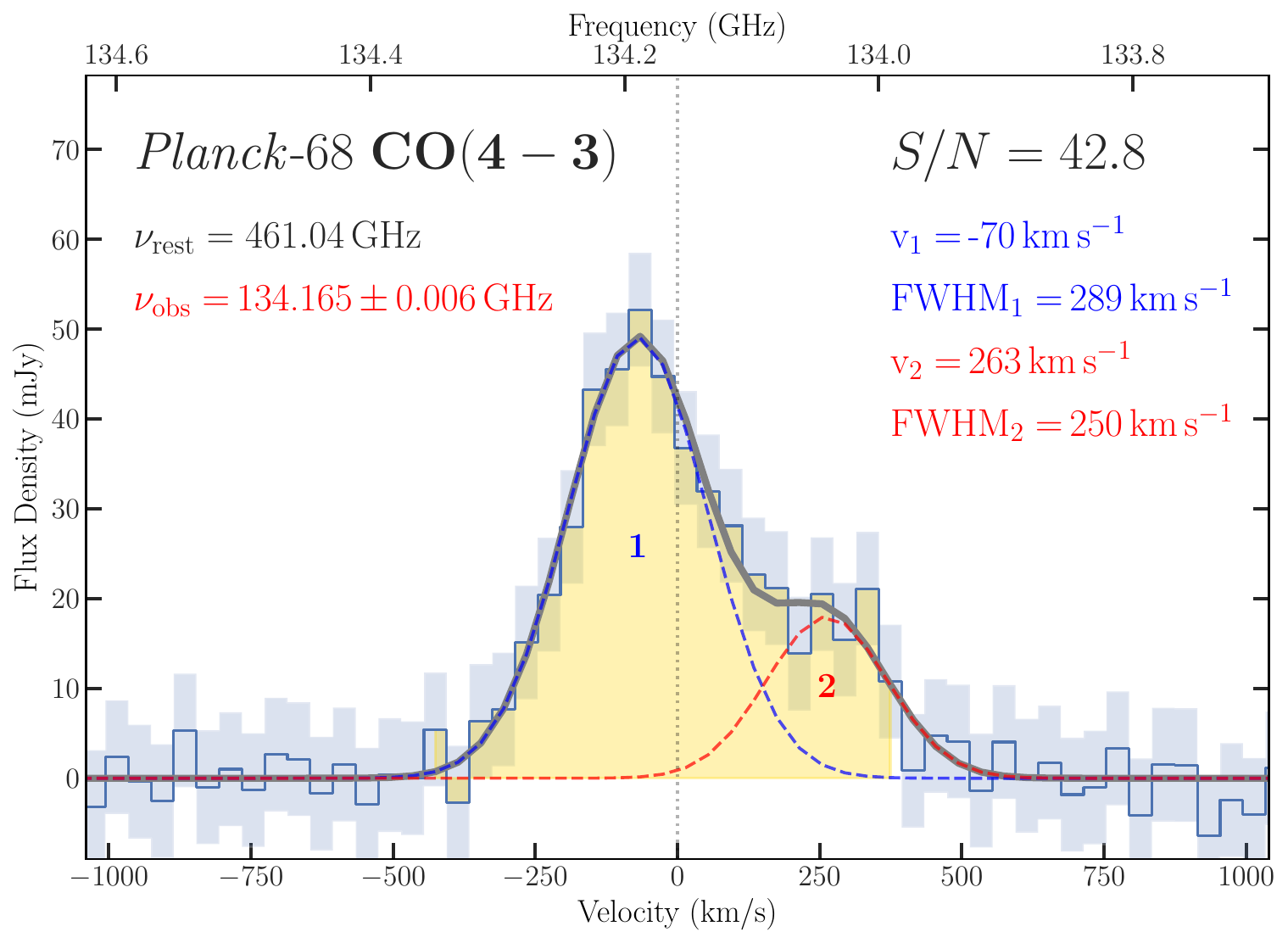}\end{subfigure}\hfill
\begin{subfigure}{0.30\textwidth}\includegraphics[width=\linewidth]{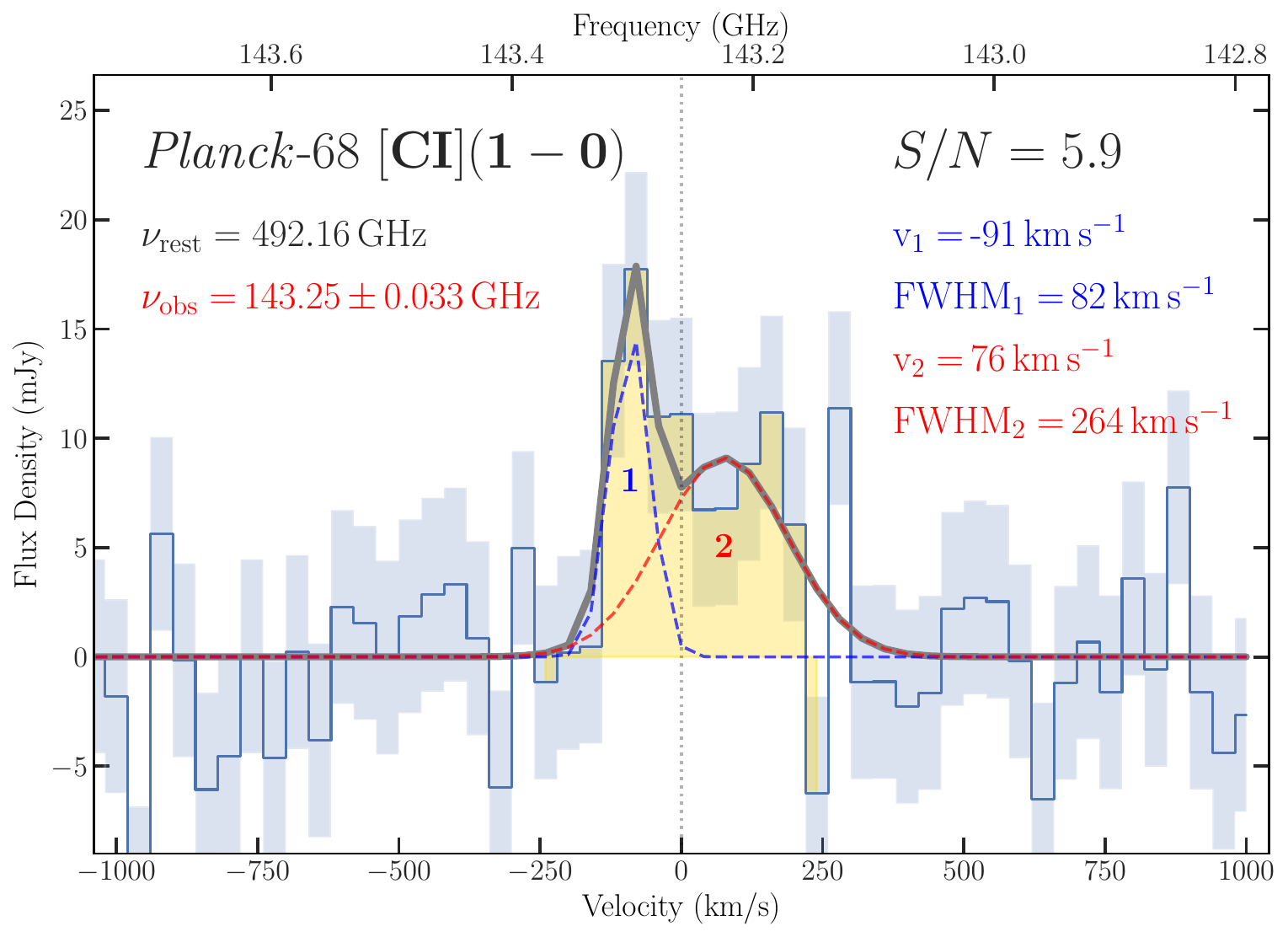}\end{subfigure}

\vspace{0.05cm}

\begin{subfigure}{0.30\textwidth}\includegraphics[width=\linewidth]{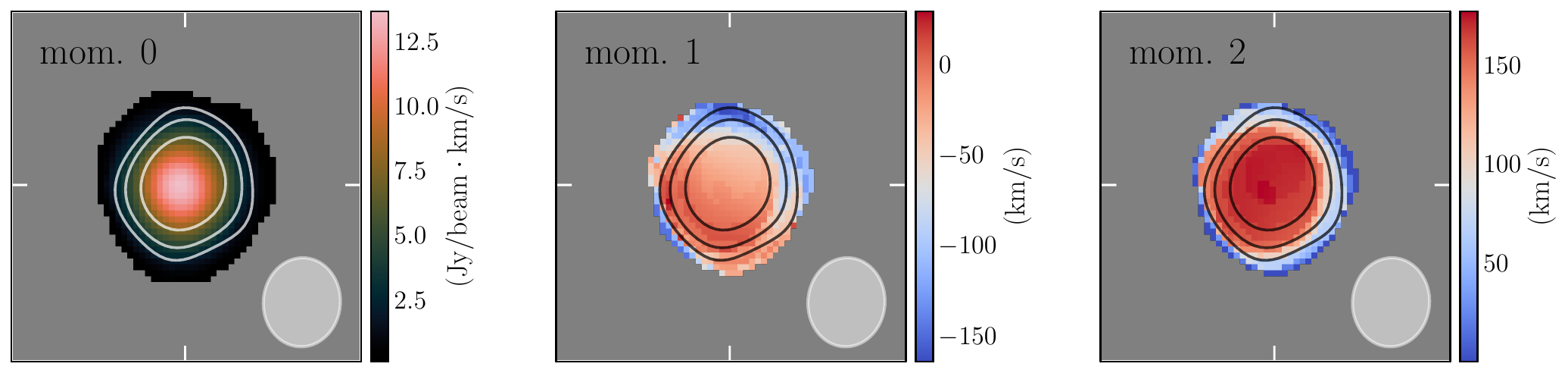}\end{subfigure}\hfill
\begin{subfigure}{0.30\textwidth}\includegraphics[width=\linewidth]{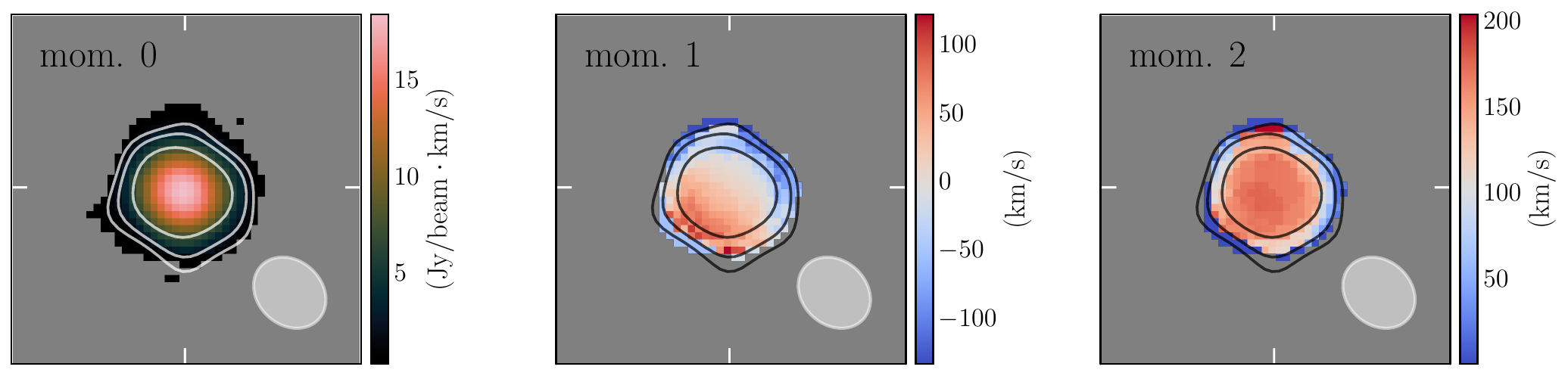}\end{subfigure}\hfill
\begin{subfigure}{0.30\textwidth}\includegraphics[width=\linewidth]{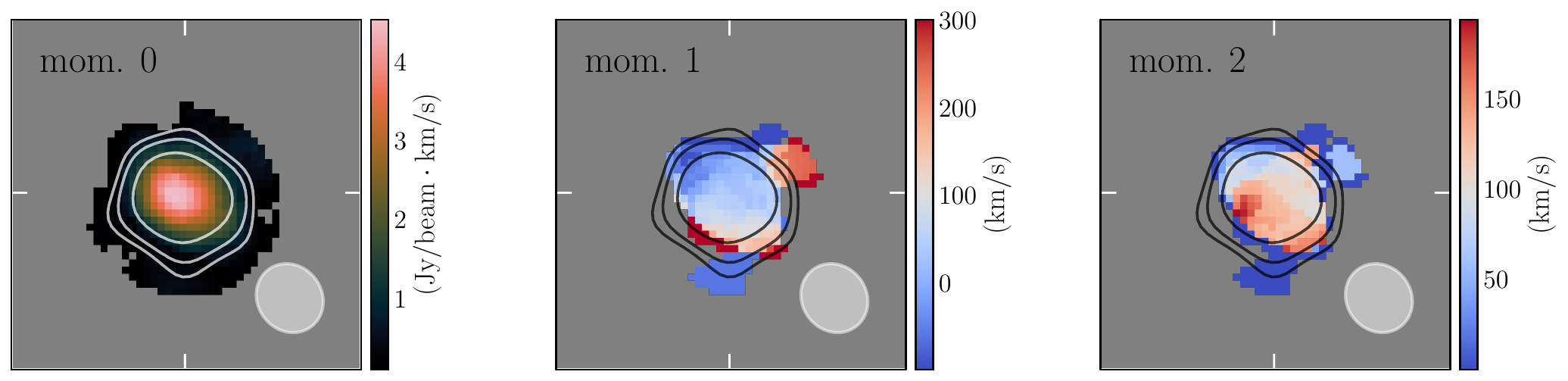}\end{subfigure}

\caption{Planck-68 line analysis: CO(3--2), CO(4--3) and [CI](1--0). Top panel: line profiles with derived spectroscopic quantities. The bottom panels show the moment 0th, 1st and 2nd maps. The white circles on the bottom right of the lower panels represent the synthesised beam, while the contours refer to the continuum emission in the same band at 3, 5 and 10$\sigma$.}\label{fig:Planck68_all}
\vspace{0.2cm}

\begin{subfigure}{0.46\linewidth}\includegraphics[width=\linewidth]{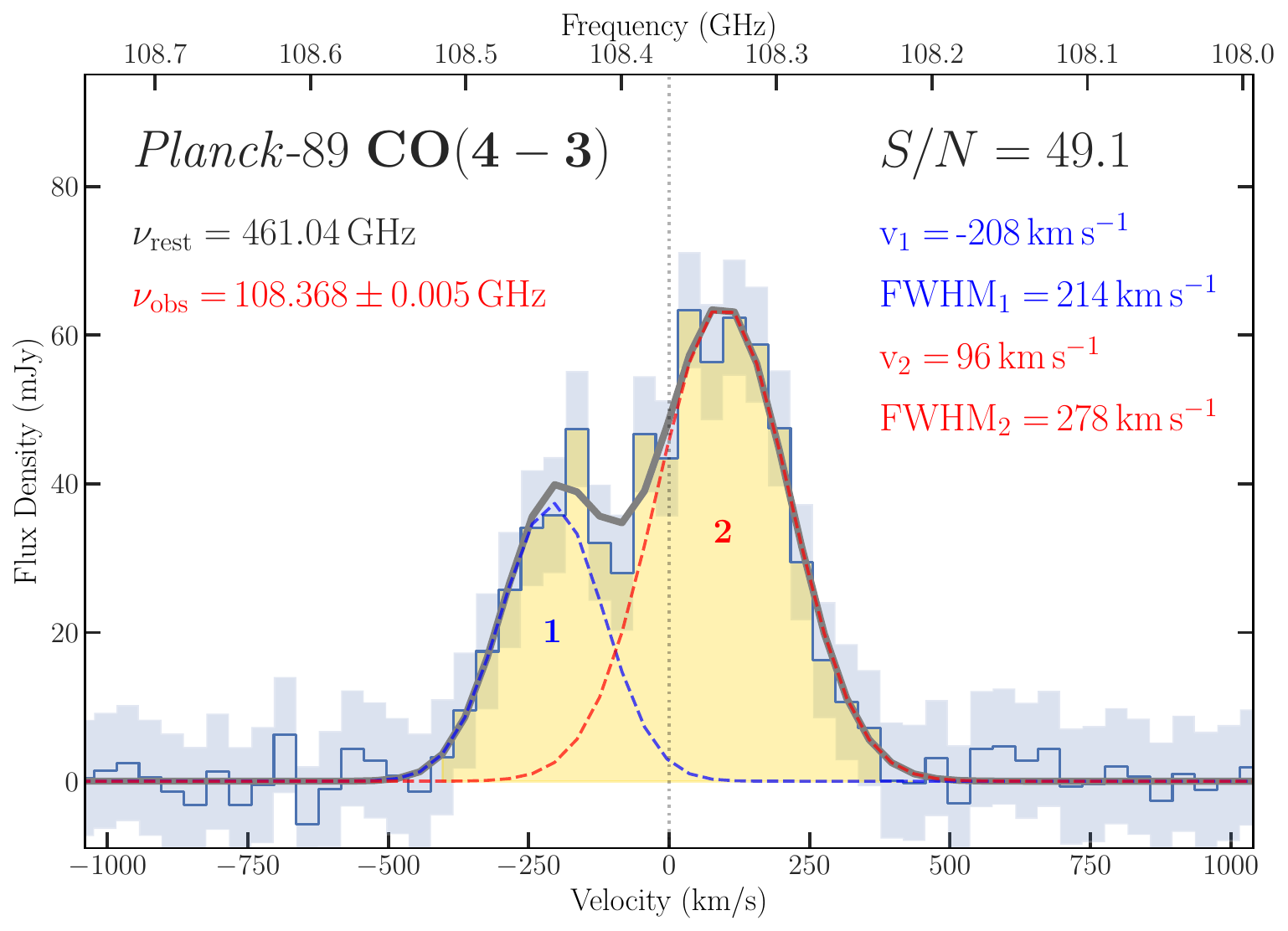}\end{subfigure}\hfill
\begin{subfigure}{0.46\linewidth}\includegraphics[width=\linewidth]{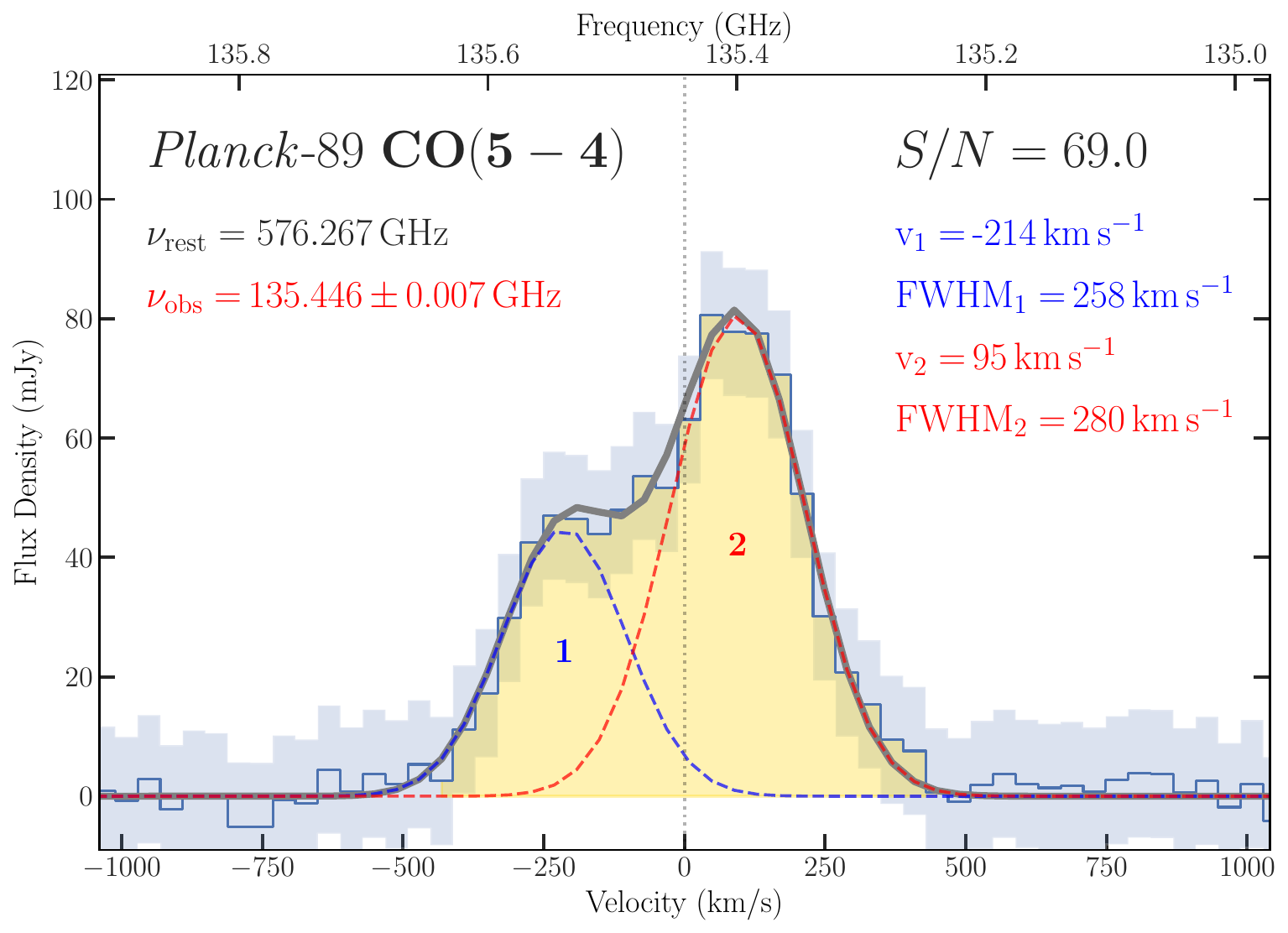}\end{subfigure}

\vspace{0.05cm}

\begin{subfigure}{0.46\linewidth}\includegraphics[width=\linewidth]{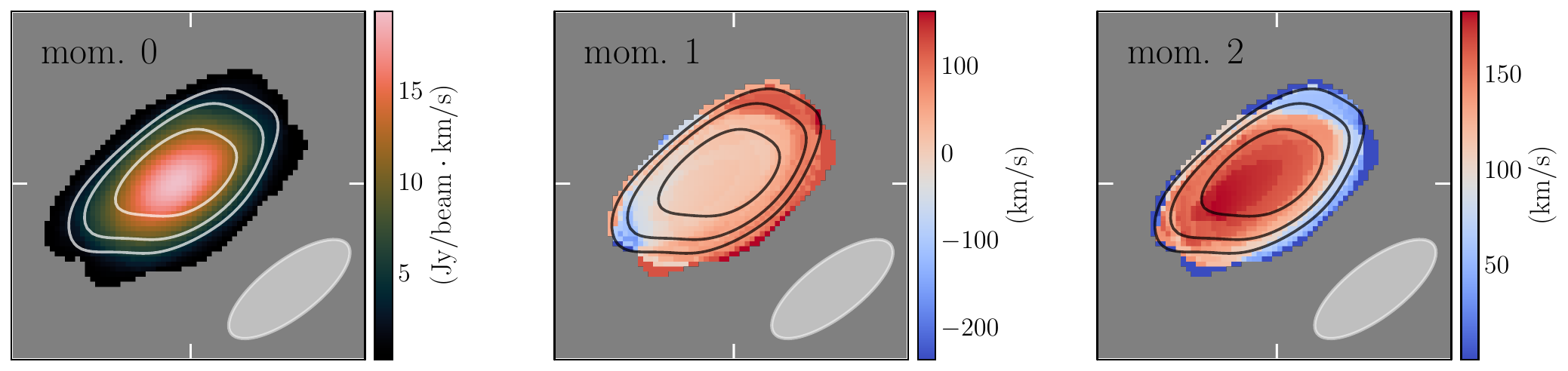}\end{subfigure}\hfill
\begin{subfigure}{0.46\linewidth}\includegraphics[width=\linewidth]{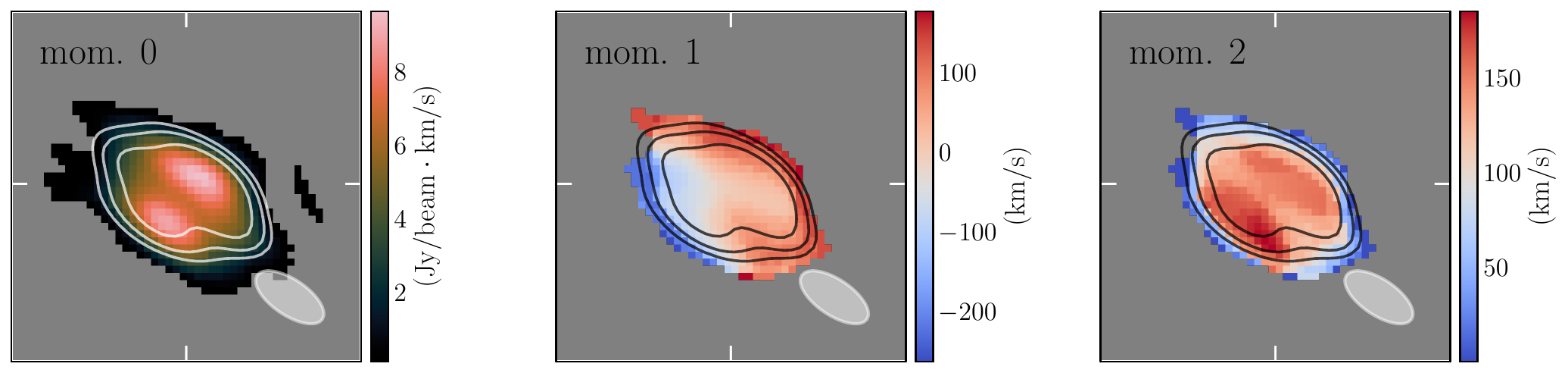}\end{subfigure}

\caption{Same as Fig.~\ref{fig:Planck68_all}, but for Planck-89.}\label{fig:Planck89_all}
\vspace{0.2cm}

\begin{subfigure}{0.46\linewidth}\includegraphics[width=\linewidth]{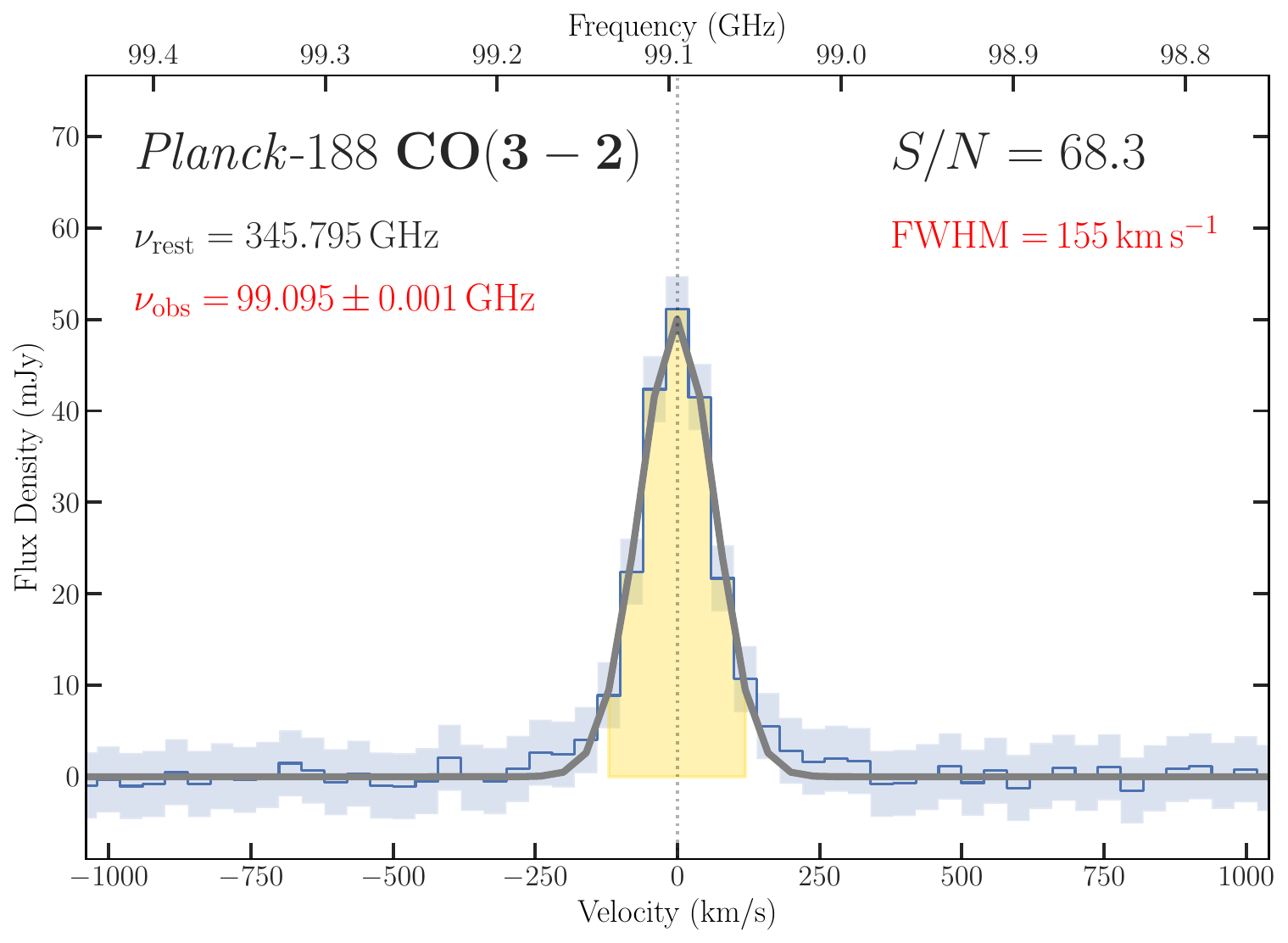}\end{subfigure}\hfill
\begin{subfigure}{0.46\linewidth}\includegraphics[width=\linewidth]{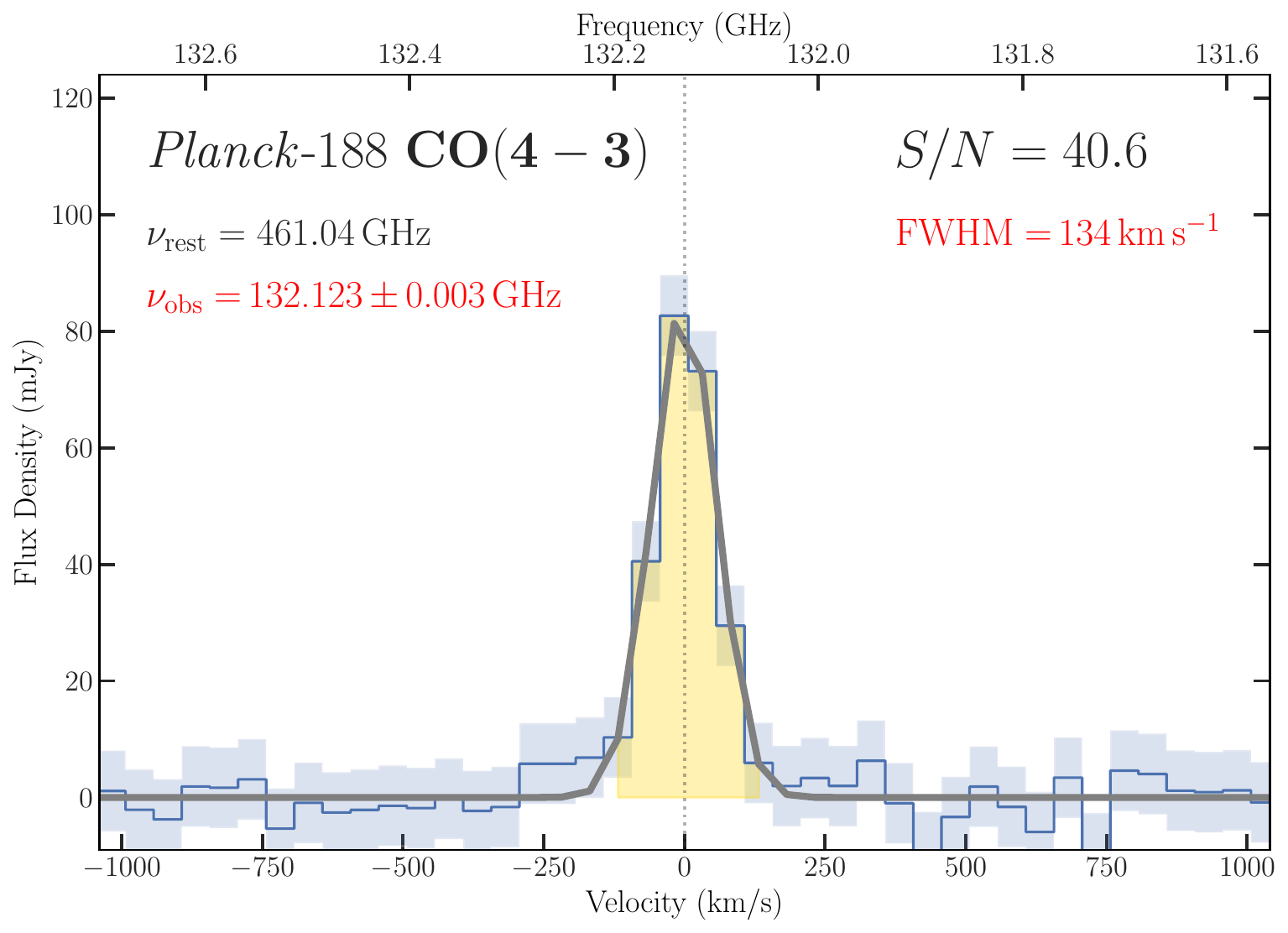}\end{subfigure}

\vspace{0.05cm}

\begin{subfigure}{0.46\linewidth}\includegraphics[width=\linewidth]{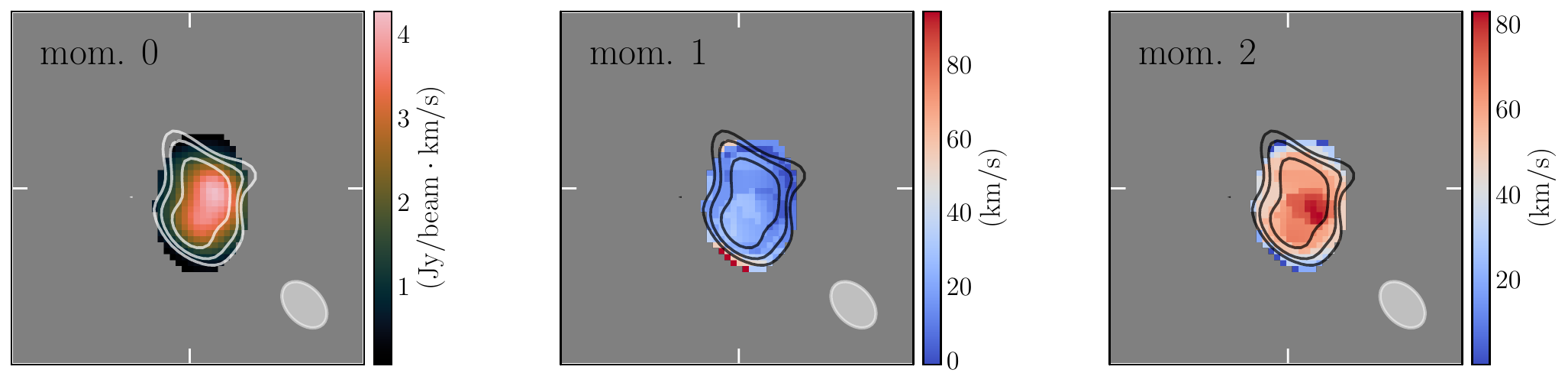}\end{subfigure}\hfill
\begin{subfigure}{0.46\linewidth}\includegraphics[width=\linewidth]{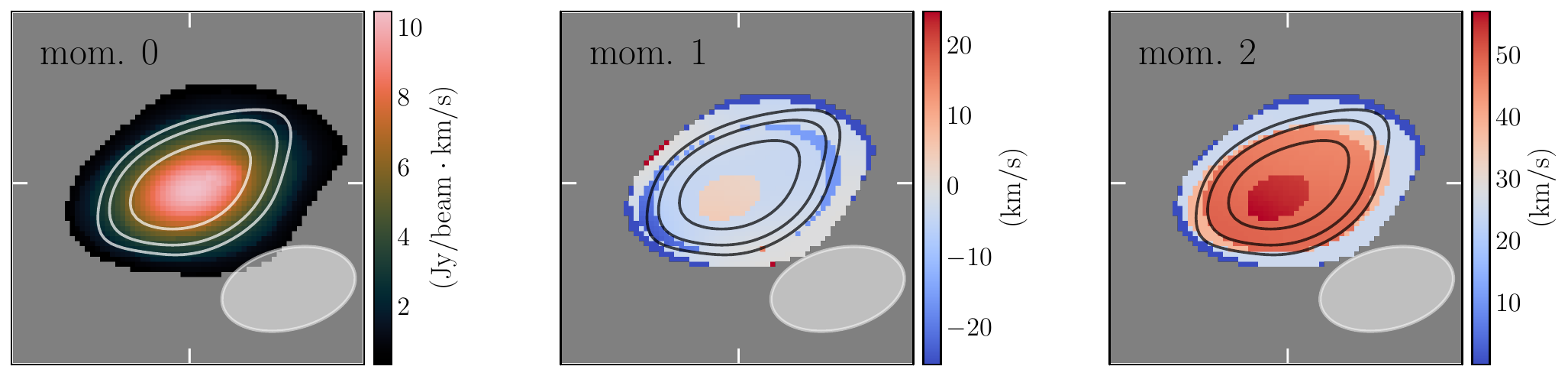}\end{subfigure}

\caption{Same as Fig.~\ref{fig:Planck68_all}, but for Planck-188.}\label{fig:Planck188_all}

\end{figure*}
\section{Lens modelling pipeline and parameters}\label{app:lens_parameters}

In this appendix, we report in detail the results of the lens modelling performed on the only source with resolved lens features in the NOEMA sample: Planck-41. 

Our analysis follows the ``Source, Light, and Mass'' (\texttt{SLaM}) pipeline strategy implemented in \texttt{PyAutoLens}. This framework adopts a chained modelling approach, in which a sequence of increasingly complex models is fitted through a series of non-linear searches. Each step uses the posterior probability distributions of the previous stage to initialise the priors of the subsequent one, allowing an efficient and stable exploration of the high-dimensional parameter space characteristic of strong-lensing analyses. 
In the first step, we performed a simplified parametric fit aimed at establishing a robust initial model and constraining the global properties of the lens mass distribution. In this phase, the lens mass was modelled as an isothermal profile with the centre fixed at the PanSTARRS $r$-band peak position, while the source light was described by a \citet{Sersic1963} profile, treated as a linear light component in \texttt{PyAutoLens}, where only the intensity is solved linearly while the S\'ersic index $n$ and the structural parameters are free. Fixing the mass centre at this stage reduces parameter degeneracies and allows the model to converge efficiently toward the correct region of the parameter space. This initial step provides well-constrained posterior distributions that are subsequently used to initialise the priors of the second modelling stage.

In the second step, we refined the mass model by allowing for a more general \texttt{PowerLaw} density profile, in which the logarithmic slope of the mass distribution is a free parameter. In this phase, the lens mass centre was allowed to vary, and all relevant parameters were optimised simultaneously. The source light model was carried over from the previous step, ensuring continuity and stability in the chained optimisation. Parameter estimation and posterior sampling were performed using the nested sampling algorithm \texttt{Nautilus} \citep{nautilus}, which provides an efficient exploration of the posterior distributions and robust Bayesian evidence estimates.

The foreground lens galaxy was ultimately modelled as a Singular Isothermal Ellipsoid (SIE; \citealt{kormann94}), implemented within \texttt{PyAutoLens} as an elliptical power law mass profile. The model is fully described by the Einstein radius $\theta_{\rm E}$, the lens center coordinates $(x^{\rm lens}, y^{\rm lens})$, the ellipticity components $(e_x, e_y)$, and the logarithmic slope $\gamma$. The ellipticity components were parametrised in terms of the axis ratio $q$ and the position angle $\phi$ via
\begin{equation}
    e_x = f \cdot \cos(2\phi) \quad \text{and} \quad e_y = f \cdot \sin(2\phi),
\end{equation}
where $f = (1 - q)/(1 + q)$. The lens redshift was fixed to the photometric value throughout the modelling.

The intrinsic light distribution of the background source was modelled using a parametric core-S\'ersic profile.
This profile is characterised by the effective radius $R_{\rm eff}$, the S\'ersic index $n$, the centroid coordinates $(x^{\rm source}, y^{\rm source})$, and the ellipticity components $(\epsilon_1, \epsilon_2)$.
It extends the standard S\'ersic model by introducing an inner core through three additional parameters: the break radius $R_{\rm break}$, which marks the transition between the inner core and the outer S\'ersic envelope; the inner slope $\beta$, which sets the logarithmic gradient of the surface brightness within $R_{\rm break}$ ($\beta = 0$ corresponding to a flat core, increasing values to a progressively steeper cusp); and the sharpness parameter $\alpha$, which controls the abruptness of the transition between the two regimes.
These three parameters were kept fixed at their standard values ($\alpha = 3.0$, $\beta = 0.25$, $R_{\rm break} = 0\farcs025$) to avoid degeneracies with $R_{\rm eff}$ and $n$ at the angular resolution of the NOEMA data, since the inner core is unresolved and its detailed shape is not constrained by the observations. This parametric description offers a physically motivated and computationally efficient representation of the source light, well-suited for capturing the global morphology of the lensed galaxy, given the angular resolution and signal-to-noise ratio of the data.

The median values of the posterior distributions for the lens and source parameters, along with their 68\% confidence intervals, are reported in Table~\ref{tab:lens_results}. We quote median values rather than maximum-likelihood estimates, as they are less sensitive to local likelihood maxima and provide a more representative summary of the inferred parameter distributions. 

\begin{table}[t]
\centering
\caption{Lens model parameters for Planck-41.}
\label{tab:lens_results}
\renewcommand{\arraystretch}{1.1}
\begin{tabular}{lr}
\hline\hline
\multicolumn{2}{c}{Planck-41} \\
\hline\hline
\multicolumn{2}{c}{Lens Mass Model (SIE $\rightarrow$ PowerLaw)} \\
\hline
$y_{\rm lens}$ (arcsec)$^*$
  & $-0.01^{+0.05}_{-0.04}$ \\ 
$x_{\rm lens}$ (arcsec)$^*$
  & $-0.03^{+0.05}_{-0.04}$ \\ 
$e_x$
  & $0.22 \pm 0.03$ \\ 
$e_y$
  & $0.18 \pm 0.02$ \\
$\theta_{\rm E}$ (arcsec)
  & $3.74^{+0.17}_{-0.15}$ \\ 
$\gamma$
  & $1.87^{+0.09}_{-0.08}$ \\ 
$z_{\rm lens}$
  & $0.291$ \\ 
\hline
\multicolumn{2}{c}{Source Light Model (core-S\'ersic)} \\
\hline
$y_{\rm source}$ (arcsec)$^*$
  & $-0.65^{+0.04}_{-0.05}$ \\ 
$x_{\rm source}$ (arcsec)$^*$
  & $-0.36 \pm 0.04$ \\ 
$\epsilon_{\rm 1}$
  & $-0.34^{+0.10}_{-0.09}$ \\ 
$\epsilon_{\rm 2}$
  & $0.41^{+0.08}_{-0.09}$ \\ 
$R_{\rm eff}$ (arcsec)
  & $0.76^{+0.26}_{-0.23}$ \\ 
$n$
  & $3.8^{+0.7}_{-0.8}$ \\
$\alpha$
  & $3.0$ \\ 
$\beta$
  & $0.25$ \\ 
$R_{\rm break}$ (arcsec)
  & $0.025$ \\ 
$z_{\rm source}$
  & $2.3481$ \\
\hline
\multicolumn{2}{c}{Derived Physical Parameters} \\
\hline
$\theta_{\rm E}$ (kpc)$^{(a)}$   
    & $16.34^{+0.74}_{-0.67}$       \\ 
$M_{\rm E}$ ($10^{11}\,M_\odot$)  
    & $20.4^{+0.1}_{-0.1}$ \\ 
$\mu_{\rm cont}$
  & $10.8^{+1.2}_{-1.3}$ \\ 
$R_{\rm eff}$ (kpc)$^{(b)}$
  & $6.4^{+2.2}_{-1.9}$ \\ 
\hline
\end{tabular}
\vspace{3pt}
\begin{flushleft}
\footnotesize
Median values of the posterior distributions with their 68\% confidence level errors. Values without uncertainties are kept fixed in the fit.\\
(*) The $x$ and $y$ coordinates are the arcsec offset with respect to the initial fixed mass distribution position (R.A. = 17:45:15.7730, Dec = +40:31:02.606)\\
(a) Einstein radius in kpc computed at the lens redshift.\\
(b) Effective S\'ersic radius in kpc computed at the source redshift.
\end{flushleft}
\end{table}

\end{appendix}

\end{document}